\documentclass[3p,fleqn]{elsarticle}

\usepackage{lineno,hyperref,xparse}
\hypersetup{
  colorlinks   = true,    
  urlcolor     = blue,    
  linkcolor    = blue,    
  citecolor    = blue     
}

\journal{Journal of \LaTeX\ Templates}
\usepackage{graphicx,epsfig,epstopdf}
\usepackage[numbers]{natbib}
\usepackage{array}
\newcolumntype{M}[1]{>{\centering\arraybackslash}m{#1}}
\usepackage{multirow}
\usepackage{float}
\usepackage{amssymb,amsmath,amsthm,empheq,bm}
\usepackage{xcolor}
\usepackage{tkz-euclide}
\usepackage{arcs}
\usepackage{xfrac}
\usepackage{tikz}
\usepackage{siunitx}
\makeatother
\usepackage{caption}
\usepackage{subcaption}

\usepackage{cleveref}
\Crefname{figure}{Fig.}{Figs.}
\crefname{equation}{equation}{equations}
\crefrangeformat{equation}{equations #3(#1)#4--#5(#2)#6}
\Crefrangeformat{figure}{Figs. #3#1#4--#5#2#6}
\crefrangeformat{figure}{figs. #3#1#4--#5#2#6}
\newcommand{\al}{\alpha}
\newcommand{\be}{\beta}

\newcommand{\dt}{\delta t}

\newcommand{\overbar}[1]{\mkern 1.5mu\overline{\mkern-1.5mu#1\mkern-1.5mu}\mkern 1.5mu}

\makeatletter
\def\ps@pprintTitle{%
 \let\@oddhead\@empty
 \let\@evenhead\@empty
 \def\@oddfoot{}%
 \let\@evenfoot\@oddfoot}
\makeatother

\begin{document}

\begin{frontmatter}

\title{A new transient higher order compact scheme for computation of flow and heat transfer in nonuniform polar grids}

\author[add1,add2]{Dharmaraj Deka}
\ead{dharma@tezu.ernet.in}
\author[add1]{Shuvam Sen\corref{correspondingauthor}}
\ead{shuvam@tezu.ernet.in}

\cortext[correspondingauthor]{Corresponding author}
\address[add1]{Department of Mathematical Sciences, Tezpur University, Tezpur 784028, India}
\address[add2]{Department of Mathematics, Kamrup College, Chamata 781306, India}

\begin{abstract}
In this work, a higher order compact (HOC) discretization is developed on the nonuniform polar grid. The discretization conceptualized using the unsteady convection-diffusion equation (CDE) is further extended to flow problems governed by the Navies-Stokes (N-S) equations as well as the Boussinesq equations. The scheme developed here combines the advantages of body-fitted mesh with grid clustering, thereby making it efficient to capture flow gradients on polar grids. The scheme carries a spatial convergence of order three with temporal order of convergence being almost two. Diverse flow problems are being investigated using the scheme. Apart from a verification studies we validate the scheme by time marching simulation for the benchmark problem of driven polar cavity and the problem of natural convection in the horizontal concentric annulus. In the process, a one-sided approximation for the Neumann boundary condition for vorticity is also presented. Finally, the benchmark problem of forced convection around a circular cylinder is tackled. The results obtained in this study are analyzed and compared with the well-established numerical and experimental data wherever available in the literature. The newly developed scheme is found to generate accurate solutions in each case. 
\end{abstract}


\end{frontmatter}

\let\clearpage\relax
\section{Introduction}

Fluid flow problems involving circular geometries have attracted a great deal of attention over the years owing to their theoretical significance and physical relevance. For such importance, HOC schemes have been developed on the cylindrical polar coordinates. Preliminary works to develop HOC schemes on cylindrical polar coordinate systems mostly focused on the Poisson equation and uniform grids \citep{IyengarManohar1987,JainJainKrishna1994,BorgesDaripa2001,LaiWang2001,ZhuangSun2001,Lai2002}. Here it is important to note that in polar coordinates Poisson equation involves variable coefficients and thus the above-mentioned works involved significant endeavors and were not mere extension of the works done in the Cartesian coordinates. Additionally, a valid discretization at singularity $r=0$ is necessary. Subsequently, considerable efforts could be found in the literature on HOC discretization of N-S equations in the body-fitted polar coordinate system \cite{SanyasirajuManjula2005,KalitaRay2009,RayKalita2010,YuTian2013b,DasPanditRay2023}. In the year 2005 Sanyasiraju and Manjula \citep{SanyasirajuManjula2005} proposed a higher order semicompact scheme to solve the flow past an impulsively started circular cylinder. The authors used a wider stencil to discretize a few terms of the governing equation rendering semicompactness to the scheme. This strategy also helped alleviate challenges associated with the variable coefficients of the second order derivative. Subsequently, Kalita and Ray \citep{KalitaRay2009} developed a temporally second order and spatially third order accurate HOC scheme for streamfunction-vorticity $(\psi-\omega)$ formulation of unsteady N-S equations. As the authors worked with a modified differential equation the resulting difference equation carried a higher order of singularity. The scheme was specially adapted to simulate incompressible flow past a circular cylinder directly on nonuniform polar grids. A rather straightforward adaptation of this scheme for steady CDE can be found in \citep{RayKalita2010}. In the year 2013, the pioneering work on compact difference schemes for the pure streamfunction formulation of N-S equations in polar coordinates was reported by Yu and Tian \citep{YuTian2013b}. The authors here worked with the steady biharmonic equation in polar coordinates. The scheme developed therein is of second order accuracy and carries streamfunction and its first derivatives as the unknown variables. Of late Das \textit{et al.} \citep{DasPanditRay2023} also worked with steady second order equation with variable coefficients in polar coordinates and introduced a third order accurate HOC scheme. The scheme claimed to be implemented on a nonuniform grid used an implicit form of first order derivatives and was successful in simulating steady incompressible flows. It is clear from the above discussion that classical HOC discretization of transient generalized second order PDE with variable coefficients in polar coordinates has not been attempted apart from the work of Kalita and Ray \citep{KalitaRay2009} on $\psi-\omega$ form of the N-S equations, whose extension to other governing equations such as the Boussinesq equation is not immediate. Additionally, it is intriguing to notice that developed HOC schemes in polar coordinates have been traditionally used to handle the nonuniformity of grids in radial direction only. However, heat and fluid flow problems are often associated with generation of steep gradients of streamfunction and vorticity near the bluff body or domain boundary walls, and as such grid clustering in all directions should be economical. 

In this work, we intend to develop a new transient HOC formulation that can simulate fluid as well as heat transfer problems directly in polar coordinates. The scheme conceived in the process must be generalizable to approximate second order equations with variable coefficients in polar coordinates and should be efficient for circular geometries. Starting with a second order PDE with variable coefficients, we advocate a discretization strategy amenable for extension to a nonlinear system of N-S equations and beyond. Additionally, we adopt the philosophy of nonuniform grids for accurate resolution of complex flow problems. We look forward to combining the virtues of Pad\'{e} approximation of first order derivatives on nonuniform grids with variable coefficients and compact discretization of higher order derivatives in terms of flow variables and their first order gradients. Theoretically, the scheme has a temporal convergence of second order and a spatial convergence of order three. The scheme developed here leads to stable higher order discretization in conjunction with both Dirichlet and convective boundary conditions. The formulation is validated by applying it to the problem of an unsteady Gaussian pulse governed by the linear CDE. This problem is equipped with known analytical solution, which helps in error analysis. The spatial and temporal order of convergences are also established during this process. However, to comprehend the robustness and adaptability of this new scheme, we carry out the simulations for benchmark problems of both flow and heat transfer, namely the flow inside a driven polar cavity, natural convection inside a circular annulus, and forced convection around a stationary circular cylinder. These flows are governed by N-S equations and Boussinesq equations, both the equations being cast in transient incompressible form. The $\psi-\omega$ formulation of these coupled nonlinear PDE's  are conformed for the present computation.   

The rest of this manuscript is organised in four sections. In \cref{discretization} compact discretization of transient CDE on cylindrical polar coordinate grids. \Cref{algorithm} deals with solution of algebraic system of equations. The four test cases are described in the \cref{testcase} and finally, \cref{conclusion} gives a gist of the whole work.

\section{Mathematical formulation and discretization procedure}\label{discretization}

In the nonrectangular domain $\Omega=[a_r,b_r]\times[a_\theta,b_\theta]$, the polar $(r,\theta)$ form of the CDE is 
\begin{equation}\label{CDE_pol}
  a \phi_t-\phi_{rr}-\frac{1}{r^2}\phi_{\theta \theta}+c_1(r,\theta,t) \phi_r + c_2(r,\theta,t) \phi_\theta=f(r,\theta,t),
\end{equation}
where $a>0$ is a constant; $c_1$ and $c_2$ are convection coefficients. 

\begin{figure}[!h]
\centering
  \begin{subfigure}[b]{0.45\textwidth}
   \centering
        \includegraphics[width=\linewidth]{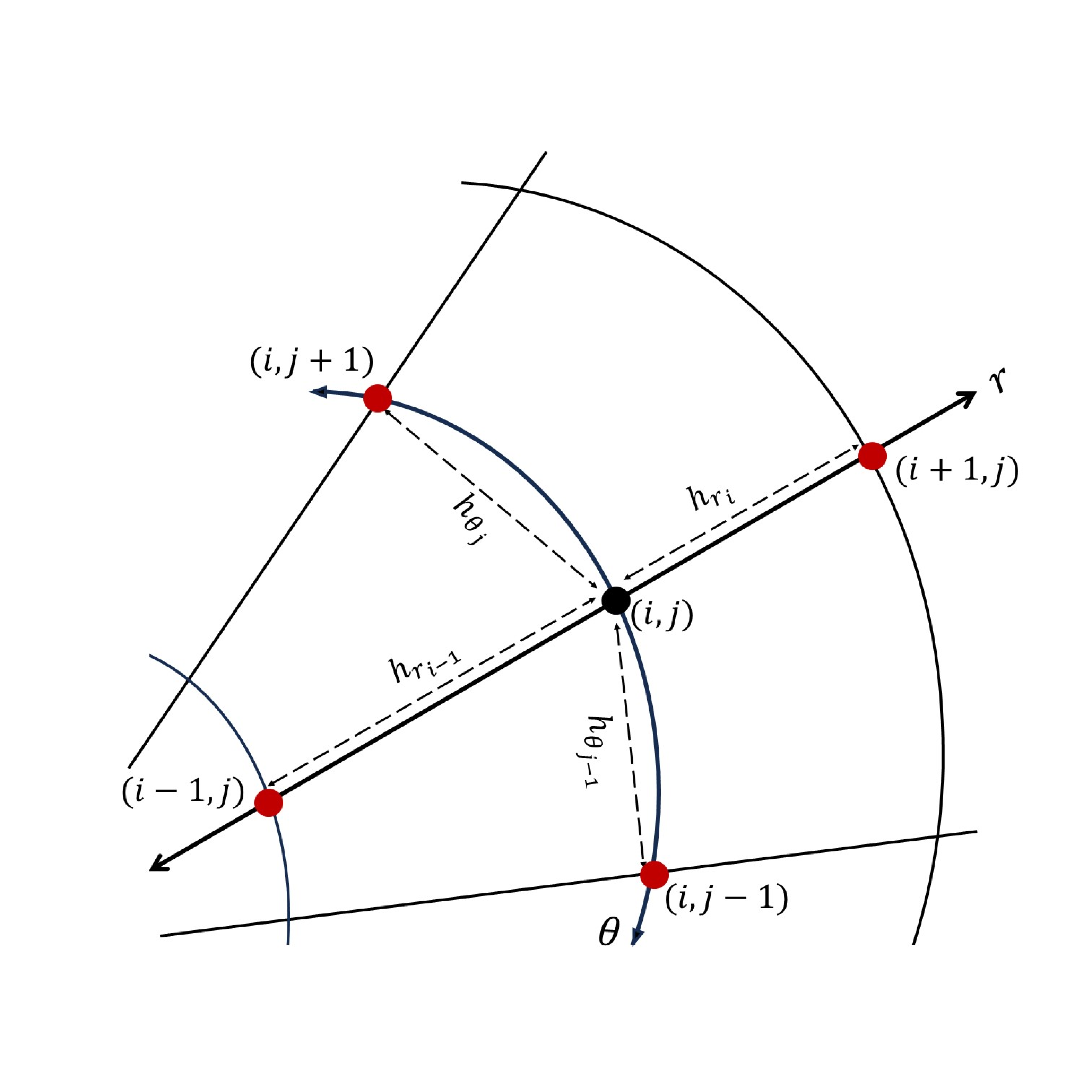}
    \end{subfigure}
  \begin{minipage}[b]{0.04\textwidth}
  \subcaption{ }\label{nonuniform1}
  \end{minipage}
  \begin{subfigure}[b]{0.45\textwidth}
   \centering
        \includegraphics[width=\linewidth]{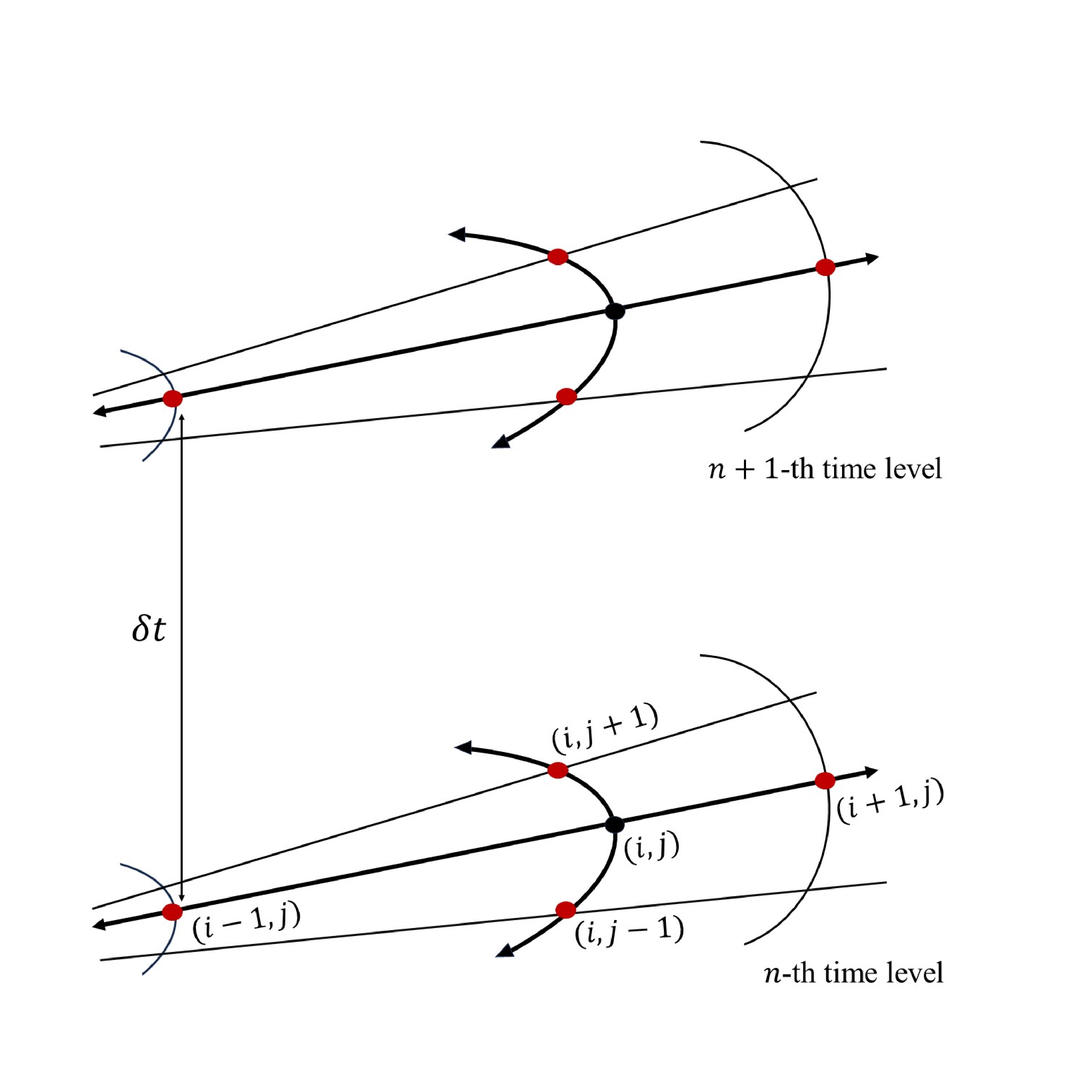}
  \end{subfigure}
  \begin{minipage}[b]{0.04\textwidth}
  \subcaption{ }\label{nonuniform2}
  \end{minipage}
\caption{(a) Projection of nonuniform computational stencil in $(r,\theta)$ plane and (b) Nonuniform computational stencil in $(r,\theta,t)$ hyper-plane. }\label{fig:1}
\end{figure}
We discretize the domain by considering $a_r=r_1<r_2<r_3<\cdots<r_{n_r}=b_r$ and $a_\theta=\theta_1<\theta_2<\theta_3<\cdots<\theta_{n_\theta}=b_\theta$ where lengths between two consecutive $r_i$'s, $i=1,2,3,\dots,n_r$ and $\theta_j$'s, $j=1,2,3,\dots,n_\theta$ may be unequal. A typical stencil can be seen in \cref{fig:1}. The figure depicts the grid points at the $n$-th and $(n+1)$-th time level as well. We define the mesh sizes along radial and tangential directions to be
\begin{align*}
  h_{r_i}&=r_{i+1}-r_i, \quad i\in{1,2,3,\dots,n_r-1}, \\
  h_{\theta_j}&=\theta_{j+1}-\theta_j, \quad j\in{1,2,3,\dots,n_\theta-1}.
\end{align*}

For the sufficiently smooth transport variable $\phi(r,\theta)$, the first and second order finite difference operator $\delta_r$ and $\delta_{rr}$ in the radial direction are defined as,
\begin{equation}\label{dr}
\delta_r \phi_{i,j}=\frac{1}{h_i+h_{i-1}} \left(\phi_{i+1,j}-\phi_{i-1,j}\right)
\end{equation}
and
\begin{equation}\label{drr}
\delta_{rr} \phi_{i,j}=\frac{2}{h_{r_i}+h_{r_{i-1}}} \left[\frac{\phi_{i+1,j}}{h_{r_i}}+\frac{\phi_{i-1,j}}{h_{r_{i-1}}}-\left(\frac{1}{h_{r_i}}+\frac{1}{h_{r_{i-1}}}\right)\phi_{i,j} \right]
\end{equation}
respectively, where $\phi_{i,j}$ denotes $\phi(r_i,\theta_j)$. The first and second order partial derivatives of $\phi$ along $r$-direction at a point $(r_i,\theta_j)$ lying inside the reference domain can be approximated using Taylor series expansion as
\begin{equation}\label{phir_taylor}
\begin{aligned}
  \partial_r \phi_{i,j}=\delta_r \phi_{i,j}-\frac{1}{2} \left(h_{r_i}-h_{r_{i-1}}\right) \partial_{rr} \phi_{i,j}-\frac{1}{6}\left(\frac{h_{r_i}^3+h_{r_{i-1}}^3}{h_{r_i}+h_{r_{i-1}}}\right)\partial_{rrr} \phi_{i,j}
  &-\frac{1}{24}\left(\frac{h_{r_i}^4-h_{r_{i-1}}^4}{h_{r_i}+h_{r_{i-1}}}\right)\partial_{rrrr} \phi_{i,j}\\ &+\mathcal{O}\left(\frac{h_{r_i}^5+h_{r_{i-1}}^5}{h_{r_i}+h_{r_{i-1}}} \right),
\end{aligned}
\end{equation}
and
\begin{equation}\label{phirr_taylor}
\begin{aligned}
  \partial_{rr}\phi_{i,j}=\delta_r^2 \phi_{i,j}-\frac{1}{3}\left(h_{r_i}-h_{r_{i-1}}\right) \partial_{rrr} \phi_{i,j} -\frac{1}{12}\left(\frac{h_{r_i}^3+h_{r_{i-1}}^3}{h_{r_i}+h_{r_{i-1}}}\right)\partial_{rrrr}\phi_{i,j}
  &-\frac{1}{60}\left(\frac{h_{r_i}^4-h_{r_{i-1}}^4}{h_{r_i}+h_{r_{i-1}}}\right)\partial_{rrrrr} \phi_{i,j}\\ &+\mathcal{O}\left(\frac{h_{r_i}^5+h_{r_{i-1}}^5}{h_{r_i}+h_{r_{i-1}}}\right).
\end{aligned}
\end{equation}

Substituting $\partial_{rrr} \phi_{i,j}$ from \cref{phir_taylor} in \cref{phirr_taylor}, we can obtain the following second order approximation for the second-order derivative given by 
\begin{equation}\label{phirr1}
\begin{aligned}
    \partial_{rr}\phi_{i,j}=\frac{h_{r_i}^2-h_{r_i}h_{r_{i-1}}+h_{r_{i-1}}^2}{h_{r_i}h_{r_{i-1}}}\delta_r^2\phi_{i,j}-\frac{2(h_{r_i}-h_{r_{i-1}})}{h_{r_i}h_{r_{i-1}}}(\delta_r \phi_{i,j}-\phi_{r_{i,j}})
    &-\frac{1}{12} h_{r_i} h_{r_{i-1}}\partial_{rrrr}\phi_{i,j}\\
    &+\mathcal{O}\left((h_{r_{i-1}}-h_{r_i})h_{r_i} h_{r_{i-1}}\right).
\end{aligned}
\end{equation}
It is easy to see that \cref{phirr1} reverts back to standard central approximation of order two on uniform grid.

Further, \cref{phir_taylor} yeilds 
\begin{equation}\label{phirr2}
\begin{aligned}
  \partial_{rr} \phi_{i,j}=\delta_r \phi_{r_{i,j}}-\frac{1}{2} \left(h_{r_i}-h_{r_{i-1}}\right) \partial_{rrr} \phi_{i,j}-\frac{1}{6}\left(\frac{h_{r_i}^3+h_{r_{i-1}}^3}{h_{r_i}+h_{r_{i-1}}}\right)\partial_{rrrr} \phi_{i,j}
  &-\frac{1}{24}\left(\frac{h_{r_i}^4-h_{r_{i-1}}^4}{h_{r_i}+h_{r_{i-1}}}\right)\partial_{rrrrr} \phi_{i,j}\\ &+\mathcal{O}\left(\frac{h_{r_i}^5+h_{r_{i-1}}^5}{h_{r_i}+h_{r_{i-1}}} \right).
\end{aligned}
\end{equation}

\Cref{phirr_taylor,phirr2}, when simplified give rise to the difference operator $(2\delta_r^2 \phi_{i,j}-\delta_r \phi_{r_{i,j}})$ that carries a first-order truncation error for the second-order derivative and is given by
\begin{equation}\label{phirr3}
  \partial_{rr}\phi_{i,j}=(2\delta_r^2 \phi_{i,j}-\delta_r \phi_{r_{i,j}})-\frac{1}{6}(h_{r_i}-h_{r_{i-1}})\partial_{rrr}\phi_{i,j}+\mathcal{O}\left(\frac{h_{r_i}^4-h_{r_{i-1}}^4}{h_{r_i}+h_{r_{i-1}}}\right).
\end{equation}
This necessitates a further approximation of $\partial_{rrr}\phi_{i,j}$ in \cref{phirr3} and is carried out using \cref{phir_taylor} to arrive at
\begin{equation}\label{phirr4}
\begin{aligned}
  \partial_{rr} \phi_{i,j}
  =&2\left(\frac{h_{r_i}^2-h_{r_i} h_{r_{i-1}}+h_{r_{i-1}}^2}{h_{r_i}^2+h_{r_{i-1}}^2} (2\delta_r^2 \phi_{i,j}-\delta_r \phi_{r_{i,j}})\right.\\
   &\left.-\frac{h_{r_i}-h_{r_{i-1}}}{h_{r_i}^2+h_{r_{i-1}}^2} (\delta_r\phi_{i,j}-\phi_{r_{i,j}}) \right)+\frac{1}{12}(h_{r_i}-h_{r_{i-1}})^2\partial_{rrrr}\phi_{i,j}\\
   &+\mathcal{O}\left(\frac{(h_{r_i}-h_{r_{i-1}})(2h_{r_i}^4-2h_{r_i}^3h_{r_{i-1}}+3h_{r_i}^2h_{r_{i-1}}^2-2h_{r_i}h_{r_{i-1}}^3+2h_{r_{i-1}}^4)}{(h_{r_i}^2+h_{r_{i-1}}^2)} \right).
\end{aligned}
\end{equation}
Here, we note that the above second-order discretization not only attains optimal fourth-order convergence on uniform grid but also its leading truncation error term $\displaystyle \frac{1}{12}(h_{r_i}-h_{r_{i-1}})^2\partial_{rrrr}\phi_{i,j}$, as seen in \cref{phirr2}, converges to zero quadratically as we move from nonuniform to uniform grid. Next, eliminating the fourth-order derivative from \cref{phirr1,phirr4}, we arrive at the following discretization of the second-order derivative with third-order truncation term on nonuniform grid as,
\begin{equation}\label{phirr5}
\begin{aligned}
    \partial_{rr}\phi_{i,j}=&2\left\{\left(\frac{2h_{r_i}h_{r_{i-1}}}{h_{r_i}^2+h_{r_{i-1}}^2}+\frac{(h_{r_i}-h_{r_{i-1}})^2}{2h_{r_i} h_{r_{i-1}}}\right)\delta_r^2\phi_{i,j}-\frac{h_{r_i} h_{r_{i-1}}}{h_{r_i}^2+h_{r_{i-1}}^2}\delta_r \phi_{r_{i,j}}\right.\\
    &\left.-\frac{(h_{r_i}-h_{r_{i-1}})}{h_{r_i}^2-h_{r_i} h_{r_{i-1}}+h_{r_{i-1}}^2}\left(\frac{h_{r_i}h_{r_{i-1}}}{h_{r_i}^2+h_{r_{i-1}}^2}+\frac{(h_{r_i}-h_{r_{i-1}})^2}{h_{r_i} h_{r_{i-1}}} \right)(\delta_r \phi_{i,j}-\phi_{r_{i,j}})\right\}\\
    &+\mathcal{O}\left(\frac{h_{r_i} h_{r_{i-1}}(h_{r-i}^3-h_{r_i}^3)}{(h_{r_i}^2+h_{r_{i-1}}^2)} \right).
\end{aligned}
\end{equation}
It is quite evident that on uniform grid the above approximation shall revert back to compact fourth-order approximation. Compared to discretization in \cref{phirr4}, approximation in \cref{phirr5} posses a higher-order truncation term and the order of approximation is optimal on nonuniform compact stencil.

Again following a similar course of action in the $\theta$-direction we get
\begin{equation}\label{phithth}
\begin{aligned}
    \partial_{\theta\theta}\phi_{i,j}=&2\left\{\left(\frac{2h_{\theta_j}h_{\theta_{j-1}}}{h_{\theta_j}^2+h_{\theta_{j-1}}^2}+\frac{(h_{\theta_j}-h_{\theta_{j-1}})^2}{2h_{\theta_j} h_{\theta_{j-1}}}\right)\delta_\theta^2\phi_{i,j}-\frac{h_{\theta_j} h_{\theta_{j-1}}}{h_{\theta_j}^2+h_{\theta_{j-1}}^2}\delta_\theta \phi_{\theta_{i,j}}\right.\\
    &\left.-\frac{(h_{\theta_j}-h_{\theta_{j-1}})}{h_{\theta_j}^2-h_{\theta_j} h_{\theta_{j-1}}+h_{\theta_{j-1}}^2}\left(\frac{h_{\theta_j}h_{\theta_{j-1}}}{h_{\theta_j}^2+h_{\theta_{j-1}}^2}+\frac{(h_{\theta_j}-h_{\theta_{j-1}})^2}{h_{\theta_j} h_{\theta_{j-1}}} \right)(\delta_\theta \phi_{i,j}-\phi_{\theta_{i,j}})\right\}\\
    &+\mathcal{O}\left(\frac{h_{\theta_j} h_{\theta_{j-1}}(h_{\theta_j}^3-h_{\theta_{j-1}}^3)}{(h_{\theta_j}^2+h_{\theta_{j-1}}^2)} \right).
\end{aligned}
\end{equation}

Making use of these spatial approximations, the CDE (\ref{CDE_pol}) around the node $(i,j)$ can be expressed in the following 
semidiscrete form,
\begin{equation}\label{semi}
a \left.\frac{\partial \phi}{\partial t}\right|_{i,j} + [\mathcal{A} \phi]_{i,j}= f_{i,j}.
\end{equation}

The operator $\mathcal{A}$ is defined as, 
\begin{equation}\label{operator}
\begin{aligned}
        [\mathcal{A} \phi]_{i,j}=&-\left(2 A_1 \delta_r^2+\frac{2 A_2}{r_i^2} \delta_{\theta}^2 +\frac{A_3}{h_{r_{i-1}}}\delta_r+\frac{A_4}{r_i^2 h_{\theta_{j-1}}} \delta_{\theta} \right) \phi_{i,j}+\left(\frac{A_3}{h_{r_{i-1}}}+A_5 \delta_r+c_1 \right) \phi_{r_{i,j}}\\
        &+\left(\frac{A_4}{r_i^2 h_{\theta_{j-1}}}+\frac{A_6}{r_i^2} \delta_{\theta}+c_2 \right) \phi_{\theta_{i,j}} 
\end{aligned}
\end{equation}
where the coefficients are given as,
\begin{equation*}
\begin{aligned}
&A_1=\dfrac{2\al_r}{1+\al_r^2}+\dfrac{(1-\al_r)^2}{2\al_r}, \\
&A_2=\dfrac{2\al_\theta}{1+\al_\theta^2}+\dfrac{(1-\al_\theta)^2}{2\al_\theta},\\
&A_3=\dfrac{2(1-\al_r)}{(1-\al_r+\al_r^2)}\left\{\dfrac{\al_r}{(1+\al_r^2)}+\dfrac{(1-\al_r)^2}{\al_r} \right\}, \\ &A_4=\dfrac{2(1-\al_\theta)}{(1-\al_\theta+\al_\theta^2)}\left\{\dfrac{\al_\theta}{(1+\al_\theta^2)}+\dfrac{(1-\al_\theta)^2}{\al_\theta}\right\}, \\
&A_5=\dfrac{2}{1+\al_r^2} \\ 
&A_6=\dfrac{2}{1+\al_\theta^2}.
\end{aligned}
\end{equation*}

The operators $\delta_r$, $\delta_\theta$, $\delta_r^2$ and $\delta_\theta^2$ along with Crank-Nicolson approximation yields the fully discretized form of \cref{CDE_pol} as
\begin{equation}\label{scheme_new}
\begin{aligned}
&\left(\frac{2a}{\dt}+B_1\right)\phi_{i,j}^{n+1}-B_2\phi_{i+1,j}^{n+1}-B_3\phi_{i-1,j}^{n+1}-B_4\phi_{i,j+1}^{n+1}-B_5 \phi_{i,j-1}^{n+1}\\
=&\left(\frac{2a}{\dt}-B_1\right)\phi_{i,j}^n+B_2\phi_{i+1,j}^n+B_3\phi_{i-1,j}^n+B_4\phi_{i,j+1}^n+B_5\phi_{i,j-1}^n\\
&-B_6\left(\phi_{r_{i+1,j}}^{n+1}-\phi_{r_{i-1,j}}^{n+1}\right)-\left(B_7+c_1\right)\phi_{r_{i,j}}^{n+1}-B_8\left(\phi_{\theta_{i,j+1}}^{n+1}-\phi_{\theta_{i,j-1}}^{n+1}\right)-\left(B_9+c_2\right)\phi_{\theta_{i,j}}^{n+1}\\
&-B_6\left(\phi_{r_{i+1,j}}^n-\phi_{r_{i-1,j}}^n\right)-\left(B_7+c_1\right)\phi_{r_{i,j}}^n-B_8\left(\phi_{\theta_{i,j+1}}^n-\phi_{\theta_{i,j-1}}^n\right)-\left(B_9+c_2\right)\phi_{\theta_{i,j}}^n\\
&+\left(f_{i,j}^{n+1}+f_{i,j}^n\right).
\end{aligned}
\end{equation}
where,
\begin{align*}
B_1=&4\left(\frac{A_1}{\al_r h_{r_{i-1}}^2}+\frac{A_2}{\al_\theta h_{\theta_{j-1}}^2 r_i^2}\right),\\
B_2=&\frac{1}{(1+\al_r) h_{r_{i-1}}^2}\left(\frac{4A_1}{\al_r}+A_3\right), \qquad \quad \quad B_3=\frac{1}{(1+\al_r)h_{r_{i-1}}^2}\left(4A_1-A_3\right),\\
B_4=&\frac{1}{(1+\al_\theta) h_{\theta_{j-1}}^2 r_i^2}\left(\frac{4A_2}{\al_\theta}+A_4\right),  \quad \quad \quad B_5=\frac{1}{(1+\al_\theta)h_{\theta_{j-1}}^2 r_i^2}\left(4A_2-A_4\right),\\
B_6=&\frac{A_5}{(1+\al_r) h_{r_{i-1}}}, \qquad \qquad \qquad \qquad \qquad B_7=\frac{A_3}{h_{r_{i-1}}},\\
B_8=&\frac{A_6}{(1+\al_\theta) h_{\theta_{j-1}} r_i^2}, \qquad \qquad \qquad \qquad \quad B_9=\frac{A_4}{h_{\theta_{j-1}} r_i^2}.
\end{align*}

Here, the coefficients $B_i$'s, $i\in\{1,2,3,\dots,9\}$ of the algebraic expression depends entirely on the values of grid spacing and they remain unchanged throughout the computation once the grid is set up. Therefore, for situations with constant convection and diffusion coefficients the current scheme also offers the inherent benefit of dealing with a resulting system of equations containing constant coefficients only. Further, one can easily notice that the discrete operator $\mathcal{A}$ also contains the radial and tangential derivatives of the flow variables, which need to be approximated up to the appropriate order of accuracy. In this aspect, we generalize the Pad\'e scheme \citep{Lele1992} to obtain the following approximations of the spatial derivatives on nonuniform polar grids.       
\begin{equation}\label{phir_pade}
\left(1+\frac{h_{r_i}h_{r_{i-1}}}{6}\delta_{rr}\right)\phi_{r_{i,j}}=\left(\delta_r-\frac{h_{r_i}-h_{r_{i-1}}}{2}\delta_{rr}\right)\phi_{i,j}+\mathcal{O}\left((h_{r_i}-h_{r_{i-1}})(3h_{r_i}^2-h_{r_i} h_{r_{i-1}}+3h_{r_{i-1}}^2)\right).
\end{equation}
and
\begin{equation}\label{phith_pade}
\left(1+\frac{h_{\theta_j}h_{\theta_{j-1}}}{6}\delta_{\theta\theta}\right)\phi_{\theta_{i,j}}=\left(\delta_\theta-\frac{h_{\theta_j}-h_{\theta_{j-1}}}{2}\delta_{\theta\theta}\right)\phi_{i,j}+\mathcal{O}\left((h_{\theta_j}-h_{\theta_{j-1}})(3h_{\theta_j}^2-h_{\theta_j} h_{\theta_{j-1}}+3h_{\theta_{j-1}}^2)\right).
\end{equation}      
Subsequently, we obtain the corresponding algebraic system for \cref{phir_pade,phith_pade} as, 
\begin{equation}\label{phir_pade2}
\phi_{r_{i+1,j}}^{n+1}+2\left(1+\al_r\right)\phi_{r_{i,j}}^{n+1}+\al_r \phi_{r_{i-1,j}}^{n+1}=\frac{3}{\al_r h_{r_{i-1}}} \left(\phi_{i+1,j}^{n+1}-(1-\al_r^2)\phi_{i,j}^{n+1}-\al_r^2 \phi_{i-1,j}^{n+1}\right)
\end{equation}
and
\begin{equation}\label{phith_pade2}
\phi_{\theta_{i,j+1}}^{n+1}+2\left(1+\al_\theta\right) \phi_{\theta_{i,j}}^{n+1}+\al_\theta \phi_{\theta_{i,j-1}}^{n+1}=\frac{3}{\al_\theta h_{\theta_{j-1}}}\left(\phi_{i,j+1}^{n+1}-(1-\al_\theta^2) \phi_{i,j}^{n+1}-\al_\theta^2 \phi_{i,j-1}^{n+1} \right)
\end{equation}
respectively.    
   
As the present scheme carries the flow gradients as variables, it is vital to approximate them at the boundary points. Moreover, in many situations, due to the lack of exact values of the flow variables at the boundary, we must resort to the one-sided approximations of the first order derivatives. This is accomplished by introducing the following discretizations:

\noindent Along the tangential direction, $\forall 1\leq j \leq n_\theta$
\begin{subequations}\label{one-sided-r}
\begin{empheq}{align}
\phi_{r_{1,j}}=&-\frac{h_{r_1}}{h_{r_2}(h_{r_1}+h_{r_2})} \phi_{3,j}+\frac{h_{r_1}+h_{r_2}}{h_{r_1} h_{r_2}} \phi_{2,j}-\frac{2h_{r_1}+h_{r_2}}{h_{r_1}(h_{r_1}+h_{r_2})} \phi_{1,j}+\mathcal{O}\left(h_{r_2}(h_{r_1}+h_{r_2})\right),\\
\phi_{r_{n_r,j}}=&\frac{h_{r_{n_r-1}}}{h_{r_{n_r-2}}(h_{r_{n_r-1}}+h_{r_{n_r-2}})} \phi_{n_r-2,j}-\frac{h_{r_{n_r-1}}+h_{r_{n_r-2}}}{h_{r_{n_r-1}} h_{r_{n_r-2}}}\phi_{n_r-1,j}+\frac{2h_{r_{n_r-1}}+h_{r_{n_r-2}}}{h_{r_{n_r-1}}(h_{r_{n_r-1}}+h_{r_{n_r-2}})}  \phi_{n_r,j} \notag\\
                &+\mathcal{O}\left(h_{r_{n_r-1}}(h_{r_{n_r-1}}+h_{r_{n_r-2}})\right),
\end{empheq}
\end{subequations}
\noindent Along the radial direction, $\forall 1\leq i \leq n_r$
\begin{subequations}\label{one-sided-th}
\begin{empheq}{align}
\phi_{\theta_{i,1}}=&-\frac{h_{\theta_1}}{h_{\theta_2}(h_{\theta_1}+h_{\theta_2})} \phi_{i,3}+\frac{h_{\theta_1}+h_{\theta_2}}{h_{\theta_1} h_{\theta_2}} \phi_{i,2}-\frac{2h_{\theta_1}+h_{\theta_2}}{h_{\theta_1}(h_{\theta_1}+h_{\theta_2})} \phi_{i,1}+\mathcal{O}\left(h_{\theta_2}(h_{\theta_1}+h_{\theta_2})\right),\\
\phi_{\theta_{i,n_\theta}}=&\frac{h_{\theta_{n_\theta-1}}}{h_{\theta_{n_\theta-2}}(h_{\theta_{n_\theta-1}}+h_{\theta_{n_\theta-2}})} \phi_{i,n_\theta-2}-\frac{h_{\theta_{n_\theta-1}}+h_{\theta_{n_\theta-2}}}{h_{\theta_{n_\theta-1}} h_{\theta_{n_\theta-2}}}\phi_{i,n_\theta-1}+\frac{2h_{\theta_{n_\theta-1}}+h_{\theta_{n_\theta-2}}}{h_{\theta_{n_\theta-1}}(h_{\theta_{n_\theta-1}}+h_{\theta_{n_\theta-2}})}  \phi_{i,n_\theta} \notag\\
                &+\mathcal{O}\left(h_{\theta_{n_\theta-1}}(h_{\theta_{n_\theta-1}}+h_{\theta_{n_\theta-2}})\right).
\end{empheq}
\end{subequations}

\section{Solution of algebraic system}\label{algorithm}

The newly developed FD approximation (\ref{scheme_new}) yields a system of equation which in matrix form can be written as,  
\begin{equation}\label{system1}
  M_1\Phi^{(n+1)}=F_1\left(\Phi^{(n)},\Phi_r^{(n)},\Phi_\theta^{(n)},\Phi_r^{(n+1)},\Phi_\theta^{(n+1)}\right),
\end{equation}
with,
\begin{eqnarray*}
  \Phi &=&\left(\phi_{1,1},\phi_{1,2},\dots,\phi_{1,n_\theta},\phi_{2,1},\phi_{2,2},\dots,\phi_{2,n_\theta},\dots,\phi_{n_r,n_\theta}\right)^T,\\
  \Phi_r&=&\left(\phi_{r_{1,1}},\phi_{r_{1,2}},\dots,\phi_{r_{1,n_\theta}},\phi_{r_{2,1}},\phi_{r_{2,2}},\dots,\phi_{r_{2,n_\theta}},\dots,\phi_{r_{n_r,n_\theta}}\right)^T\\
\text{and} \qquad \Phi_\theta&=&\left(\phi_{\theta_{1,1}},\phi_{\theta_{1,2}},\dots,\phi_{\theta_{1,n_\theta}},\phi_{\theta_{2,1}},\phi_{\theta_{2,2}},\dots,\phi_{\theta_{2,n_\theta}},\dots,\phi_{\theta_{n_r,n_\theta}}\right)^T.
\end{eqnarray*}
The coefficient matrix $M_1$ here is a sparse nonsymmetric matrix with five nonzero diagonals. As stated in \cref{discretization}, an algebraic system of equations involves only constant coefficients once the grid is laid out. Thus on a grid of size $n_r\times n_\theta$, the coefficient matrix $M_1$ only deals with constant entries. Although the system appears to have a large dimension  $3n_r n_\theta\times 3n_r n_\theta$, one only needs to handle a system of size $n_r\times n_\theta$ while employing a predictor-corrector method \citep{Senetal2013,Sen2013}.

Similarly, matrix representation of \cref{phir_pade2} and \cref{phith_pade2} are
\begin{equation}\label{system2}
  M_2\Phi_r^{(n)}=F_2\left(\Phi^{(n)}\right)
\end{equation}
and
\begin{equation}\label{system3}
  M_3\Phi_\theta^{(n)}=F_3\left(\Phi^{(n)}\right)
\end{equation}
respectively. $M_2$ and $M_3$ are tri-diagonal matrices and hence systems \eqref{system2} and \eqref{system3} are very amenable to efficient computation. The algorithmic procedure adopted at this stage is same as delineated in \cite{Senetal2013,Sen2013}.

\section{Numerical examples}\label{testcase}
To examine the accuracy and effectiveness of the present scheme on nonuniform polar grids, it has been applied to four different problems of varying complexities. Validation study is carried out by solving the convection-diffusion of the Gaussian pulse. We study the efficiency of the scheme in handling complex flow patterns by solving the driven polar cavity problem. Subsequently, we enter the field of heat transfer by tackling the problem of heat convection in an annulus. Finally, we carry out a comprehensive study of flow and heat transfer around a circular cylinder. The intention behind selecting these problems is to highlight the inherent flexibility of nonuniform grids whereby certain portions of the solution domain can be better resolved with grid clustering. Furthermore, the overwhelming number of numerical solutions present in the literature gives us the leverage to compare our numerical solution with the existing ones establishing the robustness of this newly developed scheme. All the computations are executed on an Intel i7-based PC with 3.40 GHz CPU and 32 GB RAM. The tolerance for inner and outer iterations is set to be $1.0e-10$.   

\subsection{Problem 1: Convection diffusion of Gaussian pulse}\label{ex1}
This validation study is intended to capture the unsteady convection-diffusion of a Gaussian pulse in the domain $[0,2]\times[0,\frac{\pi}{2}]$. This flow situation is governed by \cref{CDE_pol} and is equipped with the analytical solution
\begin{equation}\label{gauss}
\phi(r,\theta,t)=-\frac{1}{4t+1}\exp\left[\frac{(ar\cos{\theta}-c_1 t-0.1a)^2}{a(4t+1)}+\frac{(ar\sin{\theta}
-c_2t-0.1a)^2}{a(4t+1)}\right].
\end{equation}
\Cref{gauss} directly provides the initial condition and Dirichlet boundary conditions. At $t=0$ \cref{gauss} represents a pulse of unit height centred at (0.1,0.1) and as time advances it moves along the line $\theta=\dfrac{\pi}{4}$.

\begin{figure}[!h]
\centering
        \begin{subfigure}[b]{0.44\textwidth}
        \centering
        \includegraphics[width=\linewidth]{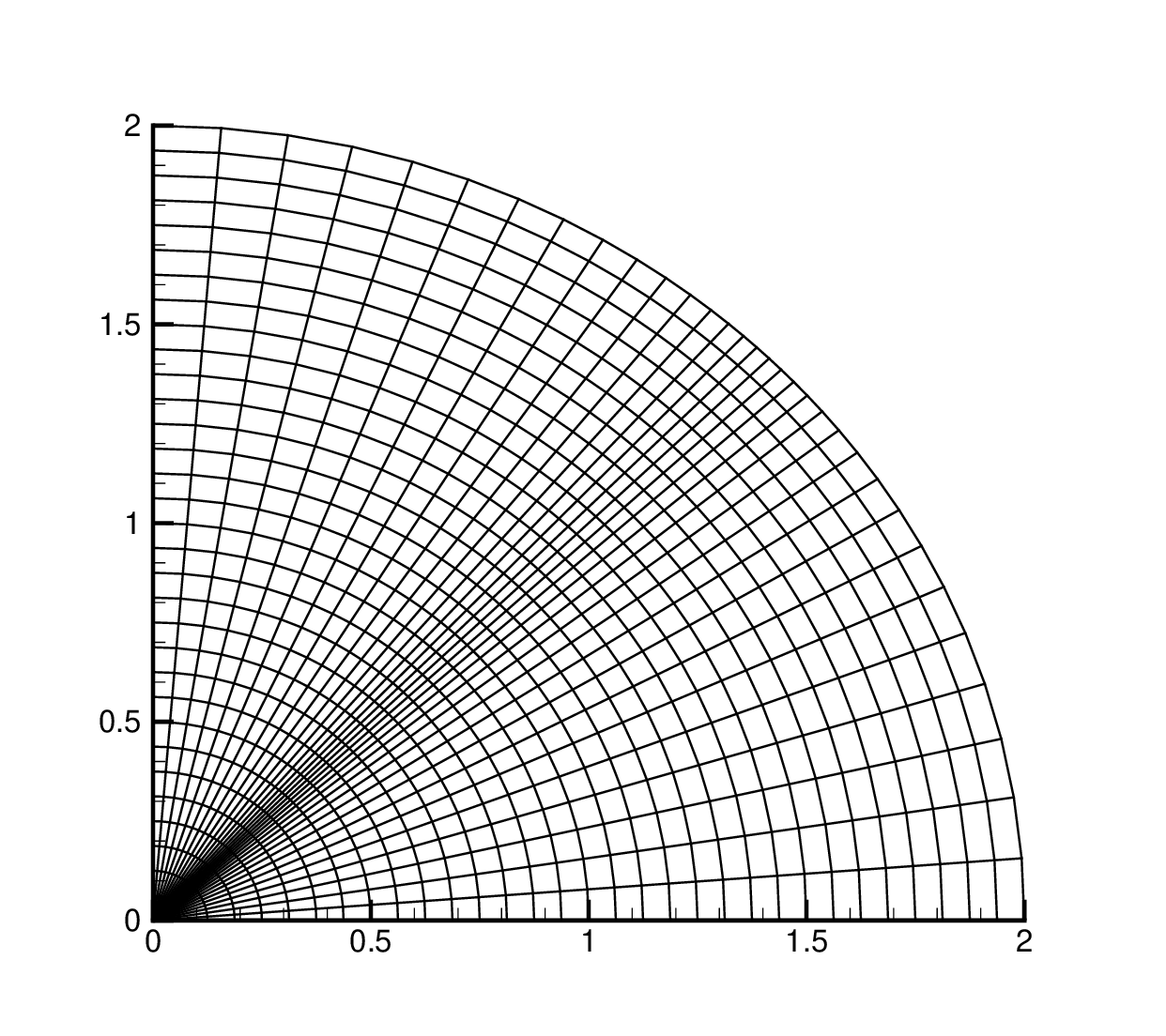}
  		\end{subfigure}
  		\begin{minipage}[b]{0.02\textwidth}
        \subcaption{ }\label{fig:P1_grid}
        \end{minipage}
        \begin{subfigure}[b]{0.44\textwidth}
                \centering
                \includegraphics[width=\linewidth]{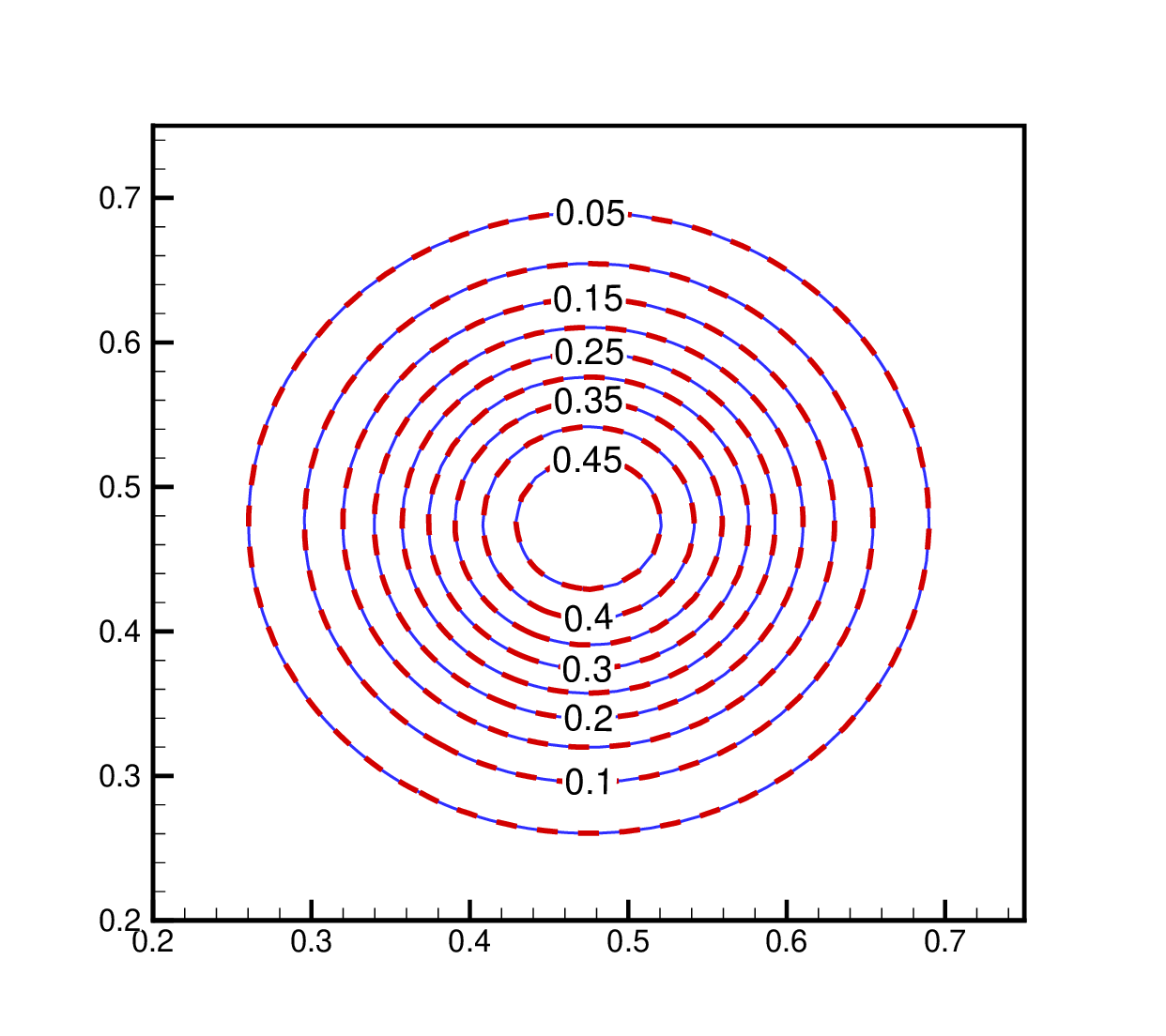}
        \end{subfigure}%
        \begin{minipage}[b]{0.02\textwidth}
        \subcaption{ }\label{fig:P1_p25}
        \end{minipage}
        \\
        \begin{subfigure}[b]{0.44\textwidth}
        \centering
        \includegraphics[width=\linewidth]{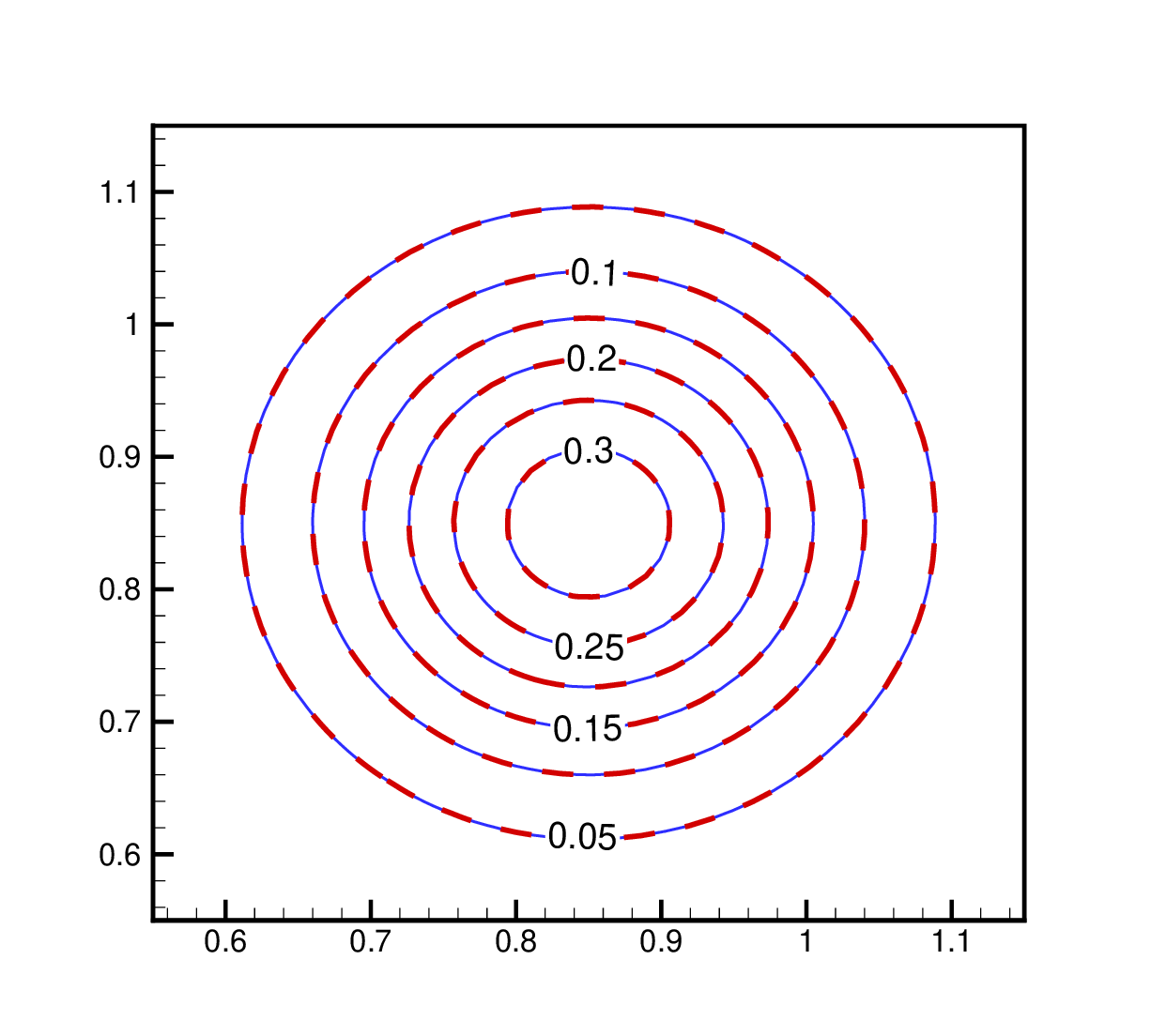}
  		\end{subfigure}
  		\begin{minipage}[b]{0.02\textwidth}
        \subcaption{ }\label{fig:P1_p50}
        \end{minipage}
        \begin{subfigure}[b]{0.44\textwidth}
        \centering
        \includegraphics[width=\linewidth]{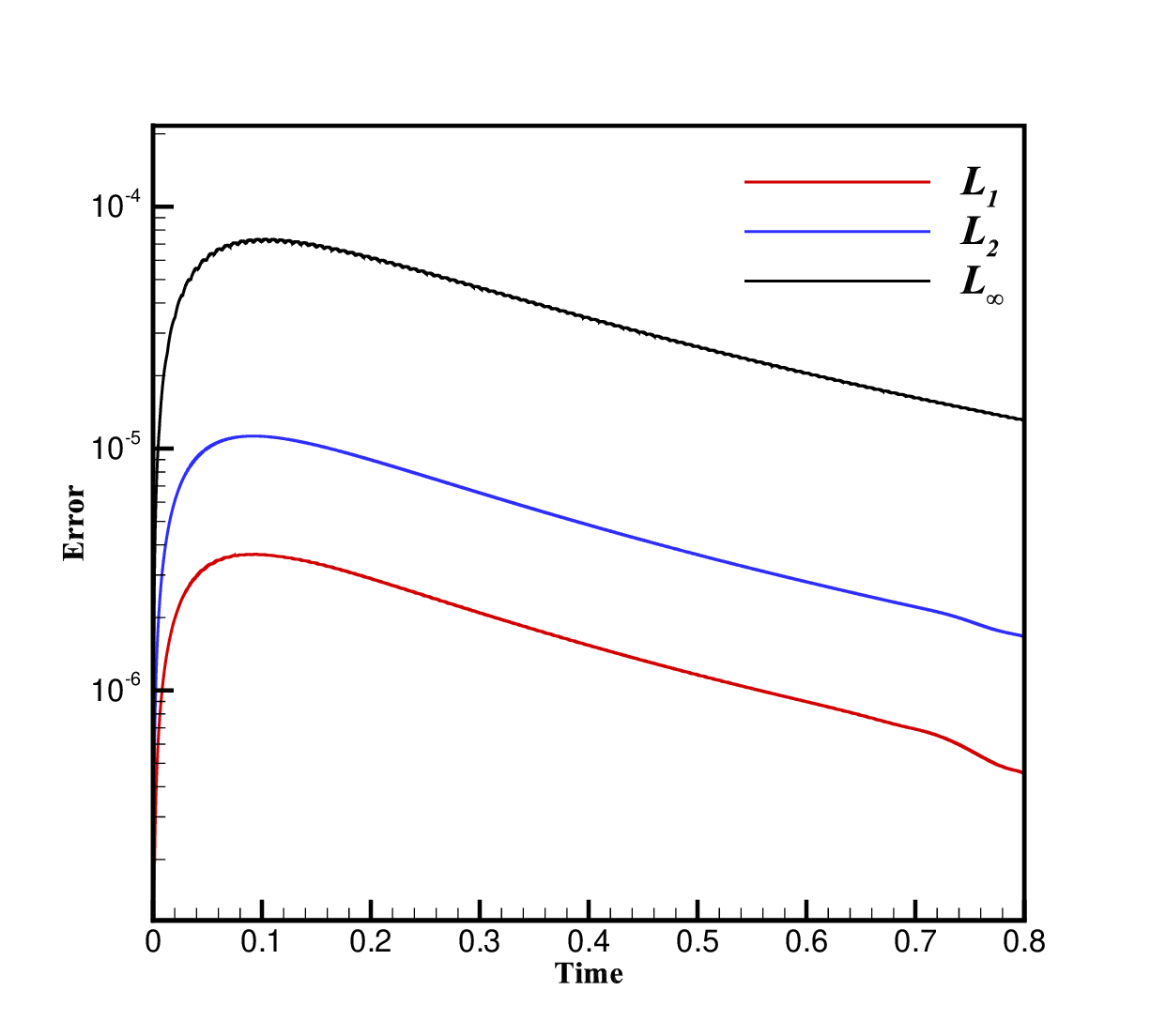}
        \end{subfigure}%
        \begin{minipage}[b]{0.02\textwidth}
        \subcaption{ }\label{fig:P1_convergence}
        \end{minipage}
        \caption{Problem 1: (a) Typical $33\times33$ grids ; exact (red) and numerical (blue) contour plots of the Gaussian pulse at (b) $t=0.25$, (c) $t=0.50$; (d) time evolution of $L_1$, $L_2$ and $L_\infty$ norm errors.}\label{fig:P1result}
\end{figure}

The convection coefficients have been set at values $c_1=c_2=150$ for the current study. Further, $a=100$ and $\dt=2.5e-5$ are kept fixed. We have used a nonuniform grid in the tangential direction that clusters in the vicinity of $\theta=\dfrac{\pi}{4}$ and uniform grid spacing has been maintained along the radial direction. The nonuniform grid is generated with the help of the trigonometric function
\begin{equation}\label{P1_grid}
\theta_j=L_\theta\left\{\frac{j}{n_\theta}+\frac{\lambda_\theta}{\Theta_\theta} \sin{\left(\frac{j\Theta_\theta}{n_\theta} \right)} \right\}.
\end{equation}
A typical $33\times33$ grid so formed has been displayed in \cref{fig:P1_grid}. The degree of clustering is decided by the clustering parameter $\lambda_\theta=0.6$. Values of the other constants have been taken as $L_\theta=\dfrac{\pi}{2}$ and $\Theta_\theta=2\pi$.

\begin{table}[!h]
\centering
\caption{Problem 1: $L_1$, $L_2$ and $L_{\infty}$ norm errors of $\phi$ at different grids and spatial rate of convergence. }\label{table:P1_1}
\vspace{0.1cm}
\renewcommand*{\arraystretch}{1}
\begin{tabular}
{ M{0.05\textwidth} M{0.1\textwidth} M{0.15\textwidth} M{0.09\textwidth} M{0.15\textwidth} M{0.09\textwidth} M{0.15\textwidth}}
\hline
Time      &              & $33\times33$  &  order   & $65\times65$   &  order  & $129\times129$  \\
\hline 
0.25      & $L_1$        &  7.337756e-4  &   4.20   &  3.990618e-5   &  4.06   &   2.390596e-6   \\
          & $L_2$        &  2.248623e-3  &   4.13   &  1.285848e-4   &  4.06   &   7.687400e-6   \\
          & $L_{\infty}$ &  1.353756e-2  &   3.92   &  8.920667e-4   &  4.07   &   5.328468e-5   \\
0.50      & $L_1$        &  3.322961e-4  &   4.16   &  1.859497e-5   &  4.04   &   1.132402e-6   \\
          & $L_2$        &  1.027676e-3  &   4.10   &  6.002334e-5   &  4.04   &   3.646823e-6   \\
          & $L_{\infty}$ &  6.769691e-3  &   4.02   &  4.174422e-4   &  3.98   &   2.643113e-5   \\
\hline 
\end{tabular}
\end{table}

\begin{table}[!h]
\centering
\caption{Problem 1: $L_1$, $L_2$ and $L_{\infty}$ norm errors of $\phi$ at different values of $\dt$ and temporal rate of convergence. }\label{table:P1_2}
\vspace{0.1cm}
\renewcommand*{\arraystretch}{1}
\begin{tabular}
{ M{0.05\textwidth} M{0.1\textwidth} M{0.15\textwidth} M{0.09\textwidth} M{0.15\textwidth} M{0.09\textwidth} M{0.15\textwidth}}
\hline
Time      &              &   $\dt=0.02$  &  order   &   $\dt=0.01$   &  order  & $\dt=0.005$     \\
\hline
0.50      & $L_1$        &  2.098479e-3  &   2.01   &  5.204632e-4   &   2.00  &   1.299927e-4   \\
          & $L_2$        &  6.556053e-3  &   2.01   &  1.632436e-3   &   2.00  &   4.077176e-4   \\
          & $L_{\infty}$ &  4.591483e-2  &   2.07   &  1.096071e-2   &   2.03  &   2.684826e-3   \\
0.80      & $L_1$        &  1.202700e-3  &   2.09   &  2.823366e-4   &   2.03  &   6.907145e-5   \\
          & $L_2$        &  4.319963e-3  &   2.05   &  1.045744e-3   &   2.02  &   2.577760e-4   \\
          & $L_{\infty}$ &  3.164084e-2  &   2.06   &  7.566229e-3   &   2.03  &   1.849941e-3   \\
\hline 
\end{tabular}
\end{table}

Comparisons between the numerical and exact solutions at two different times $t=0.25$ and $t=0.50$ are presented in \cref{fig:P1_p25,fig:P1_p50} respectively. The numerical solution is seen to be almost indistinguishable from the exact solution conforming to the efficiency of the present scheme to capture the moving pulse accurately. $L_1$, $L_2$, and $L_\infty$ norm errors at each time step for the entire computation time are presented in \cref{fig:P1_convergence}. The similar temporal decaying nature of all three norm errors justifies the convergence of the present scheme.

A quantitative assessment of the numerical solution has been carried out in \cref{table:P1_1,table:P1_2}. In \cref{table:P1_1}, we have calculated the spatial order of convergence for the computation done on three grids of different sizes $33\times33$, $65\times65$ and $129\times129$ with $\dt=2.5e-5$. For the temporal rate of convergence computations are carried out with $\dt=0.02$, $\dt=0.01$ and $\dt=0.005$ on a $129\times129$ grid. The clustering parameter has been kept at $\lambda_\theta=0.1$ to make spatial error much smaller compared to the temporal error. Computed errors at various times along with the temporal accuracy have been presented in \cref{table:P1_2}. The scheme shows a near fourth order convergence in space and a temporal accuracy of second order.     

\subsection{Problem 2: Driven polar cavity}\label{ex2}

A driven flow inside a polar cavity is also been simulated using the present scheme. This problem is governed by the transient form of N-S equations which can be given as,
\begin{subequations}\label{trNS_pol}
\begin{empheq}[left=\empheqlbrace]{align}
&\omega_t=\frac{1}{Re}\left(\omega_{rr}+\frac{1}{r}\omega_r+\frac{1}{r^2} \omega_{\theta\theta}\right)
-\left(u \omega_r +\frac{v}{r} \omega_\theta \right), \label{trNS_pol1} \\
&\omega=-\left( \psi_{rr}+\frac{1}{r} \psi_r +\frac{1}{r^2} \psi_{\theta\theta} \right) \label{trNS_pol2}
\end{empheq}
\end{subequations}
where $u(r,\theta)=\dfrac{1}{r} \psi_\theta$ and $v(r,\theta)=-\psi_r$ are the radial and tangential velocity components respectively. 

This particular problem was first studied experimentally and numerically by Fuchs and Tillmark \citep{FuchsTillmark1985} and since then it has been used extensively by researchers to validate numerical schemes, particularly for physical domains with circular arc as boundaries \citep{LeeTsuei1993,RayKalita2010,YuTian2013b,SenKalita2015,DasPanditRay2023}. 

\begin{figure}[!h]
  \begin{subfigure}[b]{0.455\textwidth}
  \centering
  \includegraphics[width=\linewidth]{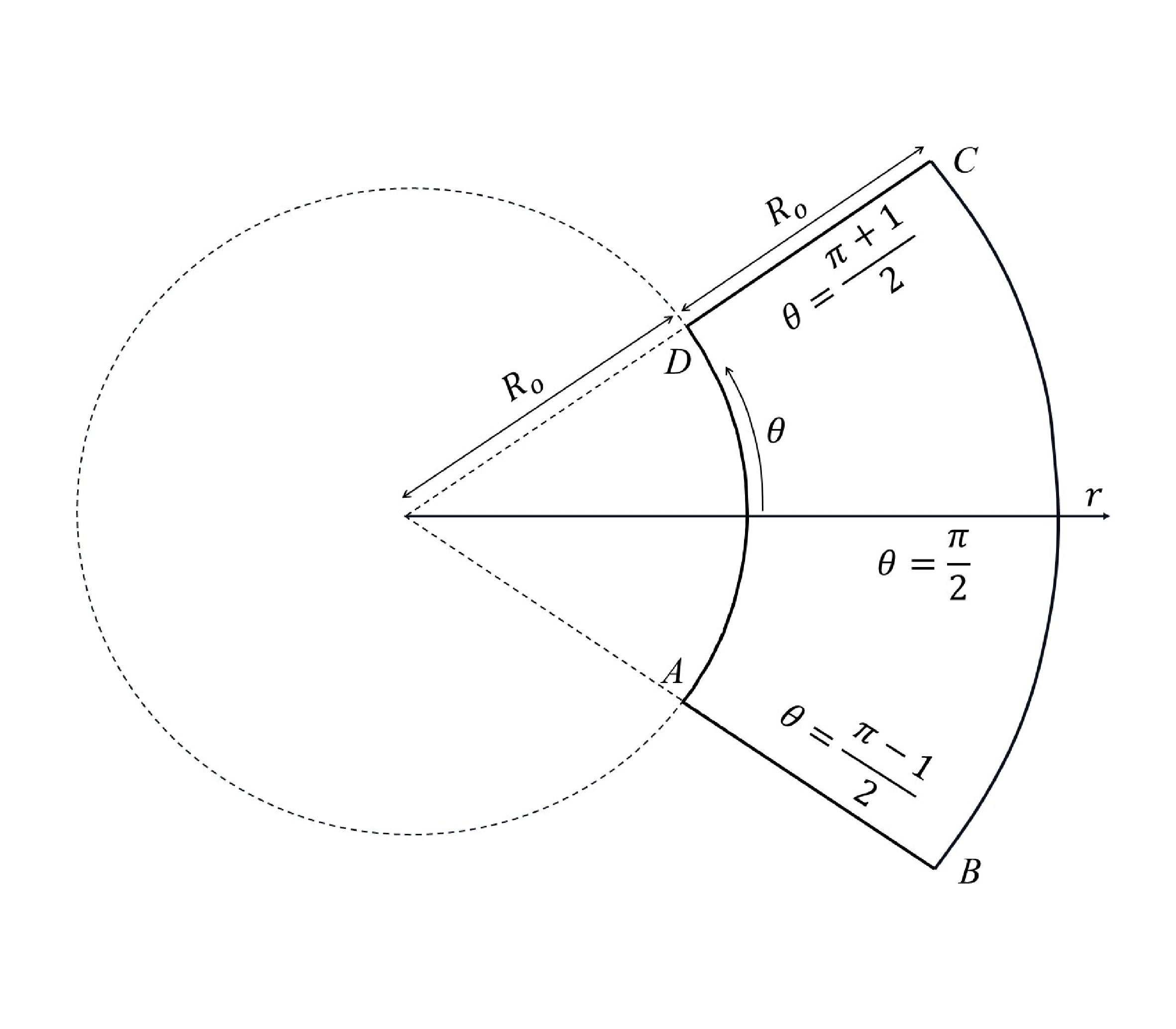}
  \end{subfigure}
  \begin{minipage}[b]{0.02\textwidth}
  \subcaption{ }\label{fig:P2_schematic}
  \end{minipage}
  \begin{subfigure}[b]{0.455\textwidth}
  \centering
  \includegraphics[width=\linewidth]{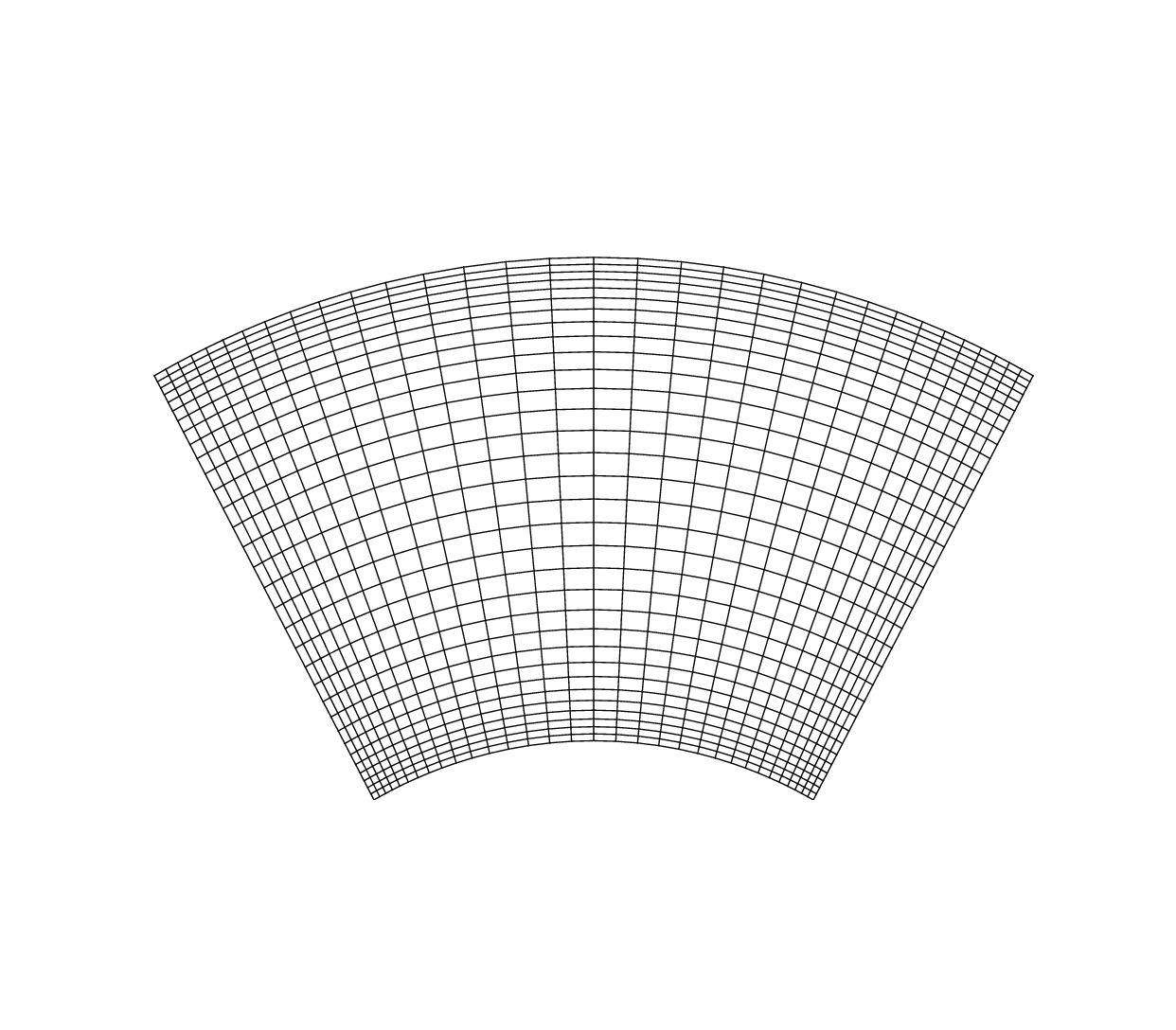}
  \end{subfigure}
  \begin{minipage}[b]{0.02\textwidth}
  \subcaption{ }\label{fig:P2_grid}
  \end{minipage}
\caption{Problem 2: (a) Schematic diagram and (b) typical $33\times33$ centrosymmetric polar grid for flow inside a polar cavity.}\label{fig:P2}
\end{figure} 

The problem is defined in an annular region $\left[1,2\right]\times\left[\dfrac{\pi-1}{2},\dfrac{\pi+1}{2}\right]$. A schematic illustration of the problem is presented in \cref{fig:P2_schematic}. The region $ABCD$ is the domain of the problem: the height $AB=DC$ of the cavity is equal to the radius $R_0$=1 of the inner circle. The inner circular arc $DA$ rotates with a unit velocity in the clockwise direction which drives the flow inside the cavity. All the other boundary walls are stationary throughout the simulation. From this, we can derive the boundary conditions for $u$ and $v$ as: $u=0$, $v=-1$ at the moving boundary $DA$ and $u=0$, $v=0$ at all the other boundaries. The streamfunction values are set to be zero at all the boundaries $(\psi=0)$. For this problem, the Reynolds number is defined as $Re=\dfrac{v_0R_0}{\nu}$, where $v_0$ is magnitude of the surface velocity of rotating side $DA$ and $\nu$ is the kinematic viscosity of the fluid. This is imperative to mention here that all boundary walls follow no-slip condition with the fluid inside the cavity, which supplies the following Neumann boundary conditions for vorticity.\\
On the right wall $AB$, $\forall 1\leq i \leq n_r$ :
\begin{equation}\label{ww_right}
\omega_{i,1}=-\frac{2}{r_i^2 h_{\theta_{1}}^2} \psi_{i,2}.
\end{equation} 
On the top wall $BC$, $\forall 1\leq j \leq n_\theta$ :
\begin{equation}\label{ww_top}
\omega_{n_r,j}=-\frac{2}{h_{r_{n_r-1}}^2}\psi_{n_r-1,j}.
\end{equation}
On the left wall $CD$, $\forall 1\leq i \leq n_r$ : 
\begin{equation}\label{ww_left}
\omega_{i,n_\theta}=-\frac{2}{r_i^2 h_{\theta_{n_\theta-1}}^2} \psi_{i,n_\theta-1}.
\end{equation}
On the bottom wall $AB$, $\forall 1\leq j \leq n_\theta$ :
\begin{equation}\label{ww_bottom}
\omega_{1,j}=-\frac{2}{h_{r_1}^2}\left(\psi_{2,j}-h_{r_1}\right)-\frac{1}{r_1}.
\end{equation}
After we compute the values of vorticity, the boundary values for its gradients are calculated using one-sided approximations as follows: \\
On the right wall $AB$, $\forall 1\leq i \leq n_r$:
\begin{equation}\label{right}
\phi_{\theta_{i,1}}=-\frac{h_{\theta_1}}{h_{\theta_2}(h_{\theta_1}+h_{\theta_2})} \phi_{i,3}+\frac{h_{\theta_1}+h_{\theta_2}}{h_{\theta_1} h_{\theta_2}} \phi_{i,2}-\frac{2h_{\theta_1}+h_{\theta_2}}{h_{\theta_1}(h_{\theta_1}+h_{\theta_2})} \phi_{i,1}+\mathcal{O}\left(h_{\theta_2}(h_{\theta_1}+h_{\theta_2})^2\right).
\end{equation}
On the top wall $BC$, $\forall 1\leq j \leq n_\theta$: 
\begin{equation}\label{top}
\begin{aligned}
\phi_{r_{n_r,j}}=&\dfrac{h_{r_{n_r-1}}}{h_{r_{n_r-2}}(h_{r_{n_r-1}}+h_{r_{n_r-2}})} \phi_{n_r-2,j}-\dfrac{h_{r_{n_r-1}}+h_{r_{n_r-2}}}{h_{r_{n_r-1}} h_{r_{n_r-2}}}\phi_{n_r-1,j}\\
&+\dfrac{2h_{r_{n_r-1}}+h_{r_{n_r-2}}}{h_{r_{n_r-1}}(h_{r_{n_r-1}}+h_{r_{n_r-2}})}  \phi_{n_r,j}+\mathcal{O}\left(h_{r_{n_r-1}}(h_{r_{n_r-1}}+h_{r_{n_r-2}})^2\right).
\end{aligned}
\end{equation}
On the left wall $CD$, $\forall 1\leq i \leq n_r$:
\begin{equation}\label{left}
\begin{aligned}
\phi_{\theta_{i,n_\theta}}=&\frac{h_{\theta_{n_\theta-1}}}{h_{\theta_{n_\theta-2}}(h_{\theta_{n_\theta-1}}
+h_{\theta_{n_\theta-2}})}\phi_{i,n_\theta-2}-\frac{h_{\theta_{n_\theta-1}}+h_{\theta_{n_\theta-2}}}{h_{\theta_{n_\theta-1}} h_{\theta_{n_\theta-2}}}\phi_{i,n_\theta-1}\\ &+\frac{2h_{\theta_{n_\theta-1}}+h_{\theta_{n_\theta-2}}}{h_{\theta_{n_\theta-1}}(h_{\theta_{n_\theta-1}}+h_{\theta_{n_\theta-2}})}  \phi_{i,n_\theta}+\mathcal{O}\left(h_{\theta_{n_\theta-1}}(h_{\theta_{n_\theta-1}}+h_{\theta_{n_\theta-2}})^2\right).
\end{aligned}
\end{equation}
On the bottom wall $DA$, $\forall 1\leq j \leq n_\theta$:
\begin{equation}\label{bottom}
\phi_{r_{1,j}}=-\frac{h_{r_1}}{h_{r_2}(h_{r_1}+h_{r_2})} \phi_{3,j}+\frac{h_{r_1}+h_{r_2}}{h_{r_1} h_{r_2}} \phi_{2,j}-\frac{2h_{r_1}+h_{r_2}}{h_{r_1}(h_{r_1}+h_{r_2})} \phi_{1,j}+\mathcal{O}\left(h_{r_2}(h_{r_1}+h_{r_2})^2\right).
\end{equation}

While working with this problem, Lee and Tsuei \citep{LeeTsuei1993} witnessed that large solution errors propagate near the rotating boundary. Besides, vortices are also created at the solid boundaries of the cavity. Therefore it is necessary to introduce a maximum number of grids in the vicinity of cavity walls as can be seen in \cref{fig:P2_grid}. This has been achieved using modified versions of the grid-generating function 
\begin{equation}\label{P2_grid}
r_i=L_r\left\{\frac{i}{n_r}+\frac{\lambda_r}{\Theta_r} \sin{\left(\frac{i\Theta_r}{n_r} \right)} \right\}.
\end{equation}
along with \cref{P1_grid} with proper choices of the constants $L_r=L_\theta=1.0$, $\lambda_r=\lambda_\theta=-0.55$ and $\Theta_r=\Theta_\theta=2\pi$. 

\begin{figure}[H]
\centering
  \begin{subfigure}[b]{0.375\textwidth}
  \centering
  \includegraphics[width=\linewidth]{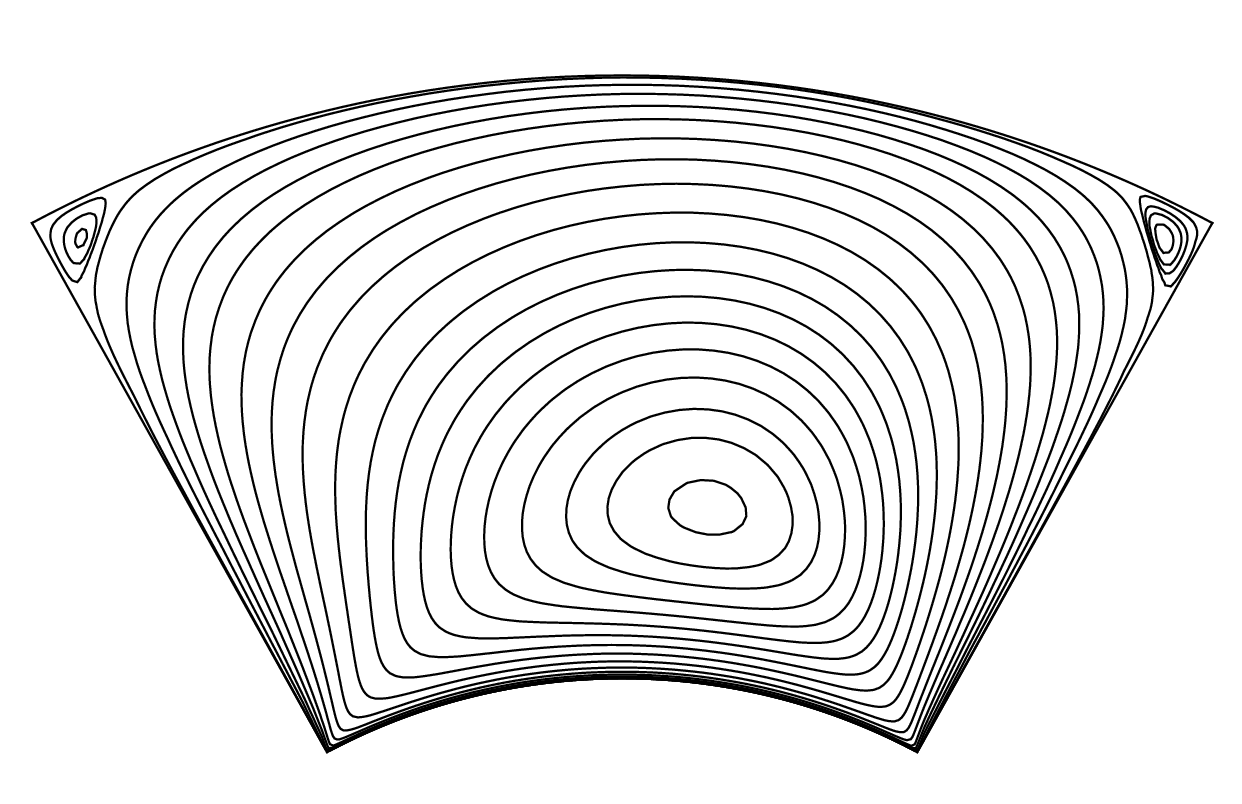}
  \end{subfigure}
  \begin{subfigure}[b]{0.375\textwidth}
  \centering
  \includegraphics[width=\linewidth]{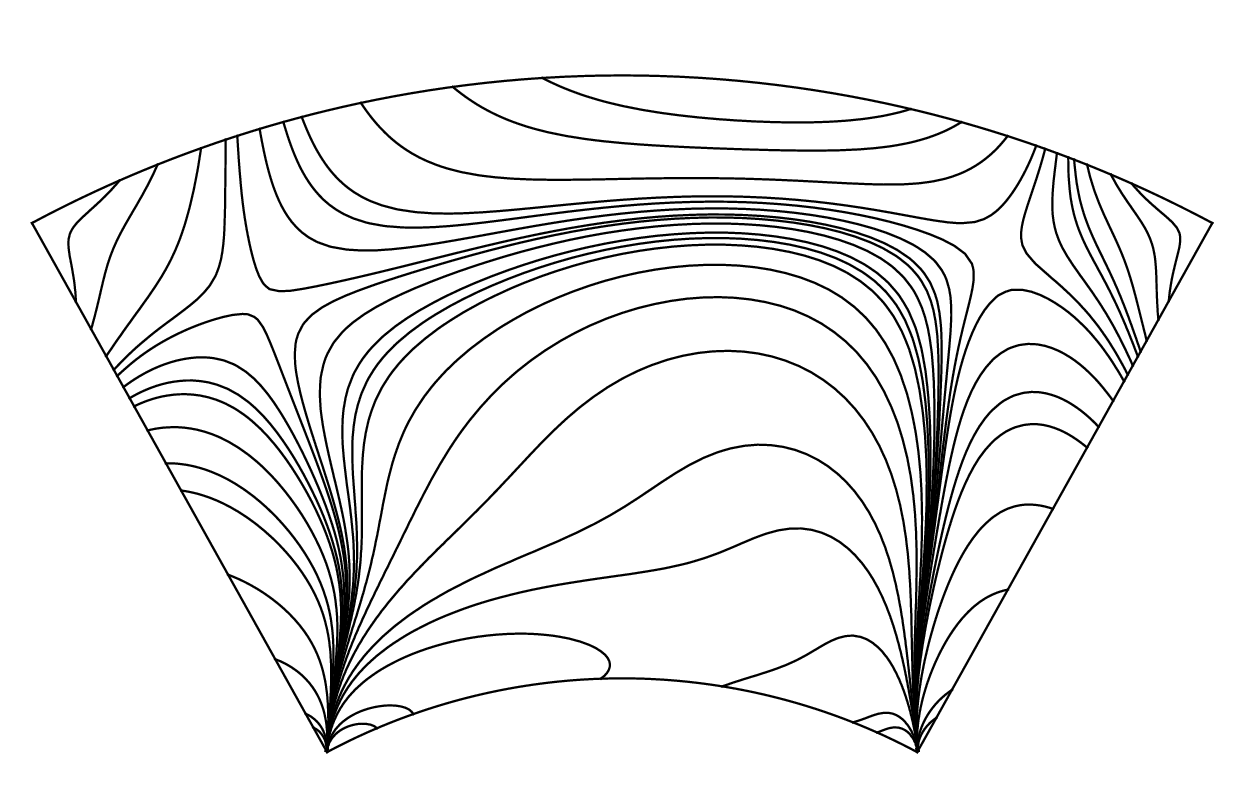}
  \end{subfigure}
  \begin{minipage}[b]{0.02\textwidth}
  \subcaption{ }\label{fig:P2_strvort1000}
  \end{minipage}
  \\
  \begin{subfigure}[b]{0.375\textwidth}
  \centering
  \includegraphics[width=\linewidth]{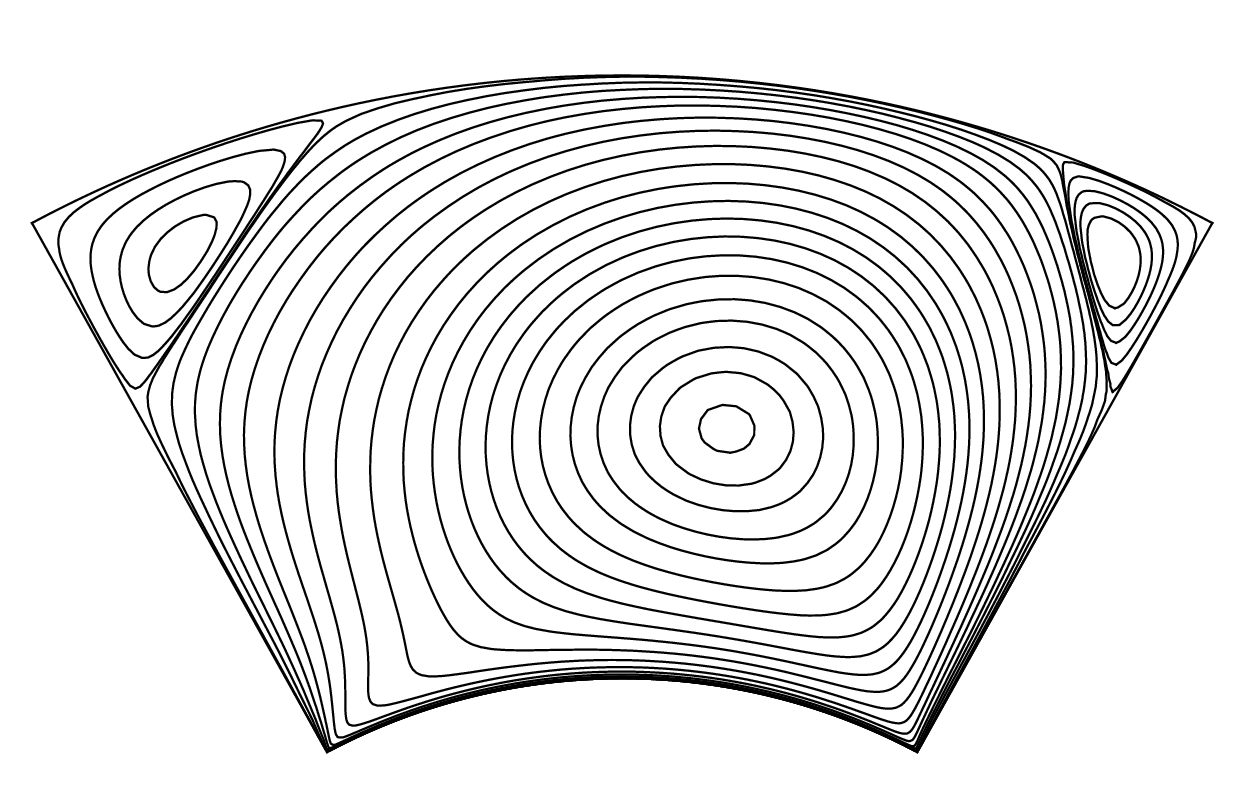}
  \end{subfigure}
  \begin{subfigure}[b]{0.375\textwidth}
  \centering
  \includegraphics[width=\linewidth]{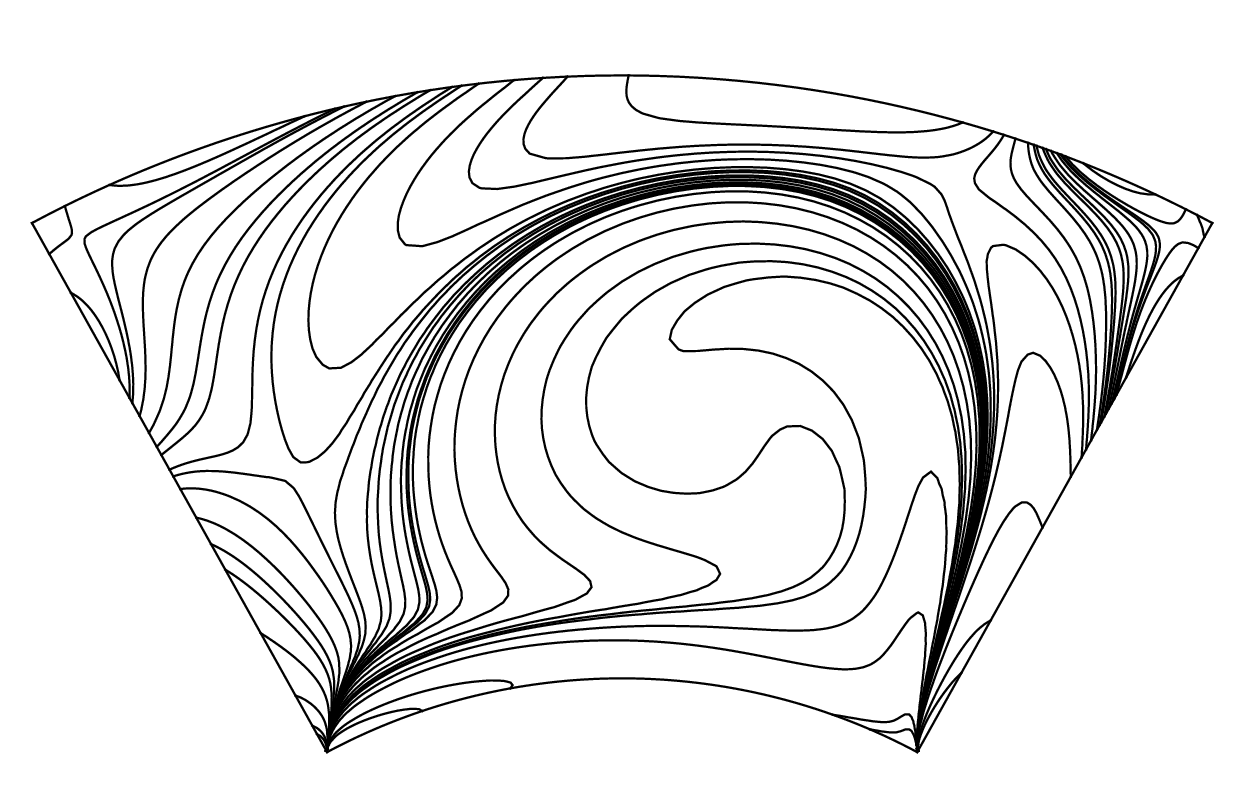}
  \end{subfigure}
  \begin{minipage}[b]{0.02\textwidth}
  \subcaption{ }\label{fig:P2_strvort1000}
  \end{minipage}
  \\
  \begin{subfigure}[b]{0.375\textwidth}
  \centering
  \includegraphics[width=\linewidth]{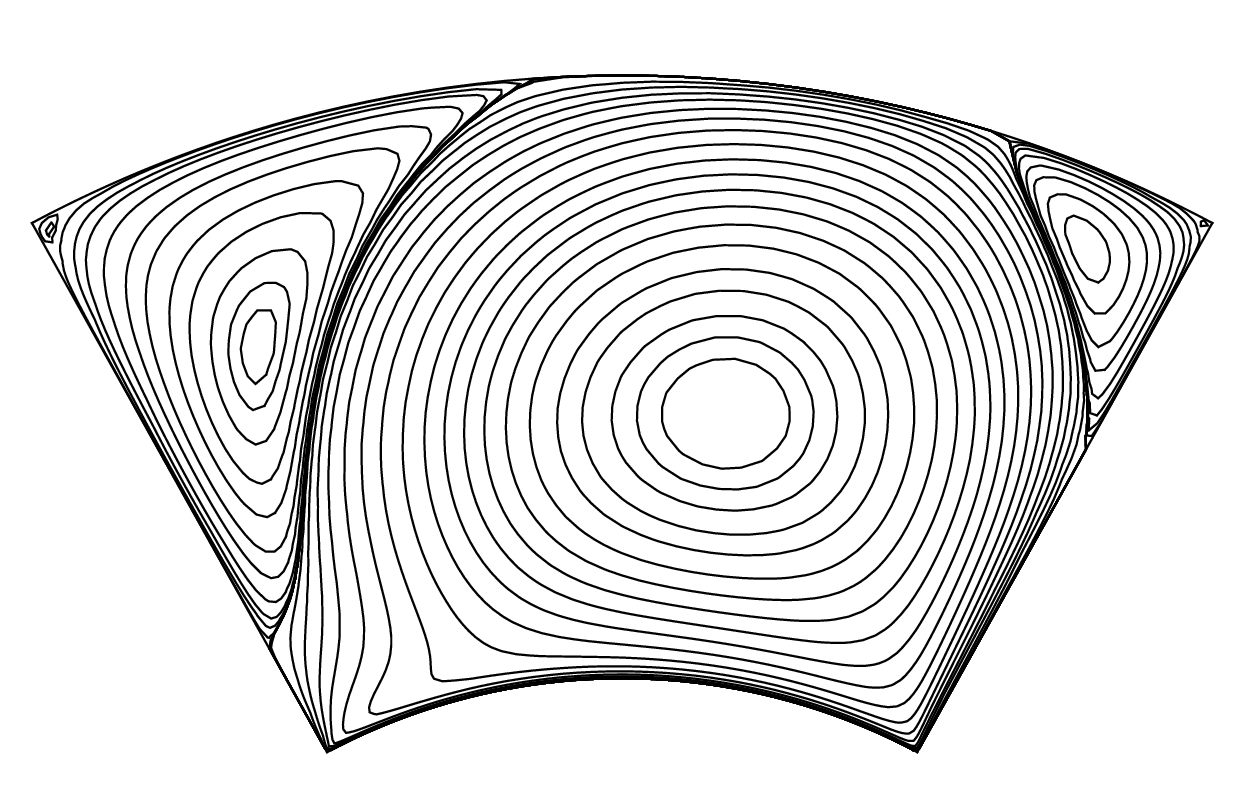}
  \end{subfigure}
  \begin{subfigure}[b]{0.375\textwidth}
  \centering
  \includegraphics[width=\linewidth]{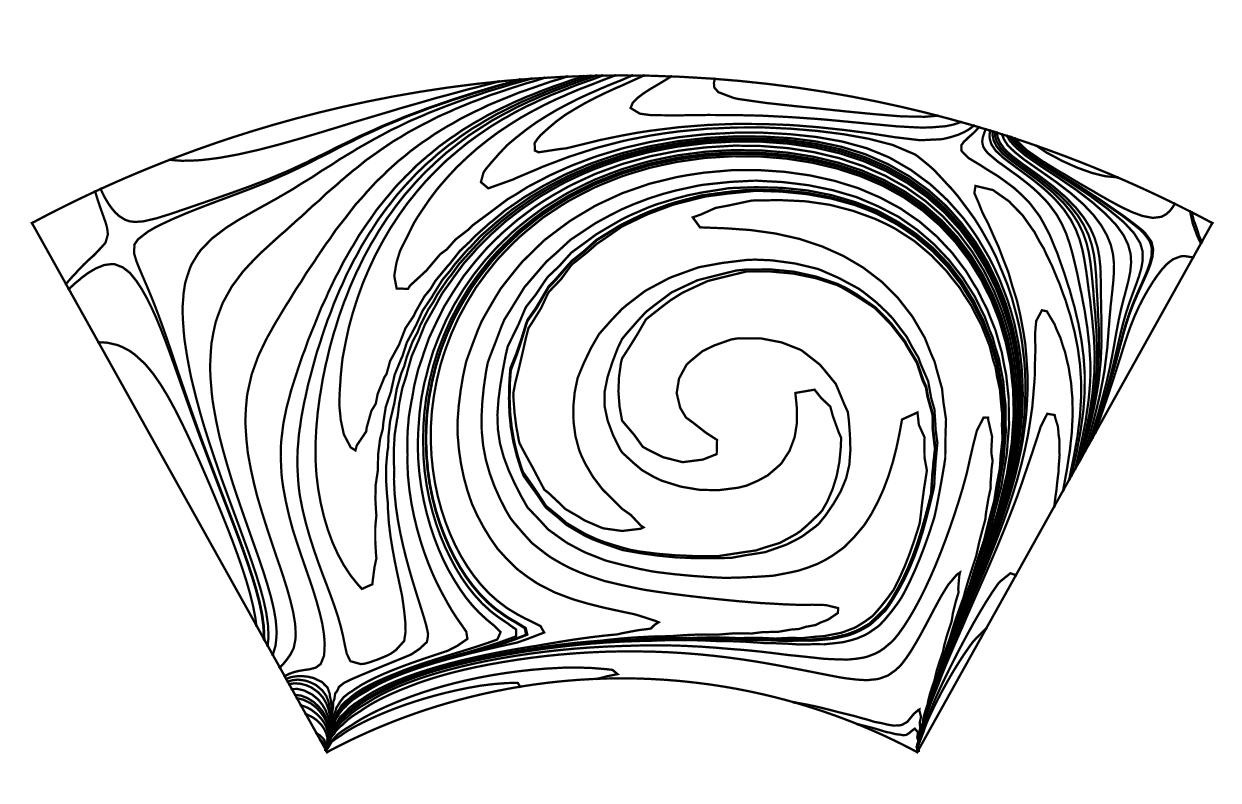}
  \end{subfigure}
  \begin{minipage}[b]{0.02\textwidth}
  \subcaption{ }\label{fig:P2_strvort1000}
  \end{minipage}
  \\
  \begin{subfigure}[b]{0.375\textwidth}
  \centering
  \includegraphics[width=\linewidth]{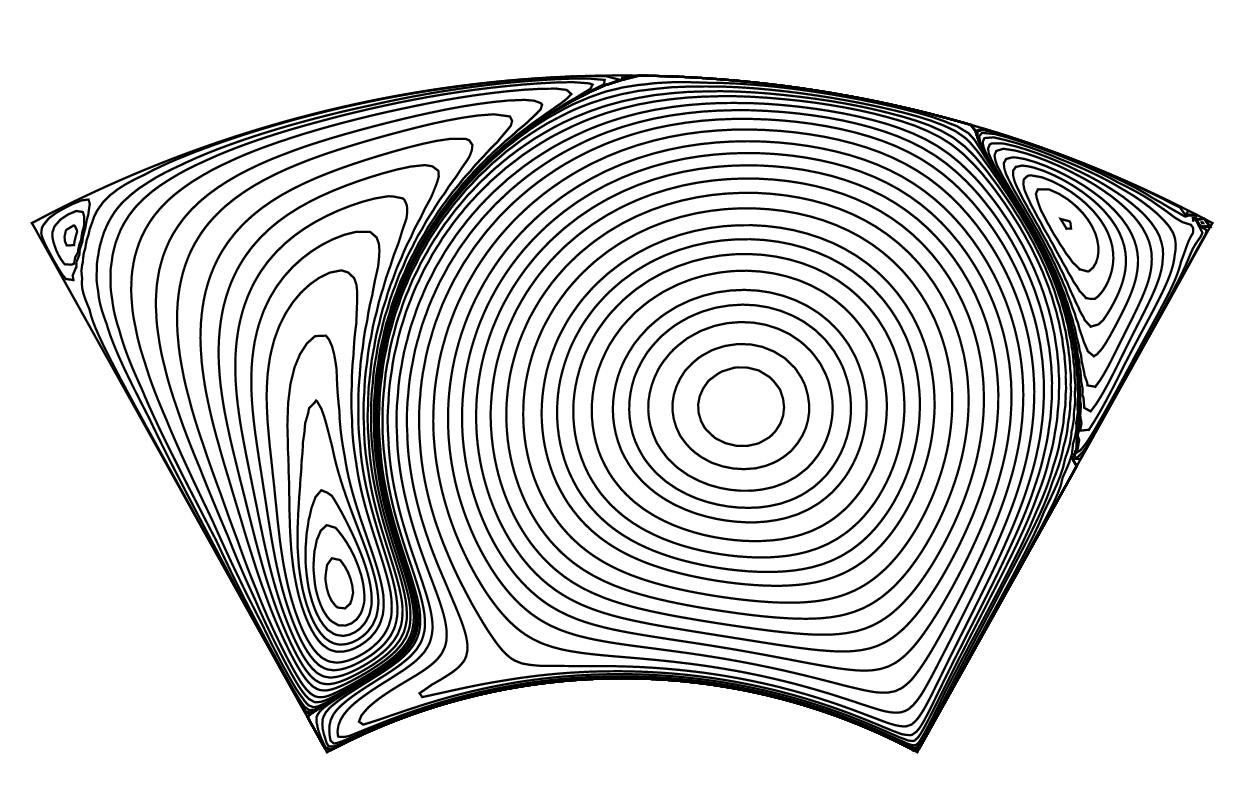}
  \end{subfigure}
  \begin{subfigure}[b]{0.375\textwidth}
  \centering
  \includegraphics[width=\linewidth]{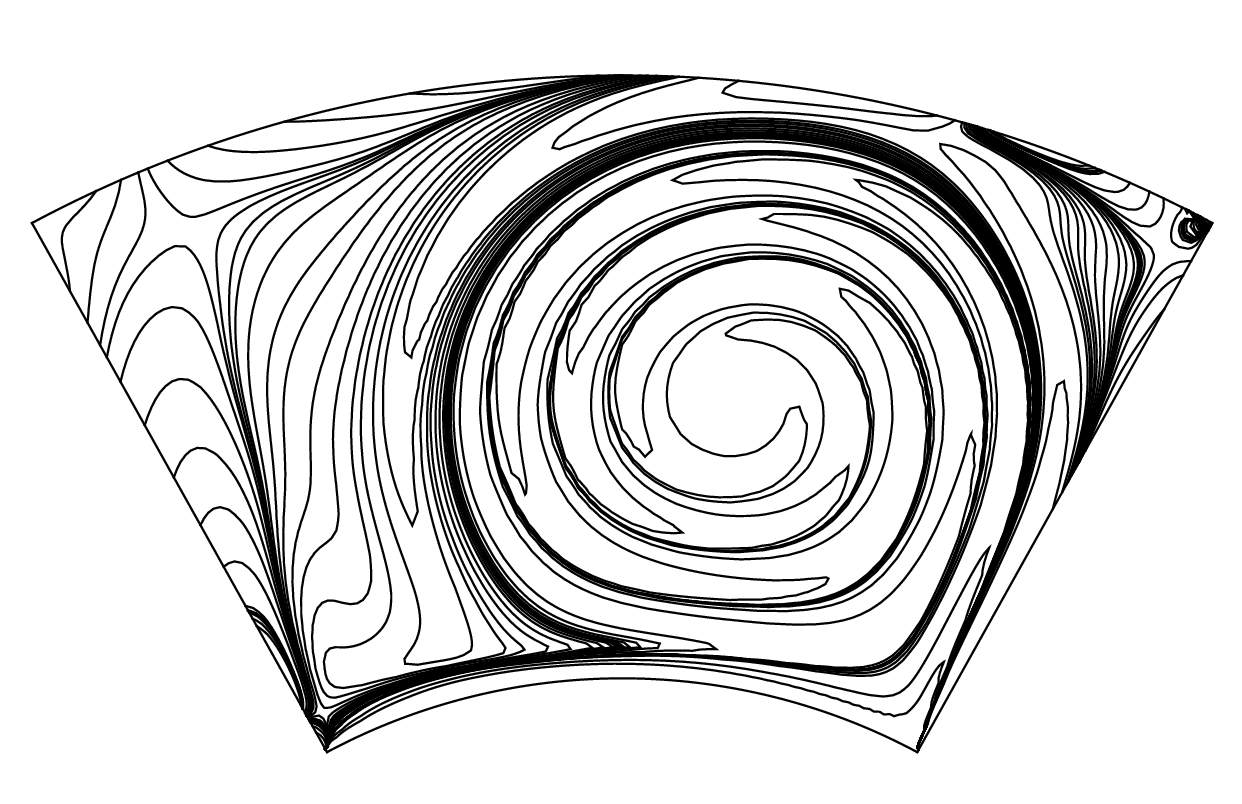}
  \end{subfigure}
  \begin{minipage}[b]{0.02\textwidth}
  \subcaption{ }\label{fig:P2_strvort2000}
  \end{minipage}
    \\
  \begin{subfigure}[b]{0.375\textwidth}
  \centering
  \includegraphics[width=\linewidth]{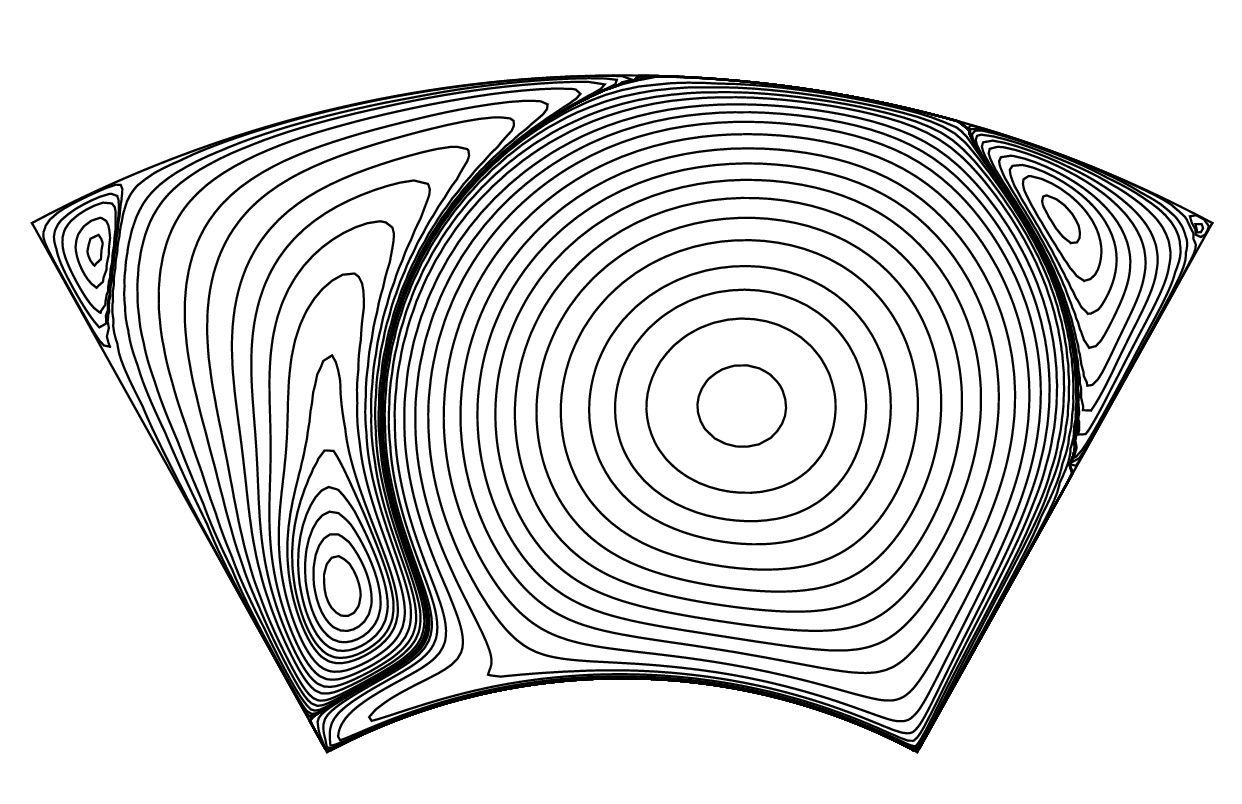}
  \end{subfigure}
  \begin{subfigure}[b]{0.375\textwidth}
  \centering
  \includegraphics[width=\linewidth]{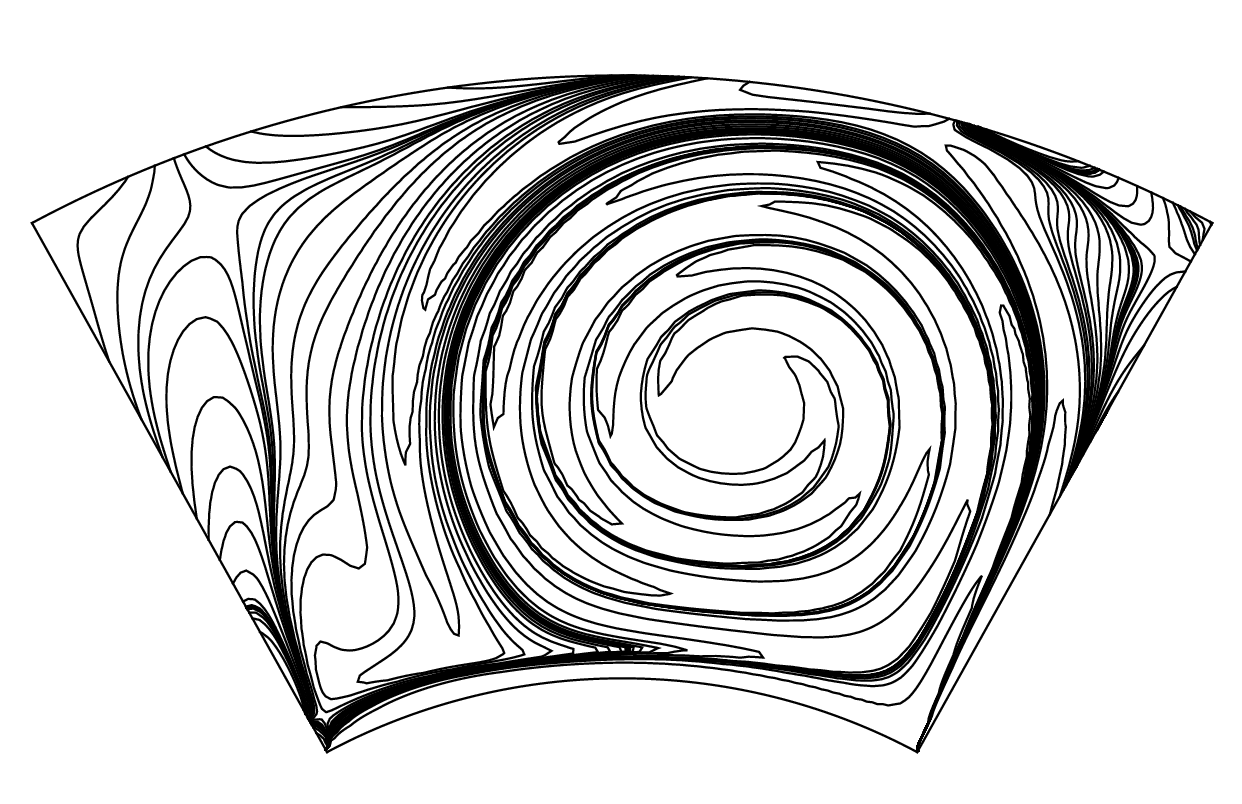}
  \end{subfigure}
  \begin{minipage}[b]{0.02\textwidth}
  \subcaption{ }\label{fig:P2_strvort3000}
  \end{minipage}
\caption{Problem 2: Contours of streamfunction (left) and vorticity (right) computed with $65\times65$ grids: (a) $Re=55$, (b) $Re=350$, (c) $Re=1000$, (d) $Re=2000$ and (e) $Re=3000$.}\label{fig:P2_strmvort}
\end{figure}

The computations for this problem are carried out for Reynolds numbers 55, 350, 1000, 2000, and 3000. The post processed steady-state streamfunction $(\psi)$ and vorticity $(\omega)$ contours for these $Re$ values are plotted in \cref{fig:P2_strmvort}. As $Re$ increases the primary vortex is seen to move gradually towards the right wall of the cavity. Similar to the case of flow inside a square cavity, with an increase in the value of $Re$ the secondary vortices at the corners opposite to the moving wall grow in size, particularly the secondary left vortex gains significantly more size as compared to the secondary right vortex and for higher values of $Re$ the former tends to occupy the entire left side of the cavity. Further, at $Re=3000$ a tertiary vortex is seen to form at the top left corner of the cavity. It is to be mentioned that all the findings are in accordance with those available in the literature \citep{LeeTsuei1993,RayKalita2010,YuTian2013b,SenKalita2015,DasPanditRay2023} both experimentally and numerically. Additionally, at higher $Re$ values the vorticity contours move away from the center of the cavity toward the cavity walls thereby developing very strong vorticity gradients in the vicinity of the cavity walls.

\begin{figure}[!h]
\centering
  \begin{subfigure}[b]{0.44\textwidth}
  \centering
  \includegraphics[width=\linewidth]{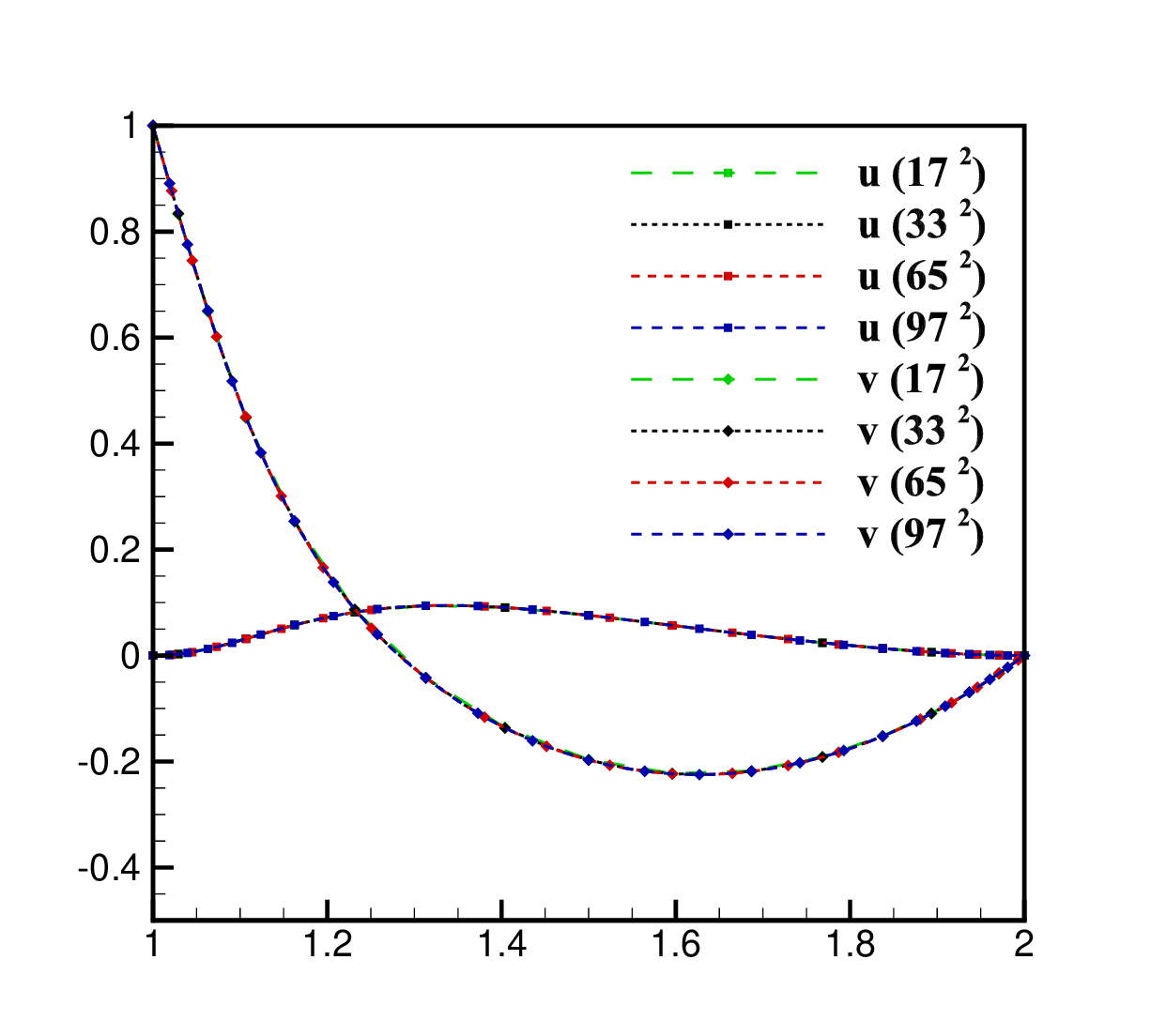}
  \end{subfigure}
  \begin{minipage}[b]{0.02\textwidth}
  \subcaption{ }\label{fig:P2_uvgrid55}
  \end{minipage}
  \begin{subfigure}[b]{0.44\textwidth}
  \centering
  \includegraphics[width=\linewidth]{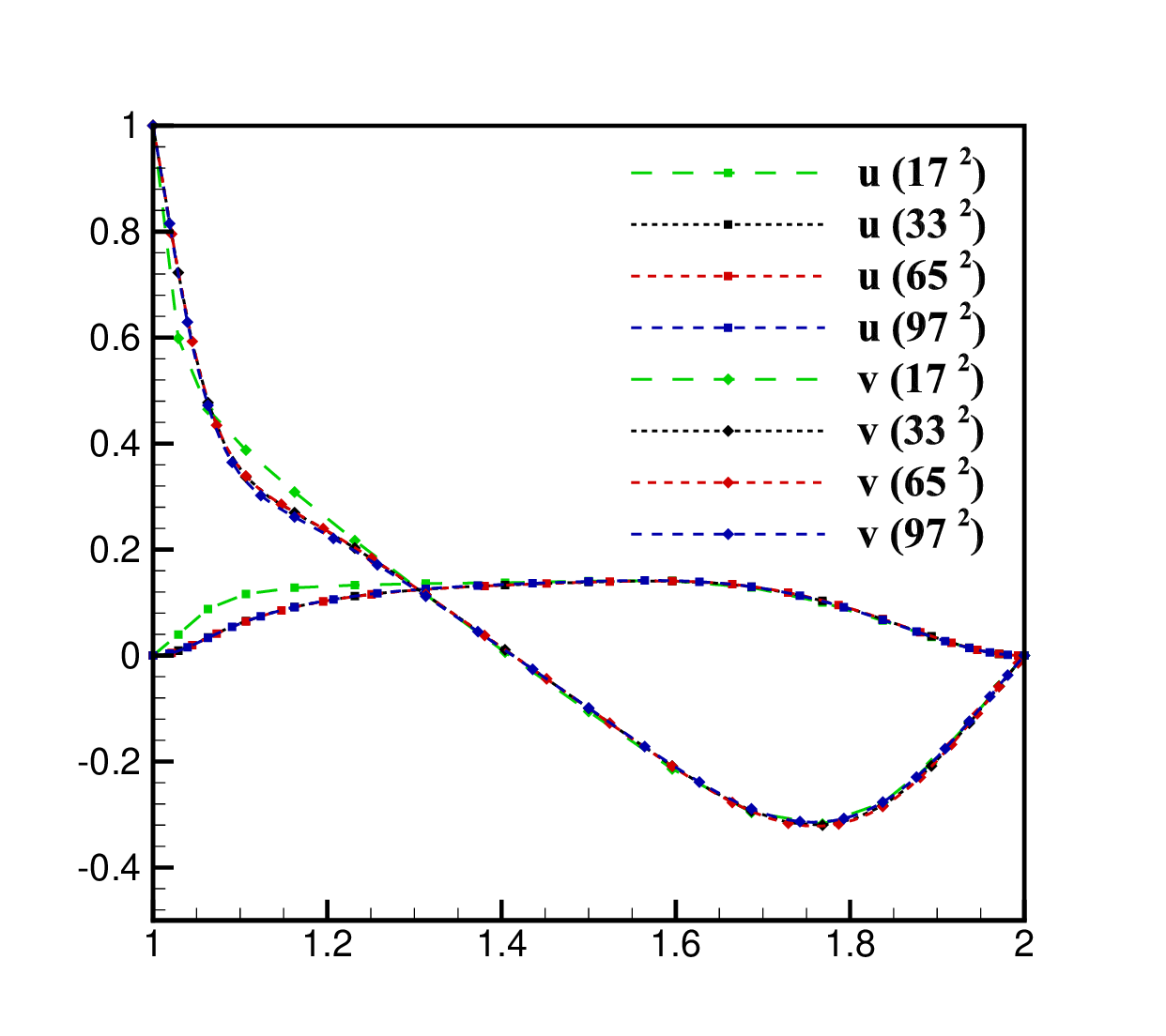}
  \end{subfigure}
  \begin{minipage}[b]{0.02\textwidth}
  \subcaption{ }\label{fig:P2_uvgrid350}
  \end{minipage}
\caption{Problem 2: The $u$- and $v$-velocity profiles along $\theta=\dfrac{\pi}{2}$ computed on grids of different sizes for (a) $Re=55$ and (b) $Re=350$.}\label{fig:P2_gridindp}
\end{figure}

\Cref{fig:P2_gridindp} presents the $u$ and $v$ velocity profiles along the radial line $\theta=\dfrac{\pi}{2}$ computed on grids of four different sizes $17\times17$, $33\times33$, $65\times65$ and $97\times97$ for $Re$'s $55$ and $350$. These figures show that the computed results are unaffected by the change in grid points and a $65\times65$ grid is sufficient to achieve a grid-independent solution for both values of the Reynolds number. 

For this problem, we have also computed the perceived rate of convergence \cite{DekaSen2021}. Despite best efforts, we could not trace the estimation of perceived order in nonuniform polar grid. The increased flow complexity in case of higher $Re$ value makes the estimation of rate of convergence computation extensive \cite{DekaSen2021}. Therefore the perceived order of convergence has been computed only for the cases $Re=55$ and 350. We employ grids of three different sizes \textit{viz.} $33\times33$, $65\times65$ and $97\times97$ and compute with $\dt=1e-4$ to minimize temporal error. The perceived order of convergence for streamfunction has been estimated in \cref{table:P2_spacestream} whereas convergence of vorticity is compiled in \cref{table:P2_spacevorticity}. From these tables, we can observe that the order of convergence for vorticity, in $L_2$-norm error, drops below quadratic because of increased flow complexity. Moreover, it is amply clear that the mesh $97\times97$ should be ideally suited to compute in conjunction with the present discretization strategy.

\begin{table}[!h]
\centering
\caption{Problem 2: $L_1$ and $L_2$ norm difference in streamfunction and perceived order of convergence in space.}\label{table:P2_spacestream}
\vspace{0.1cm}
\renewcommand*{\arraystretch}{0.7}
\begin{tabular}
{ M{0.05\textwidth} M{0.1\textwidth} M{0.30\textwidth} M{0.05\textwidth} M{0.30\textwidth} M{0.05\textwidth}}
\hline
$Re$ &Grid size & $L_1-$norm                                          & order &   $L_2-$norm                                        & order  \\
\hline
55   &  $33^2$  &                                                     &       &                                                     &        \\
     &          &$\left\Vert \psi_2 - \psi_1 \right\Vert$=4.829202e-5 &       &$\left\Vert \psi_2 - \psi_1 \right\Vert$=7.723404e-5 &        \\
     &  $65^2$  &                                                     & 3.17  &                                                     & 2.64   \\
     &          &$\left\Vert \psi_3 - \psi_1 \right\Vert$=6.470839e-5 &       &$\left\Vert \psi_3 - \psi_1 \right\Vert$=1.039171e-4 &        \\
     &  $97^2$  &                                                     &       &                                                     &        \\
\hline
350  &  $33^2$  &                                                     &       &                                                     &        \\
     &          &$\left\Vert \psi_2 - \psi_1 \right\Vert$=1.710546e-4 &       &$\left\Vert \psi_2 - \psi_1 \right\Vert$=2.594515e-4 &        \\
     &  $65^2$  &                                                     & 2.61  &                                                     & 2.41   \\
     &          &$\left\Vert \psi_3 - \psi_1 \right\Vert$=2.469988e-4 &       &$\left\Vert \psi_3 - \psi_1 \right\Vert$=3.867575e-4 &        \\
     &  $97^2$  &                                                     &       &                                                     &        \\
\hline
\end{tabular}
\end{table}

\begin{table}[!h]
\centering
\caption{Problem 2: $L_1$ and $L_2$ norm difference in vorticity and perceived order of convergence in space.}\label{table:P2_spacevorticity}
\vspace{0.1cm}
\renewcommand*{\arraystretch}{0.7}
\begin{tabular}
{ M{0.05\textwidth} M{0.1\textwidth} M{0.30\textwidth} M{0.05\textwidth} M{0.30\textwidth} M{0.05\textwidth}}
\hline
$Re$ & Grid size& $L_1-$norm                                              & order &   $L_2-$norm                                            & order  \\
\hline
55   &  $33^2$  &                                                         &       &                                                         &        \\
     &          &$\left\Vert \omega_2 - \omega_1 \right\Vert$=3.057162e-1 &       &$\left\Vert \omega_2 - \omega_1 \right\Vert$=2.983460e0  &        \\
     &  $35^2$  &                                                         & 2.26  &                                                         & 1.52   \\
     &          &$\left\Vert \omega_3 - \omega_1 \right\Vert$=4.672710e-1 &       &$\left\Vert \omega_3 - \omega_1 \right\Vert$=5.259851e0  &        \\
     &  $97^2$  &                                                         &       &                                                         &        \\
\hline
350  &  $33^2$  &                                                         &       &                                                         &        \\
     &          &$\left\Vert \omega_2 - \omega_1 \right\Vert$=3.919087e-1 &       &$\left\Vert \omega_2 - \omega_1 \right\Vert$=3.232429e0  &        \\
     &  $65^2$  &                                                         & 2.52  &                                                         & 1.66   \\
     &          &$\left\Vert \omega_3 - \omega_1 \right\Vert$=5.739092e-1 &       &$\left\Vert \omega_3 - \omega_1 \right\Vert$=5.534059e0  &        \\
     &  $97^2$  &                                                         &       &                                                         &        \\
\hline
\end{tabular}
\end{table}

\begin{figure}[!h]
\centering
  \begin{subfigure}[b]{0.44\textwidth}
  \centering
  \includegraphics[width=\linewidth]{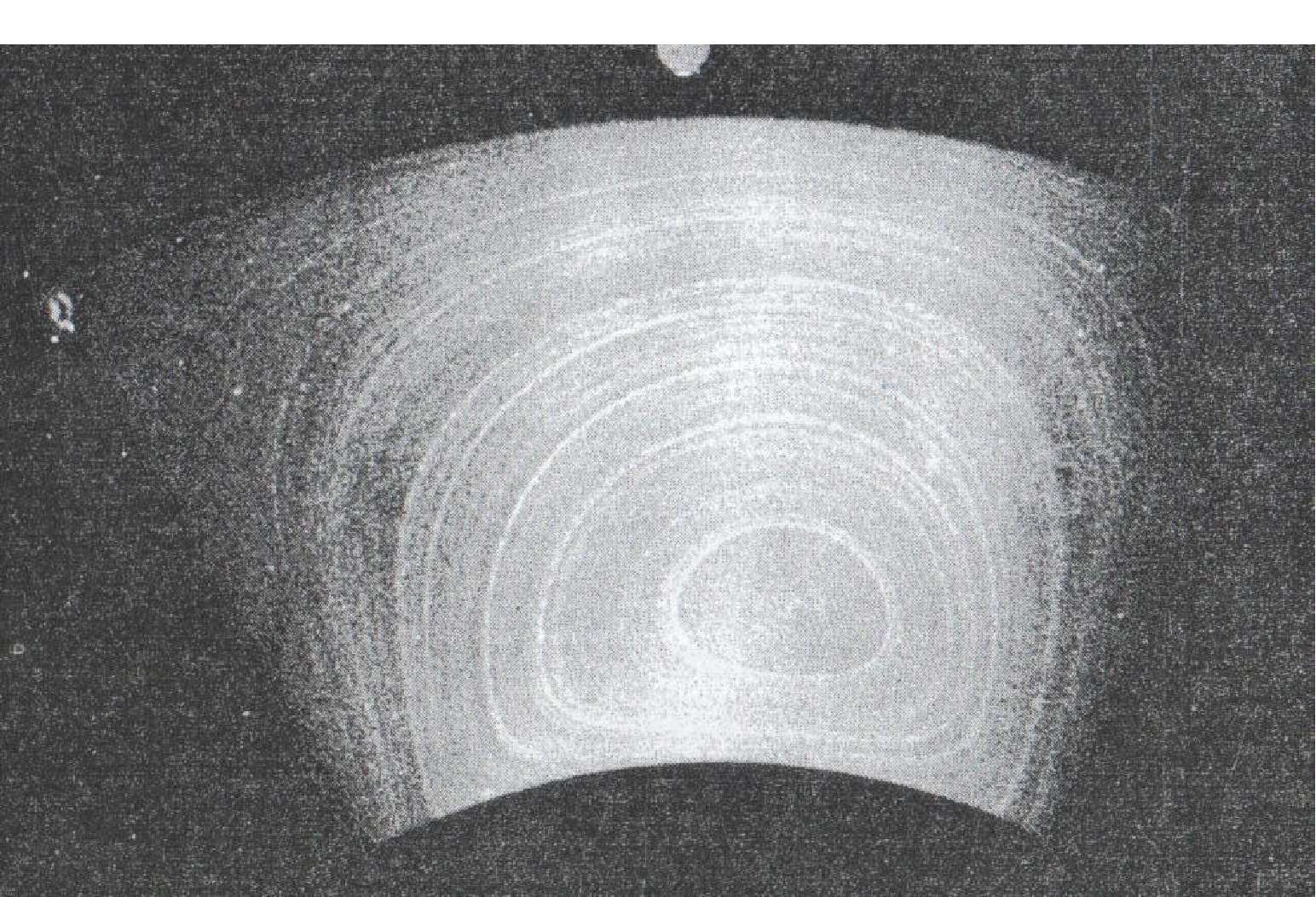}
  \end{subfigure}
  \begin{subfigure}[b]{0.44\textwidth}
  \centering
  \includegraphics[width=\linewidth]{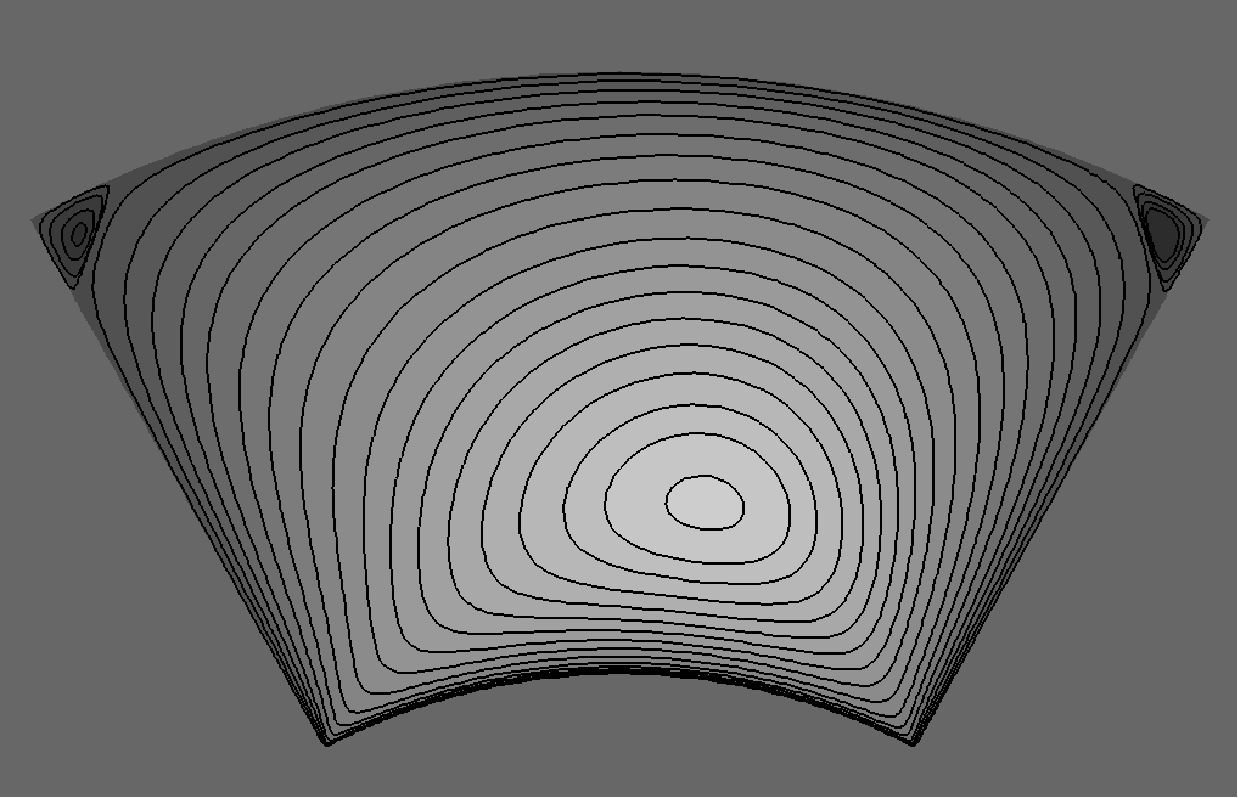}
  \end{subfigure}
  \begin{minipage}[b]{0.02\textwidth}
  \subcaption{ }\label{fig:P2_stream55}
  \end{minipage}
  \\
  \begin{subfigure}[b]{0.44\textwidth}
  \centering
  \includegraphics[width=\linewidth]{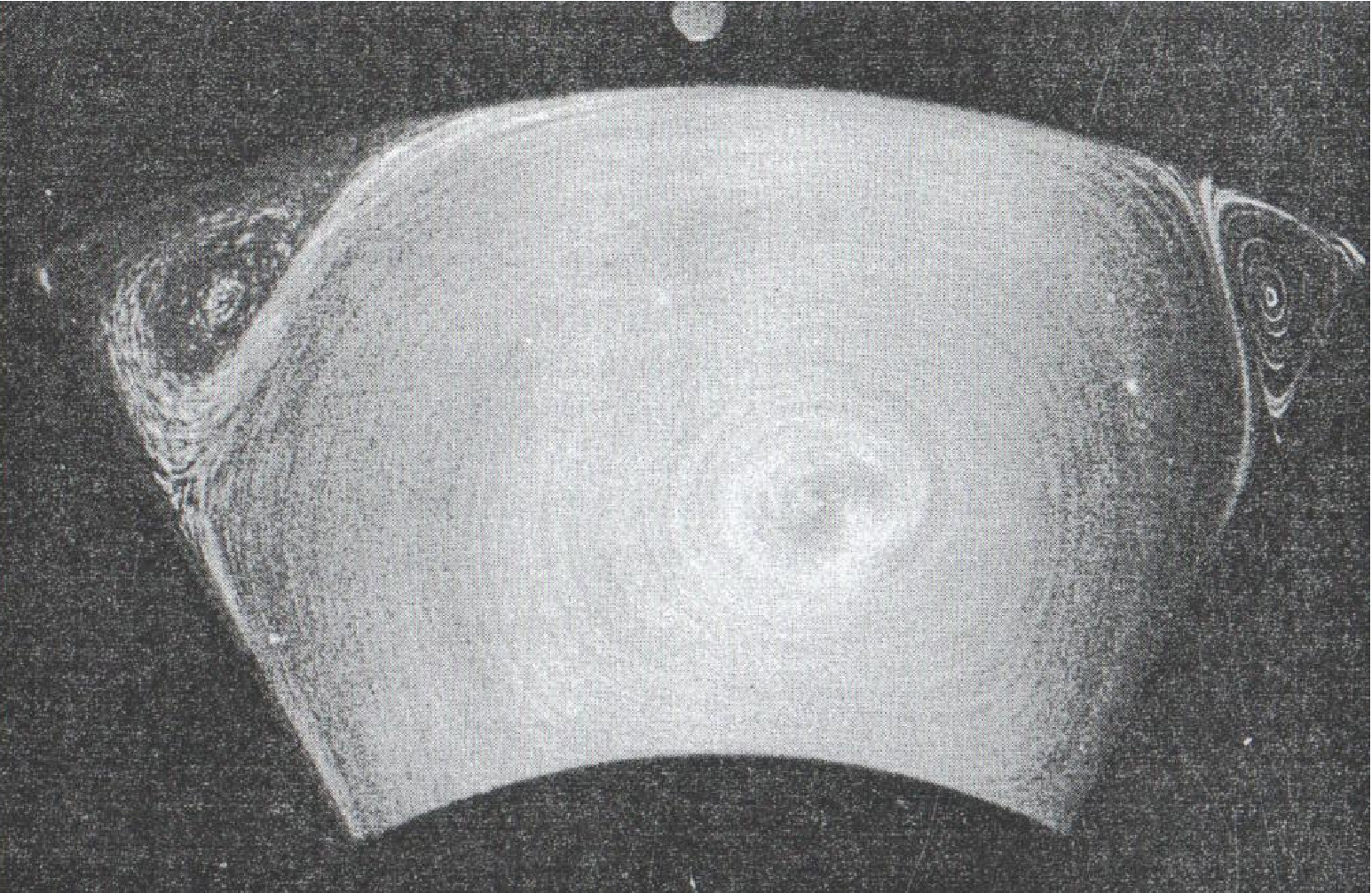}
  \end{subfigure}
  \begin{subfigure}[b]{0.44\textwidth}
  \centering
  \includegraphics[width=\linewidth]{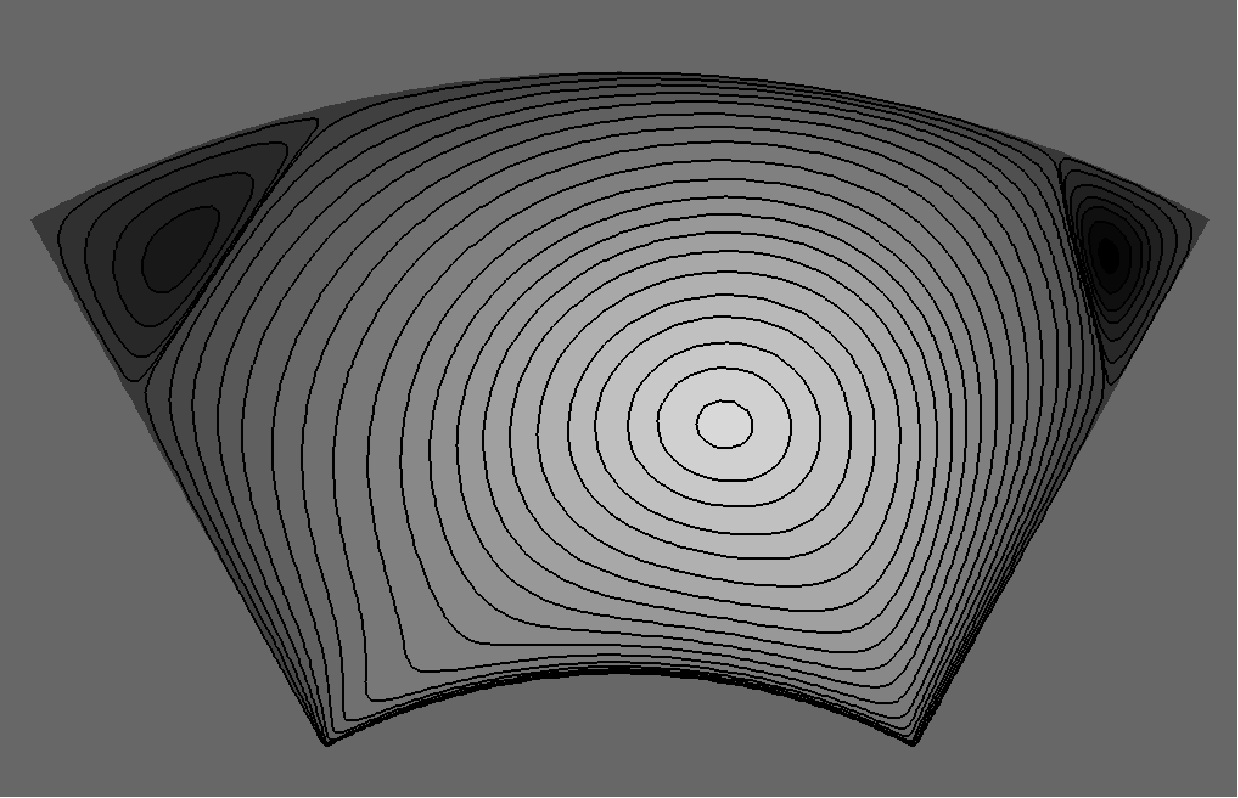}
  \end{subfigure}
  \begin{minipage}[b]{0.02\textwidth}
  \subcaption{ }\label{fig:P2_stream350}
  \end{minipage}
\caption{Problem 2: Comparison of experimental \citep{FuchsTillmark1985} (left) and numerical (right) steady-state streamfunction contours of driven polar cavity problem for (a) $Re=55$ and (b) $Re=350$.}\label{fig:P2_expnum}
\end{figure}

\begin{figure}[!h]
\centering
  \begin{subfigure}[b]{0.44\textwidth}
  \centering
  \includegraphics[width=\linewidth]{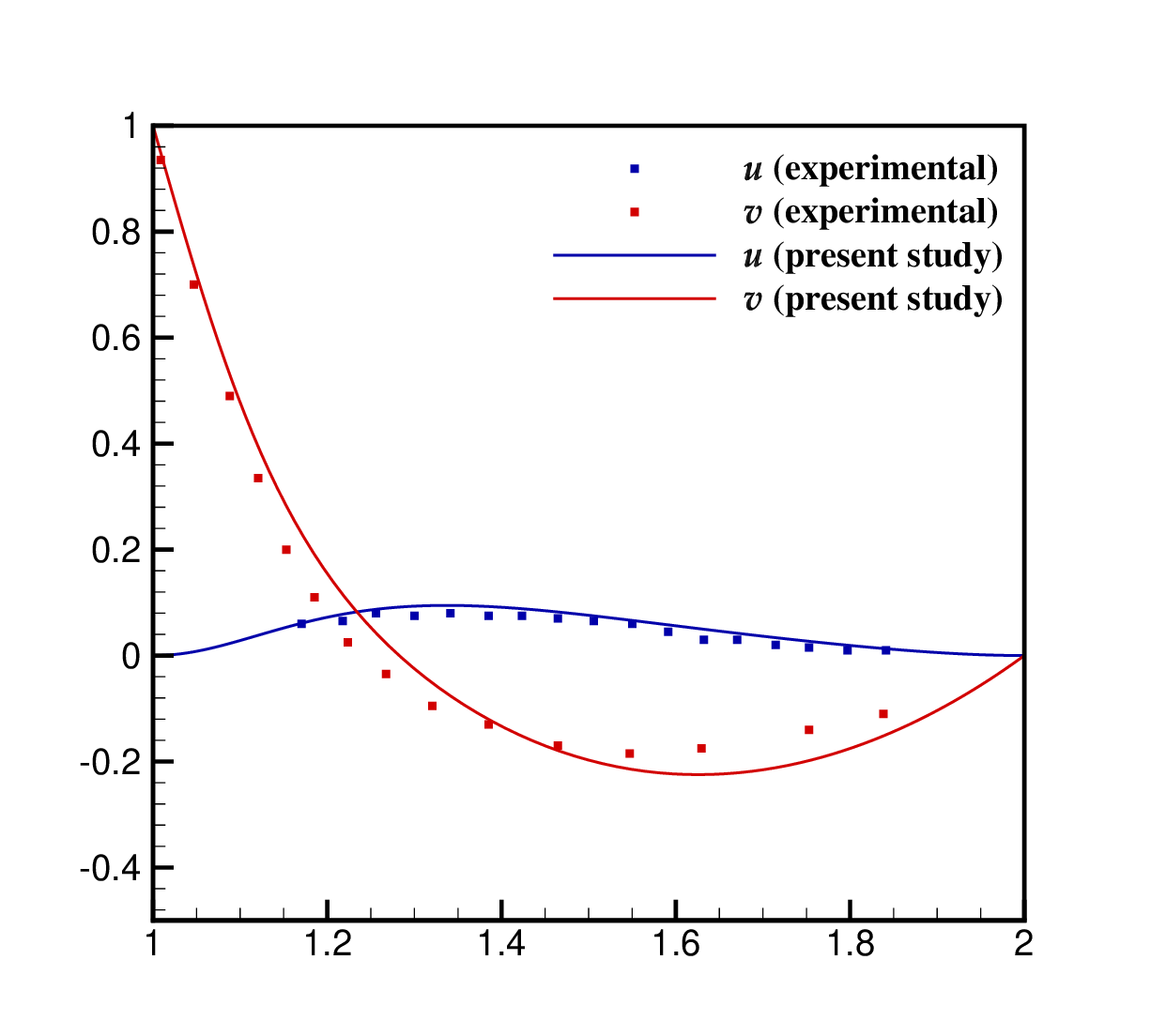}
  \end{subfigure}
  \begin{minipage}[b]{0.02\textwidth}
  \subcaption{ }\label{fig:P2_uv55}
  \end{minipage}
  \begin{subfigure}[b]{0.44\textwidth}
  \centering
  \includegraphics[width=\linewidth]{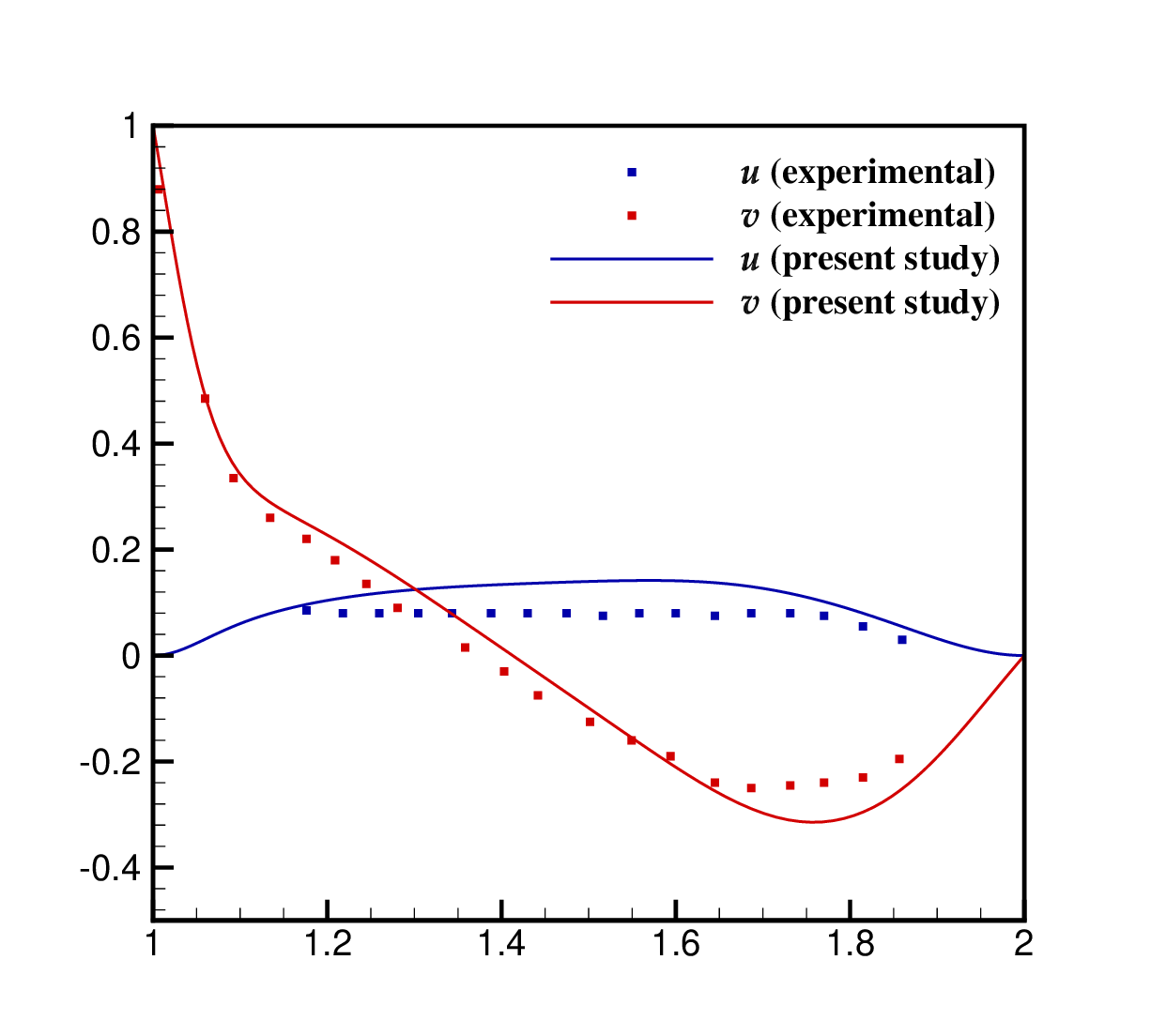}
  \end{subfigure}
  \begin{minipage}[b]{0.02\textwidth}
  \subcaption{ }\label{fig:P2_uv350}
  \end{minipage}
\caption{Problem 2: Comparisons of steady-state $u$- and $v$-velocity profiles along the radial line $\theta=\frac{\pi}{2}$ with \citep{FuchsTillmark1985} for (a) $Re=55$ and (b) $Re=350$.}\label{fig:P2_uvcomp}
\end{figure}

\begin{figure}[!h]
\centering
  \begin{subfigure}[b]{0.44\textwidth}
  \centering
  \begin{tikzpicture}
  \draw (0,0) node[inner sep=0] {\includegraphics[width=\linewidth]{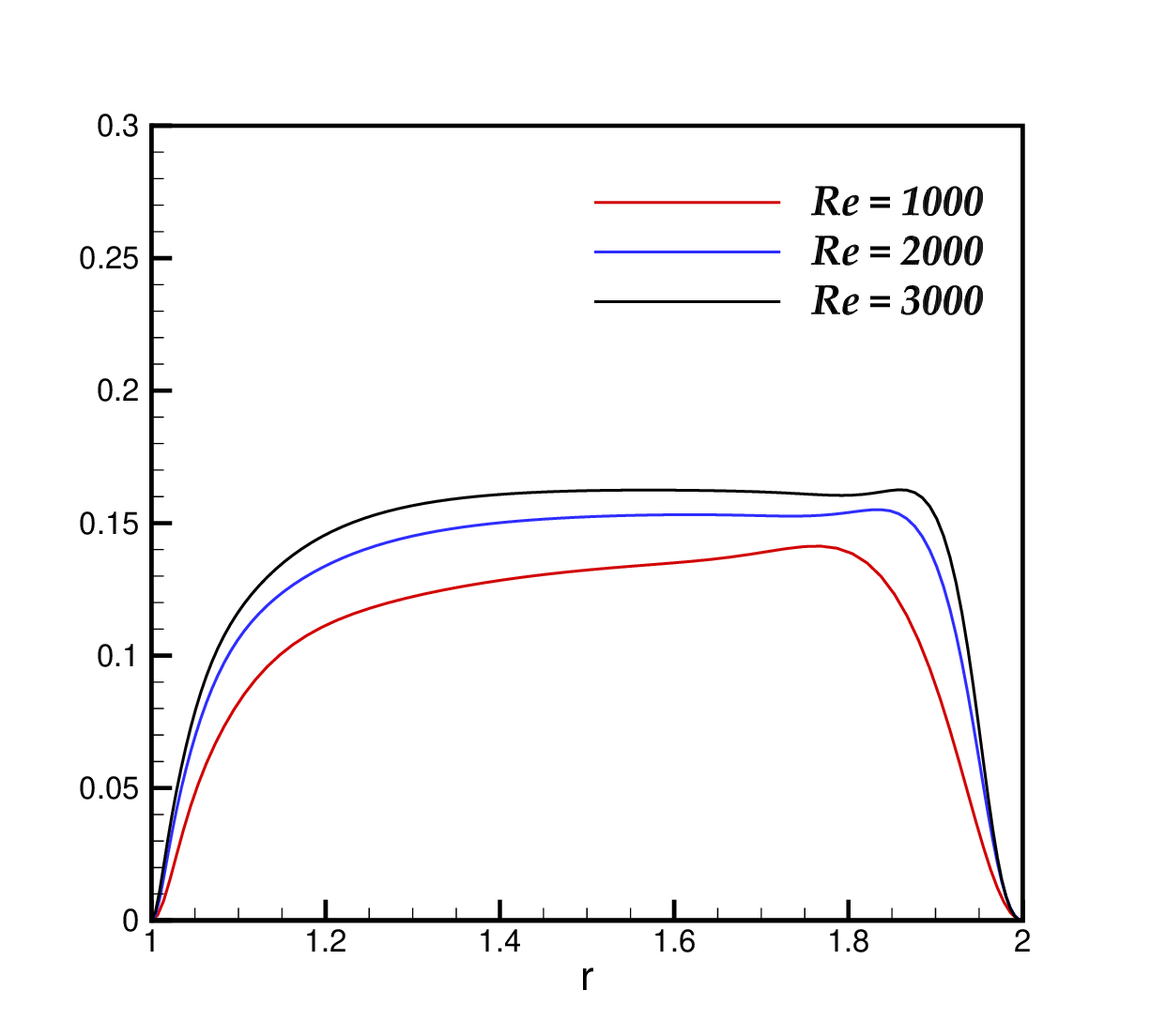}};
  \draw (-3.35,0) node {\small \rotatebox[origin=c]{90}{$u$}};
  \end{tikzpicture}
  \end{subfigure}
  \begin{minipage}[b]{0.02\textwidth}
  \subcaption{ }\label{fig:P2_u_all}
  \end{minipage}
  \begin{subfigure}[b]{0.44\textwidth}
  \centering
  \begin{tikzpicture}
  \draw (0,0) node[inner sep=0] {\includegraphics[width=\linewidth]{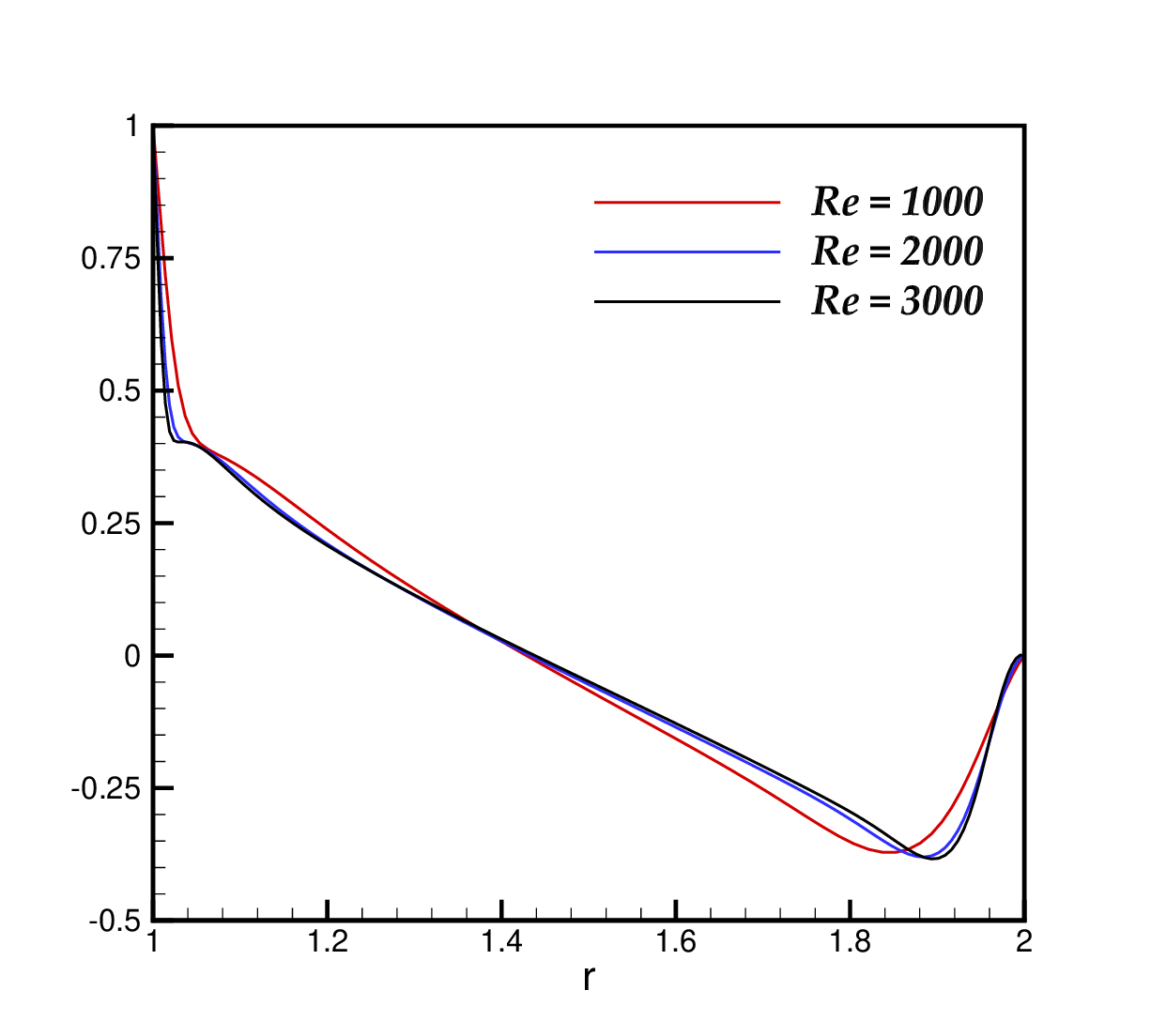}};
  \draw (-3.35,0) node {\small \rotatebox[origin=c]{90}{$v$}};
  \end{tikzpicture}
  \end{subfigure}
  \begin{minipage}[b]{0.02\textwidth}
  \subcaption{ }\label{fig:P2_v_all}
  \end{minipage}
\caption{Problem 2: Steady-state (a) $u$- and (b) $v$-velocity profiles along the radial line $\theta=\dfrac{\pi}{2}$ for different $Re$ values.}\label{fig:P2_uv_all}
\end{figure}

\begin{figure}[!h]
\centering
  \begin{subfigure}[b]{0.44\textwidth}
  \centering
  \includegraphics[width=\linewidth]{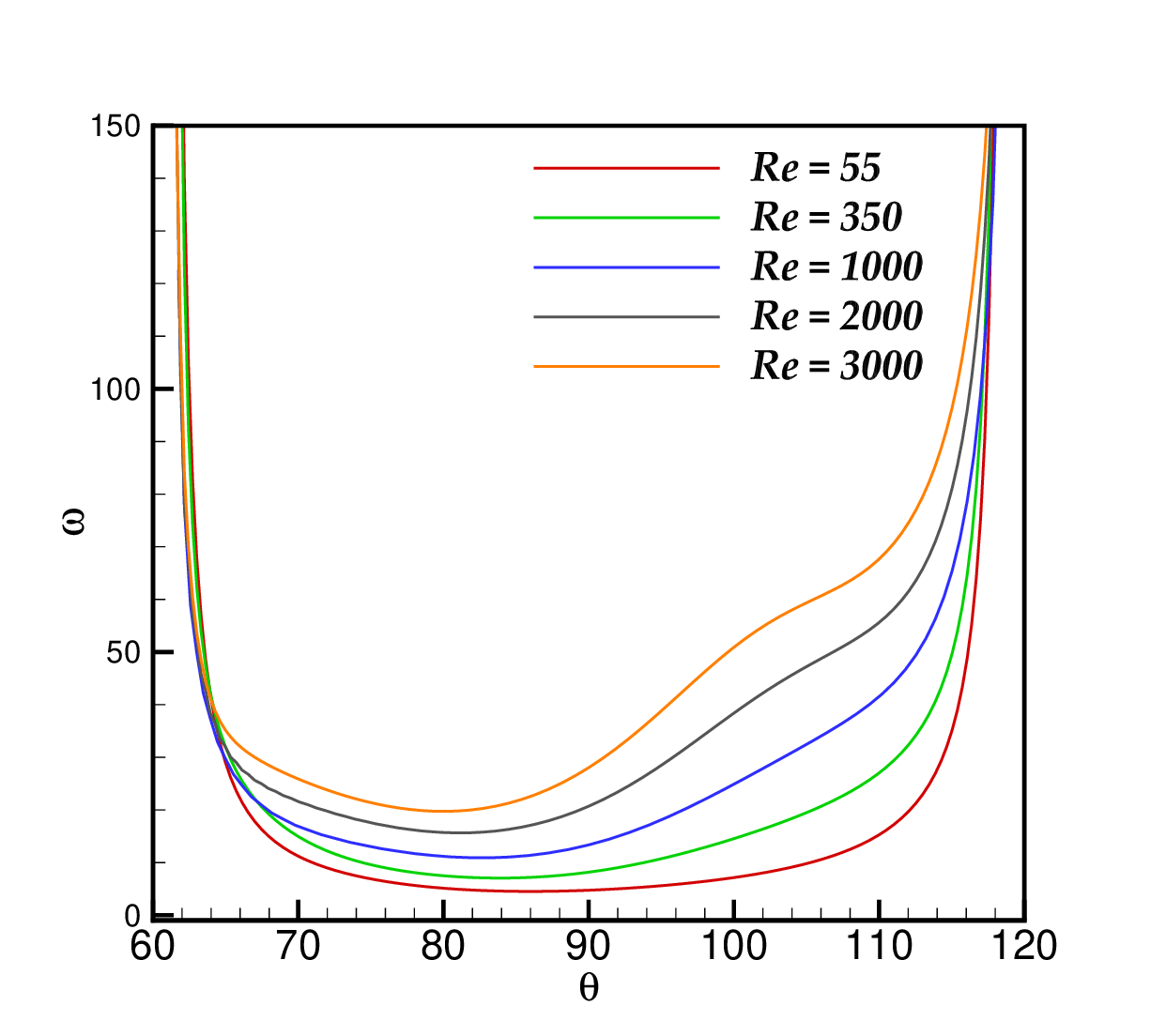}
  \end{subfigure}
  \begin{minipage}[b]{0.02\textwidth}
  \subcaption{ }\label{fig:P2_vort_r0}
  \end{minipage}
  \begin{subfigure}[b]{0.44\textwidth}
  \centering
  \includegraphics[width=\linewidth]{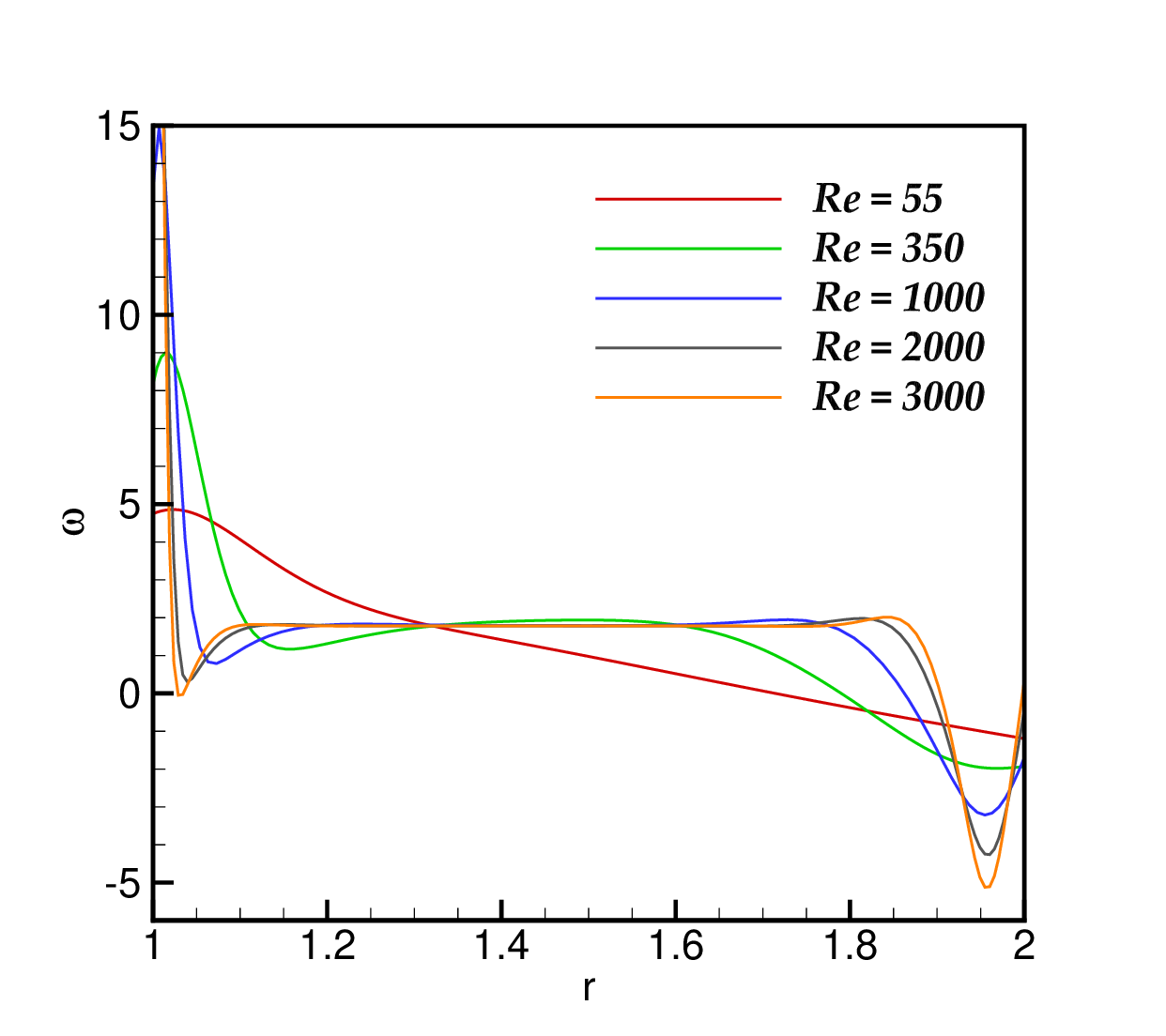}
  \end{subfigure}
  \begin{minipage}[b]{0.02\textwidth}
  \subcaption{ }\label{fig:P2_vort_th90}
  \end{minipage}
\caption{Problem 2: Vorticity contours for different $Re$ values along (a) the rotating wall and (b) the radial line $\theta=\dfrac{\pi}{2}$.}\label{fig:P2_vorticity_contour}
\end{figure}

Next, we carry out a qualitative comparison of the numerically attained results for $Re=55$ and $350$ with those of the experimental findings reported by Fuchs and Tillmark \citep{FuchsTillmark1985}. In \cref{fig:P2_expnum} we present the numerically simulated steady-state streamlines for $Re=55$ and $350$ along with those of the experimental ones. Moreover, the numerically obtained $u$ and $v$ velocity profiles along the radial line $\theta=\dfrac{\pi}{2}$ are compared to those of \citep{FuchsTillmark1985} in \cref{fig:P2_uvcomp}. Both numerical and experimental solutions show good agreement with each other for each $Re$ value. The $u$ and $v$ velocity profiles along $\theta=\dfrac{\pi}{2}$ for $Re$ values 1000, 2000, and 3000 are presented in \cref{fig:P2_uv_all}, though there is no quantitative evidence available for the same in the literature. Additionally, the variation of the vorticity along the rotating wall and the radial line $\theta=\dfrac{\pi}{2}$ are depicted in \cref{fig:P2_vorticity_contour}. It can be seen that both the radial and tangential velocities become stronger in the vicinity of the circular walls of the cavity, whereas the vorticity gradients can be seen to develop near all the boundary walls at higher values of Reynolds numbers. These observations are in line with those reported in \citep{YuTian2013b}. In \cref{table:P2_pdc1,table:P2_pdc2} we compile all our quantitative data pertaining to the primary and secondary vortices. \Cref{table:P2_pdc1,table:P2_pdc2} also carry out a comparison of strength and location of the vortices with those available in the literature for $55\leq Re \leq 3000$. It is heartening to notice that our computation compares well with the firmly established studies available in the literature. 

\begin{table}[!h]
\centering
\caption{Problem 2: Strength and location of the centre of primary vortex for different Reynolds numbers. }\label{table:P2_pdc1}
\vspace{0.1cm}
\renewcommand*{\arraystretch}{0.8}
\begin{tabular}{ M{0.1\textwidth} M{0.15\textwidth} M{0.15\textwidth} M{0.15\textwidth} M{0.15\textwidth} M{0.1\textwidth} }
\hline
$Re$ & \quad                     &   Grid Size  &$\psi_{max}$ &    $x$   &    $y$  \\
\hline
55   &  \citep{YuTian2013b}      &$513\times513$&   0.1155    &  0.141   &  1.281  \\
     &  \citep{SenKalita2015}    &$83\times119$ &   0.1155    &  0.138   &  1.285  \\
     &  \citep{DasPanditRay2023} & $81\times81$ &   0.1156    &  0.142   &  1.285  \\
     &  Present                  & $65\times65$ &   0.1155    &  0.154   &  1.283  \\  
\hline
350  &  \citep{YuTian2013b}      &$513\times513$&   0.1263    &  0.167   &  1.414  \\
     &  \citep{SenKalita2015}    &$83\times119$ &   0.1263    &  0.163   &  1.411  \\
     &  \citep{DasPanditRay2023} & $81\times81$ &   0.1266    &  0.171   &  1.414  \\
     &  Present                  & $65\times65$ &   0.1258    &  0.170   &  1.418  \\  
\hline
1000 &  \citep{YuTian2013b}      &$513\times513$&   0.1275    &  0.169   &  1.439  \\
     &  \citep{SenKalita2015}    &$83\times119$ &   0.1275    &  0.166   &  1.436  \\
     &  \citep{DasPanditRay2023} & $81\times81$ &   0.1275    &  0.174   &  1.444  \\
     &  Present                  & $65\times65$ &   0.1272    &  0.173   &  1.441  \\ 
\hline
2000 &  \citep{YuTian2013b}      &$513\times513$&   0.1253    &  0.188   &  1.447  \\
     &  Present                  & $97\times97$ &   0.1250    &  0.194   &  1.447  \\ 
    
\hline
3000 &  \citep{SenKalita2015}    &$109\times157$&   0.1240    &  0.196   &  1.447  \\
     &  Present                  & $97\times97$ &   0.1237    &  0.194   &  1.447  \\   
\hline
\end{tabular}
\end{table}

\begin{table}[!h]
\centering
\caption{Problem 2: Strength and location of the centre of secondary vortices for different Reynolds numbers. }\label{table:P2_pdc2}
\vspace{0.1cm}
\renewcommand*{\arraystretch}{0.8}
\begin{tabular}
{ M{0.05\textwidth} M{0.10\textwidth} M{0.14\textwidth} M{0.08\textwidth} M{0.08\textwidth} M{0.015\textwidth} M{0.14\textwidth} M{0.1\textwidth} M{0.07\textwidth} }
\hline
     &                           &  \multicolumn{3}{c}{Secondary Right}  &  &  \multicolumn{3}{c}{Secondary Left}  \\
\cline{3-5} \cline{7-9}
$Re$ &                           &  $\psi_{min}$   &   $x$   &   $y$     &  &  $\psi_{min}$   &   $x$    &   $y$   \\
\hline
55   &  \citep{YuTian2013b}      &   -8.690e-6     & 0.881   &  1.730    &  & -4.690e-6       & -0.880   &  1.729  \\
     &  \citep{SenKalita2015}    &   -7.964e-6     & 0.883   &  1.731    &  & -4.994e-6       & -0.876   &  1.742  \\
	 &  \citep{DasPanditRay2023} &   -8.102e-6     & 0.882   &  1.735    &  & -4.391e-6       & -0.882   &  1.735  \\
     &  Present                  &   -9.705e-6     & 0.879   &  1.726    &  & -5.212e-6       & -0.879   &  1.726  \\  
\hline
350  &  \citep{YuTian2013b}      &   -5.490e-4     & 0.794   &  1.697    &  & -3.990e-4       & -0.702   &  1.695  \\
     &  \citep{SenKalita2015}    &   -5.469e-4     & 0.796   &  1.690    &  & -4.009e-4       & -0.696   &  1.699  \\
	 &  \citep{DasPanditRay2023} &   -5.435e-4     & 0.796   &  1.689    &  & -4.010e-4       & -0.701   &  1.703  \\
     &  present                  &   -5.466e-4     & 0.806   &  1.684    &  & -3.729e-4       & -0.688   &  1.720  \\
\hline
1000 &  \citep{YuTian2013b}      &   -2.120e-3     & 0.757   &  1.713    &  & -3.460e-3       & -0.590   &  1.552  \\
	 &  \citep{SenKalita2015}    &   -2.099e-3     & 0.755   &  1.717    &  & -3.480e-3       & -0.589   &  1.549  \\
     &  \citep{DasPanditRay2023} &   -2.106e-3     & 0.762   &  1.705    &  & -3.488e-3       & -0.592   &  1.559  \\
     &  Present                  &   -2.192e-3     & 0.755   &  1.707    &  & -3.351e-3       & -0.590   &  1.603  \\
\hline
2000 &  \citep{YuTian2013b}      &   -3.080e-3     & 0.734   &  1.743    &  & -4.830e-3       & -0.522   &  1.406  \\
     &  Present                  &   -3.151e-3     & 0.734   &  1.734    &  & -4.835e-3       & -0.519   &  1.393  \\
\hline
3000 &  \citep{SenKalita2015}    &   -3.325e-3     & 0.724   &  1.757    &  & -7.089e-3       & -0.459   &  1.156  \\
     &  Present                  &   -3.346e-3     & 0.720   &  1.741    &  & -7.062e-3       & -0.452   &  1.149  \\
\hline
\end{tabular}
\end{table}

\clearpage

\subsection{Problem 3: Natural convection in horizontal concentric annulus}\label{ex3}

We further analyze the efficiency of the newly developed scheme to tackle heat transfer by implementing it to solve natural heat convection inside a horizontal concentric annulus. This problem has gained a considerable amount of attention from researchers over the years due to its relevance and applicability in various engineering and physical situations, such as heat exchangers, solar collectors, nuclear reactors, thermal energy storage systems, etc \citep{CrawfordLemlich1962,PoweCarleyCarruth1971,KuehnGoldstein1976,MackBishop1978,TsuiTremblay1984,Shahraki2002,ShiZhaoGuo2006,YangKong2019,ZhangYang2023}. It is governed by the transient Boussinesq equations which in the polar coordinate system are given as,
\begin{subequations}\label{Bous_pol}
\begin{empheq}[left=\empheqlbrace]{align}
&\omega_t=Pr\left(\omega_{rr}+\frac{1}{r}\omega_r+\frac{1}{r^2}\omega_{\theta \theta}\right)-\left(u\omega_r+\frac{v}{r}\omega_\theta\right)+RaPr\left(\mathcal{T}_r\cos{\theta}-\mathcal{T}_\theta\frac{\sin{\theta}}{r}\right),  \label{Bous_pol1} \\
&\omega=-\left(\psi_{rr}+\frac{1}{r}\psi_r+\frac{1}{r^2}\psi_{\theta\theta} \right), \label{Bous_pol2} \\
&\mathcal{T}_t=\left(\mathcal{T}_{rr}+\frac{1}{r}\mathcal{T}_r+\frac{1}{r^2}\mathcal{T}_{\theta\theta} \right)-\left(u\mathcal{T}_r+\frac{v}{r}\mathcal{T}_\theta\right). \label{Bous_pol3}
\end{empheq}
\end{subequations}
where $\mathcal{T}$ denotes the dimensionless temperature, $Pr$ and $Ra$ stand for Prandtl number and Rayleigh number respectively. The Rayleigh number $Ra$ is a dimensionless quantity defined as $Ra=\dfrac{g \be \delta \mathcal{T} L^3}{\kappa \nu}$, where $g$ is the acceleration due to gravity, $\be$ is the coefficient of thermal expansion of the fluid, $\delta \mathcal{T}$ is the temperature difference, $L$ is the length, $\kappa$ is the thermal diffusivity and $\nu$ is the kinematic viscosity of the fluid.

\begin{figure}[!ht]
\centering
  \begin{subfigure}[b]{0.44\textwidth}
  \centering
    \begin{tikzpicture}
        \draw (0,0) node[inner sep=0] {\includegraphics[width=\linewidth]{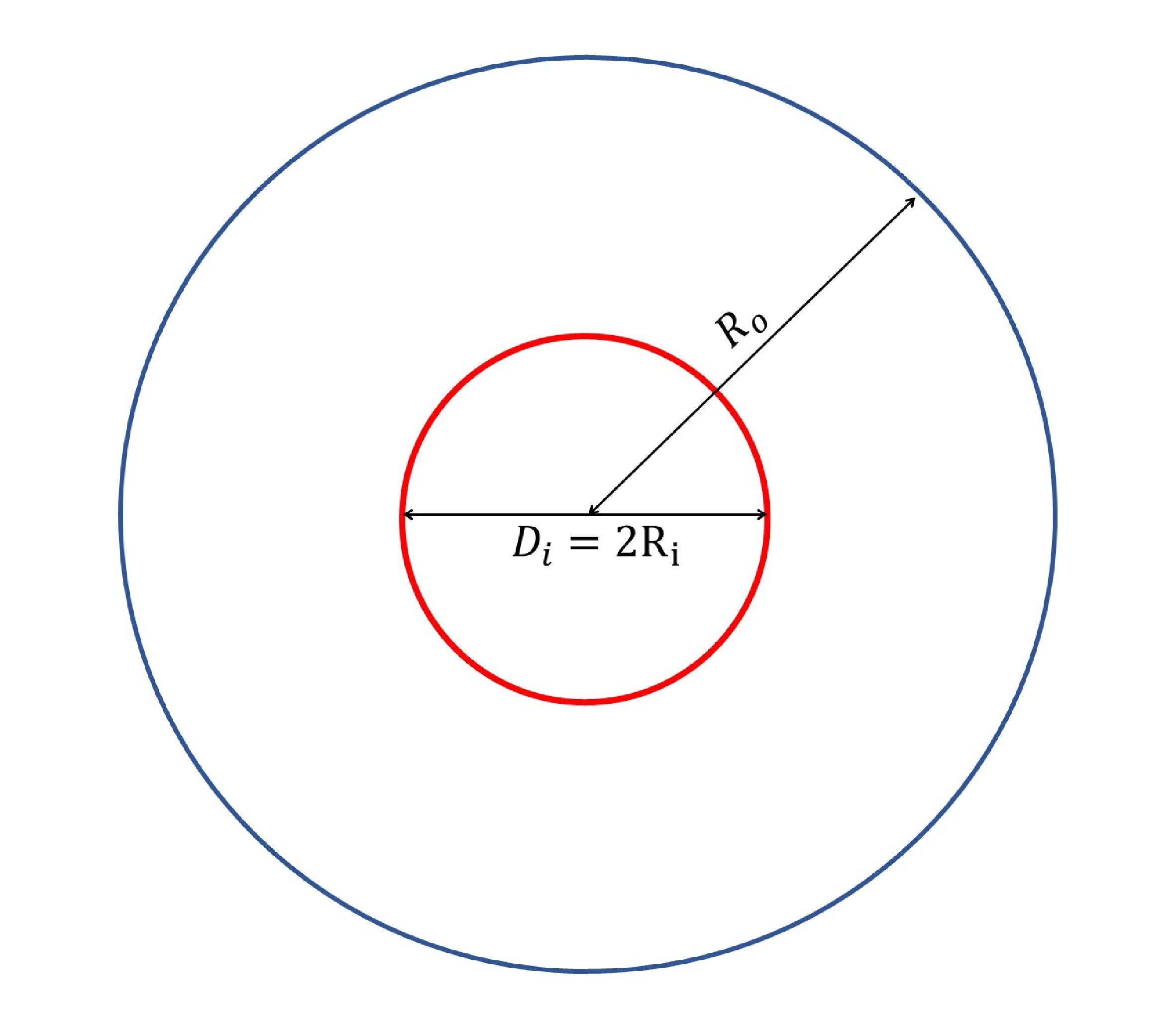}};
        \draw (0,-1.5) node {\small $\mathcal{T}_h=1$};
        \draw (0,-2.5) node {\small $\mathcal{T}_c=0$};
    \end{tikzpicture}
  \end{subfigure}
  \begin{minipage}[b]{0.02\textwidth}
  \subcaption{ }\label{fig:P3_setup}
  \end{minipage}
  \begin{subfigure}[b]{0.44\textwidth}
  \centering
  \includegraphics[width=\linewidth]{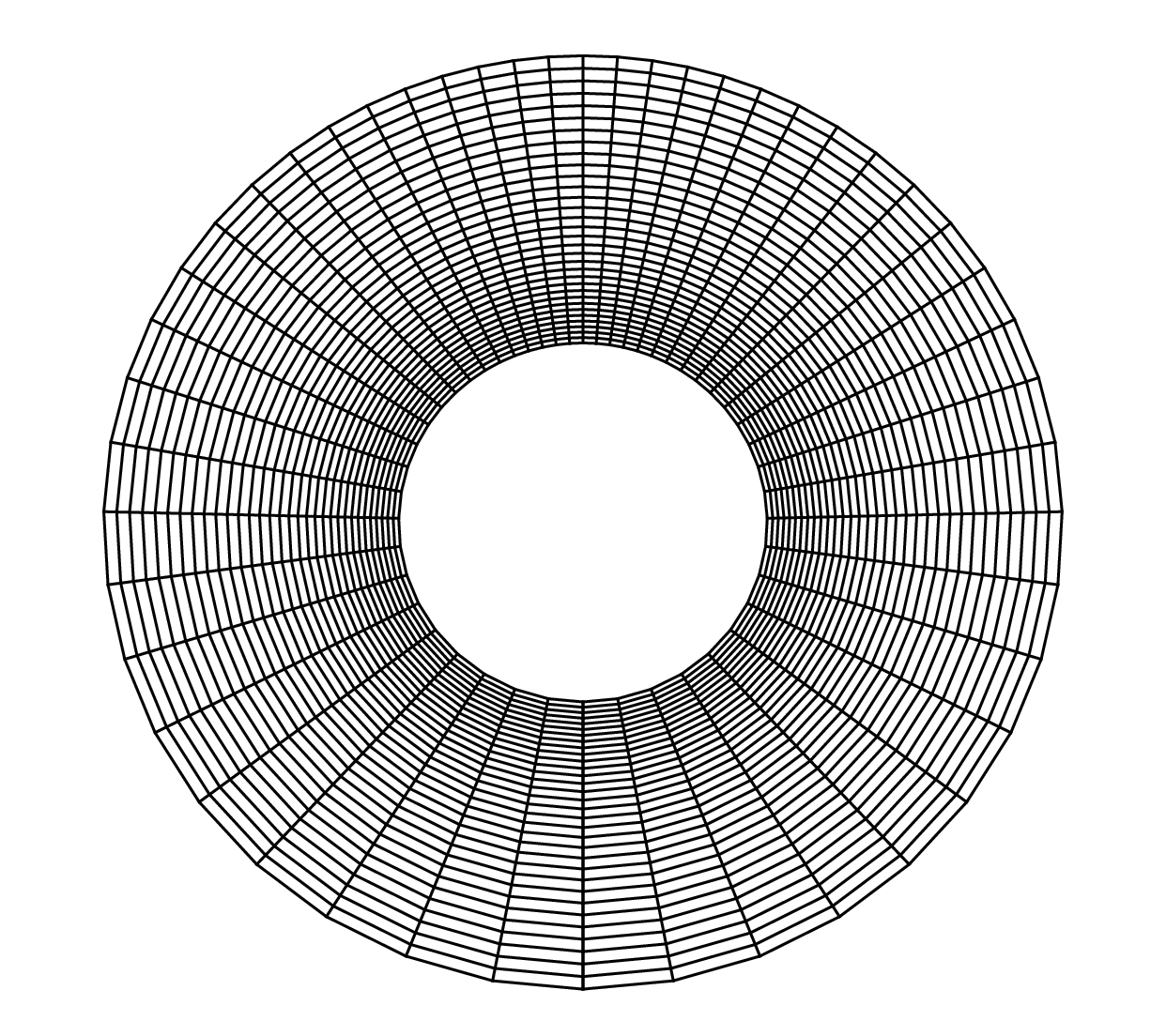}
  \end{subfigure}
  \begin{minipage}[b]{0.02\textwidth}
  \subcaption{ }\label{fig:P3_grid}
  \end{minipage}
\caption{Problem 3: (a) Problem setup and (b) typical noununiiform polar grid for heat transfer in horizontal annulus.}\label{fig:P3_setupgrid}
\end{figure}

The problem setup consists of two concentric circular walls of which the inner circle has a radius $R_i$ and the outer circle has a radius $R_o$. Two different temperatures are maintained at the inner and outer circular walls of the annulus as shown in \cref{fig:P3_setup}. The simulation is carried out for $Ra$ values $2.38e3$, $9.50e3$, $4.70e4$, $6.19e4$, and $1.02e5$. The values of the $Pr$ is set as 0.706, 0.717, and 0.718. For each case, we have considered $L/D_i$=0.8, where $L=R_o-R_i$ and $D_i=2×R_i$. The inner wall is heated to a unit nondimensional temperature $(\mathcal{T}_h=1)$ keeping the outer wall’s temperature at zero $(\mathcal{T}_c=0)$. This temperature difference serves as the driving force for the flow in this simulation. The values of streamfunction and vorticity at the inner and outer boundaries of the annulus are obtained respectively from no-slip criteria between the fluid and the walls of the annulus and one-sided approximations introduced in \cref{right,top,left,bottom}. The prior studies help us to identify that the near boundary region of the inner circle and the radial line $\theta=\dfrac{\pi}{2}$ should be observed cautiously. This motivates us to work with a mesh where grid points are clustered in these regions (see \cref{fig:P3_grid}). The grid is generated by making adequate changes in \cref{P1_grid,P2_grid} and choosing the associated parameters as $L_r=0.8$, $L_\theta=2\pi$, $\Theta_r=\pi$, $\Theta_\theta=2\pi$, $\lambda_r=-0.4$ and $\lambda_\theta=0.25$.

\begin{figure}[!h]
\centering
  \begin{subfigure}[b]{0.44\textwidth}
  \centering
  \includegraphics[width=\linewidth]{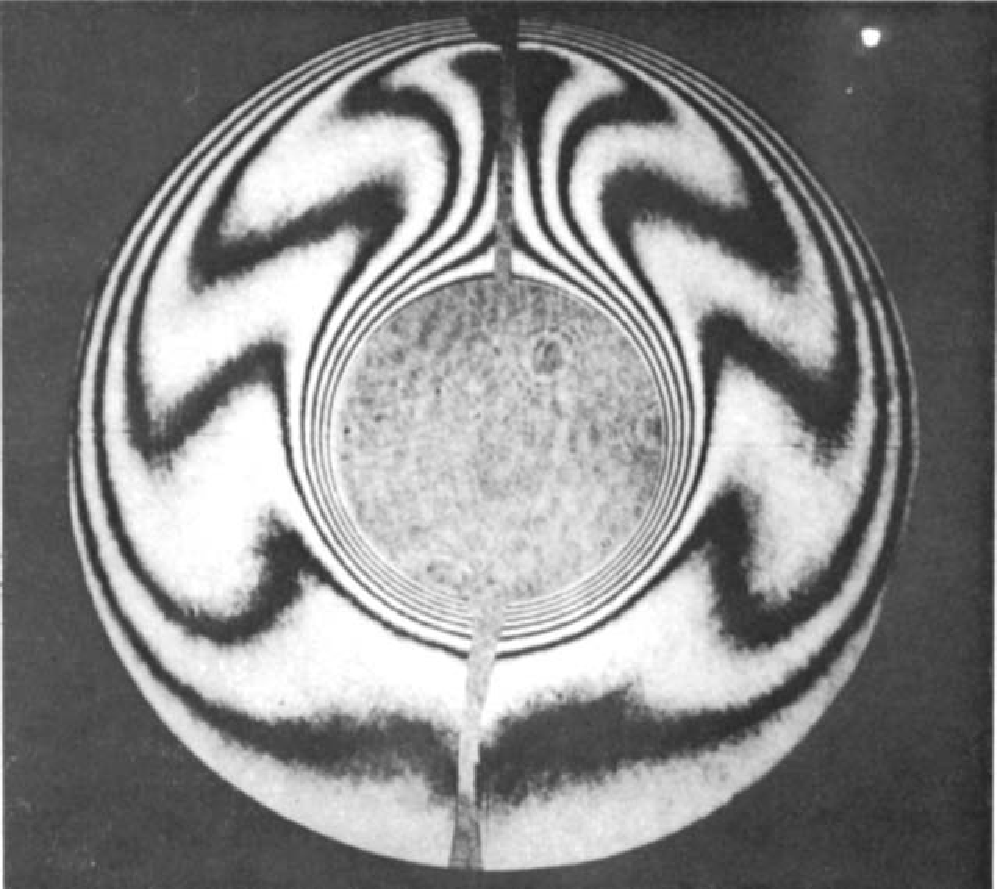}
  \end{subfigure}
  \begin{minipage}[b]{0.02\textwidth}
  \subcaption{ }\label{fig:P3_Ra47000_exp}
  \end{minipage}
  \begin{subfigure}[b]{0.44\textwidth}
  \centering
  \includegraphics[width=\linewidth]{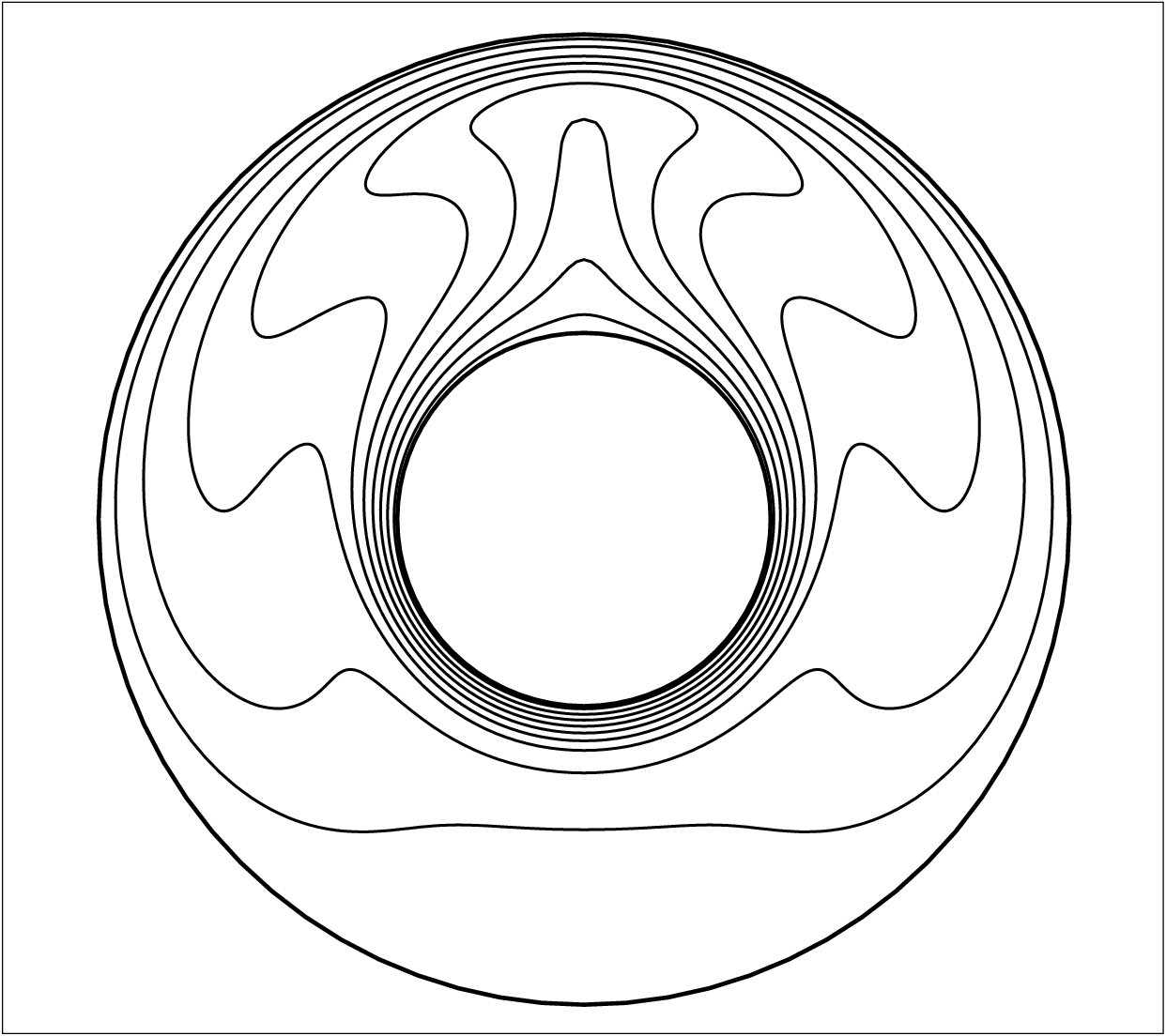}
  \end{subfigure}
  \begin{minipage}[b]{0.02\textwidth}
  \subcaption{ }\label{fig:P3_Ra47000_num}
  \end{minipage}
\caption{Problem 3: Steady-state isotherms at $Ra=4.7e4$, $Pr$=0.706: (a) experimental solution \citep{KuehnGoldstein1976}  and (b) numerical solution.}\label{fig:P3_expnum}
\end{figure} 

\begin{figure}[!h]
\centering
        \begin{subfigure}[]{0.24\textwidth}
                \centering
                \includegraphics[width=1.1\linewidth]{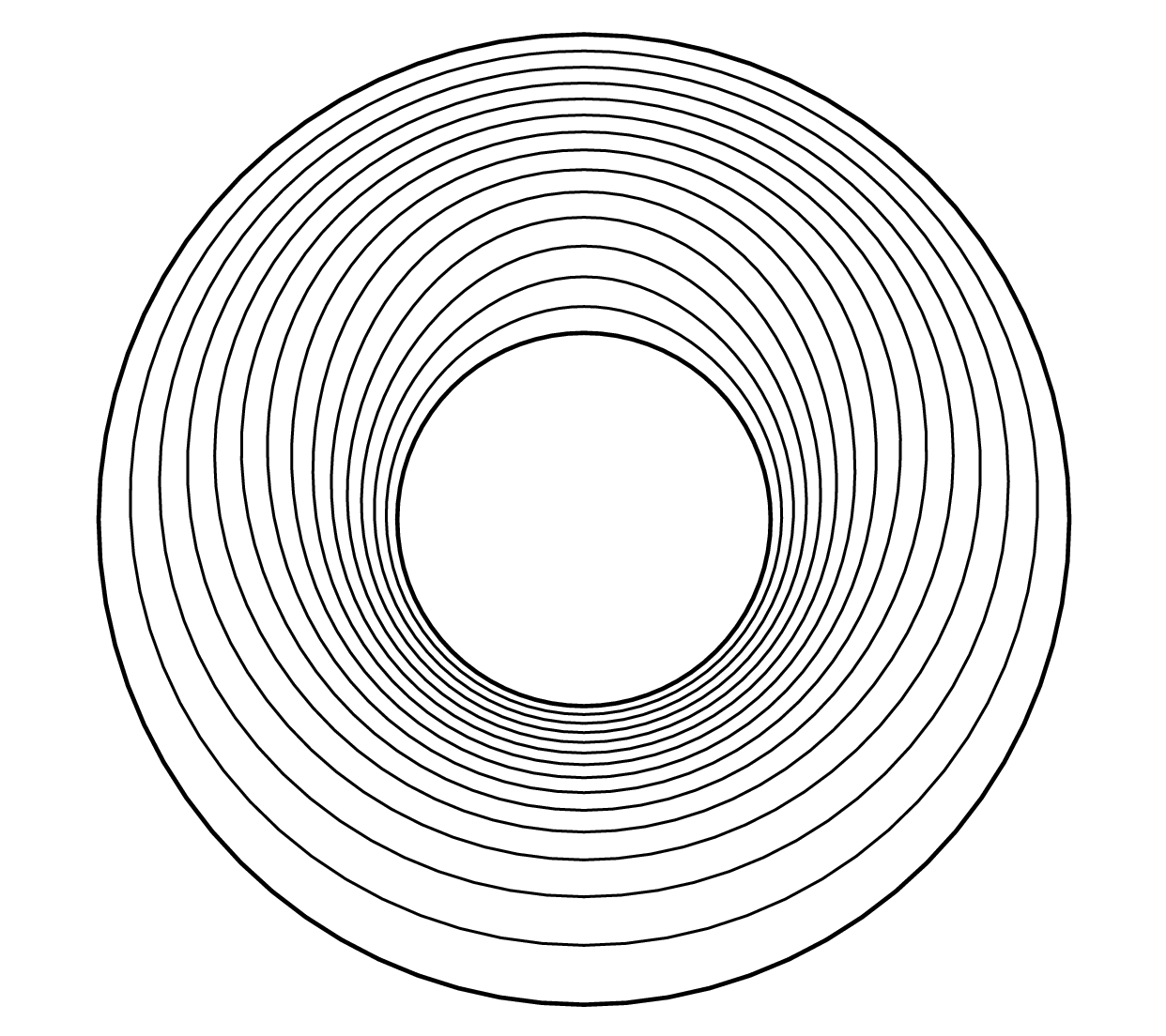}
        \end{subfigure}
        \begin{subfigure}[]{0.24\textwidth}
                \centering
                \includegraphics[width=1.1\linewidth]{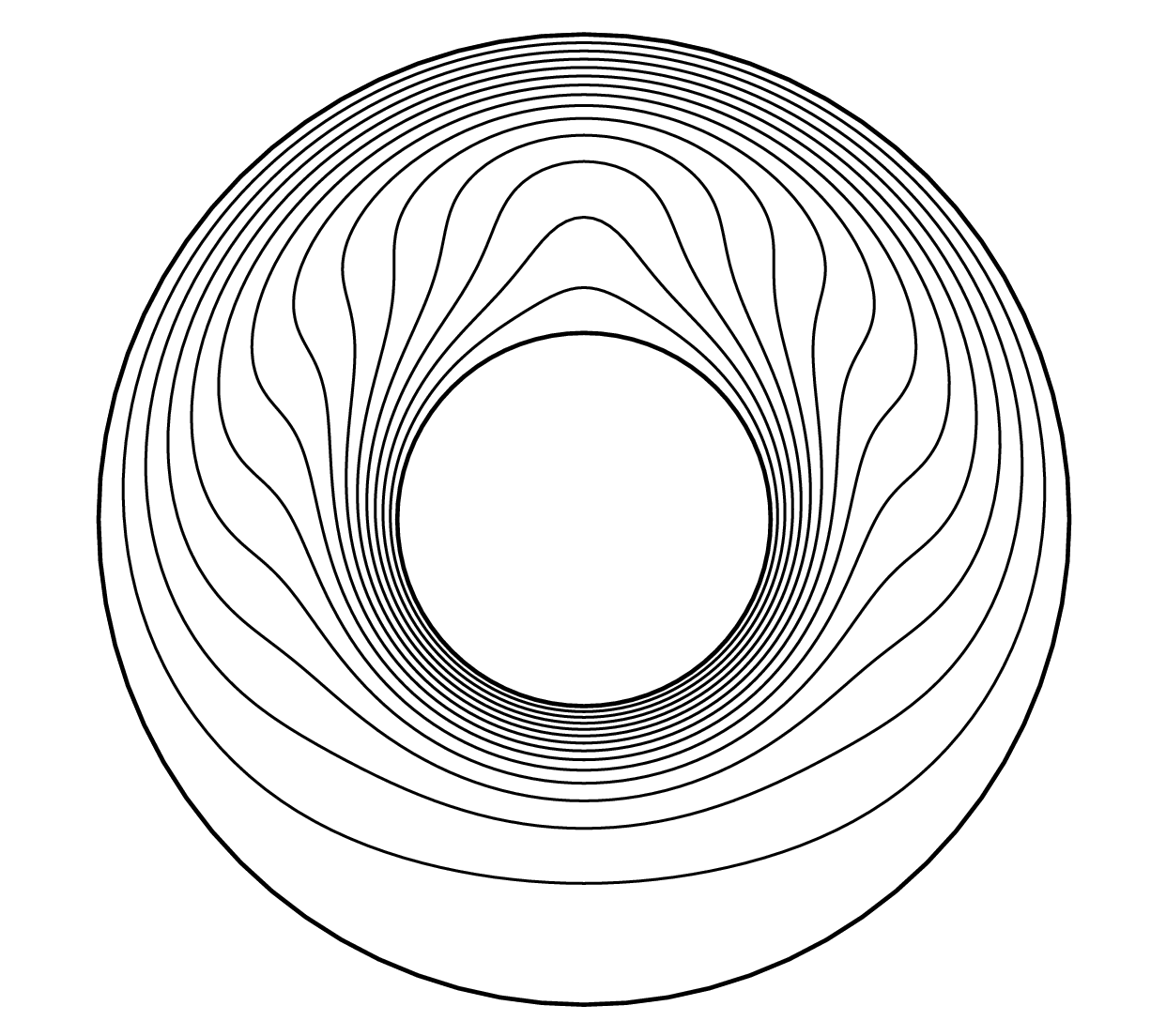}
        \end{subfigure}
        \begin{subfigure}[]{0.24\textwidth}
                \centering
                \includegraphics[width=1.1\linewidth]{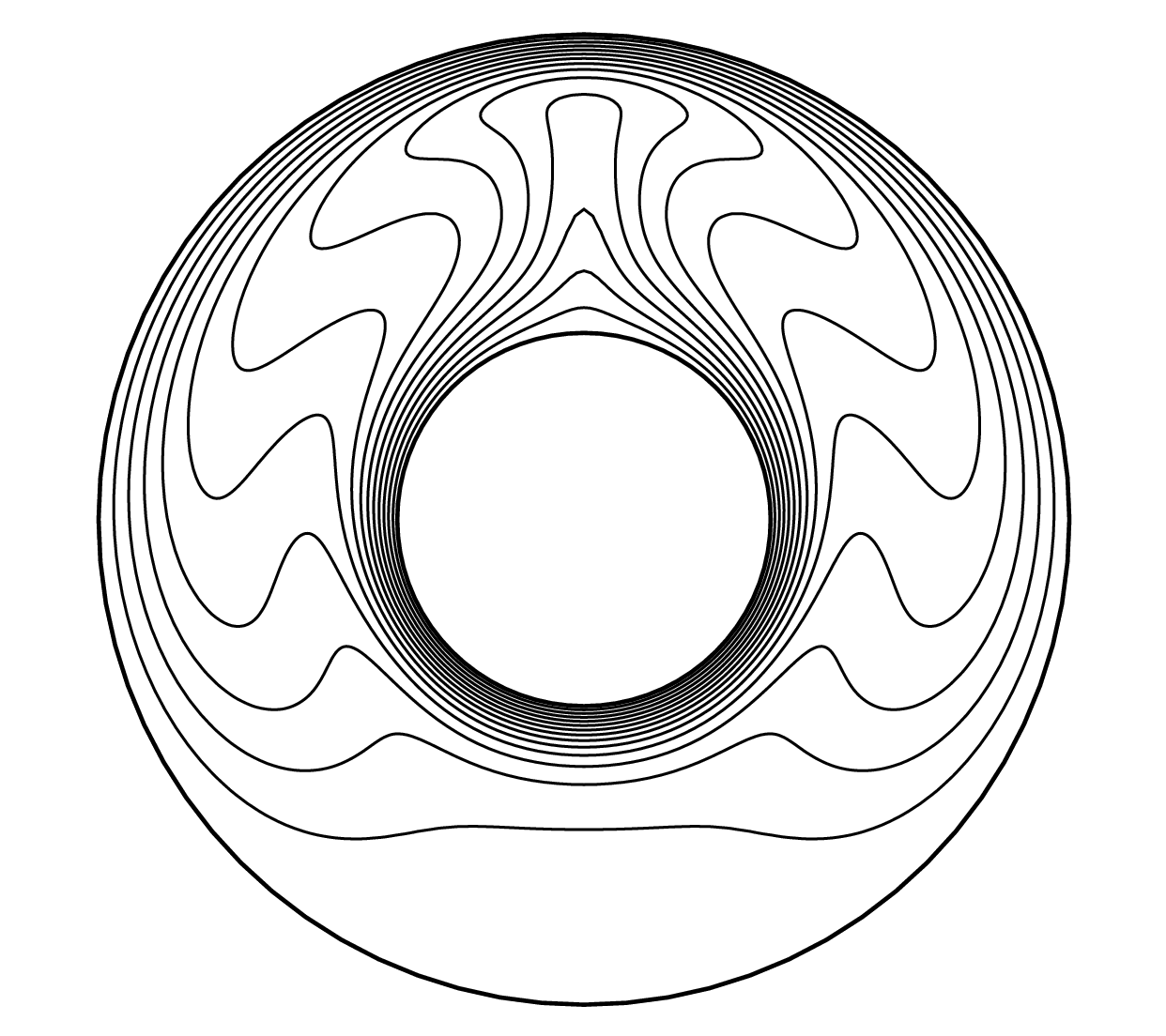}
        \end{subfigure}
        \begin{subfigure}[]{0.24\textwidth}
                \centering
                \includegraphics[width=1.1\linewidth]{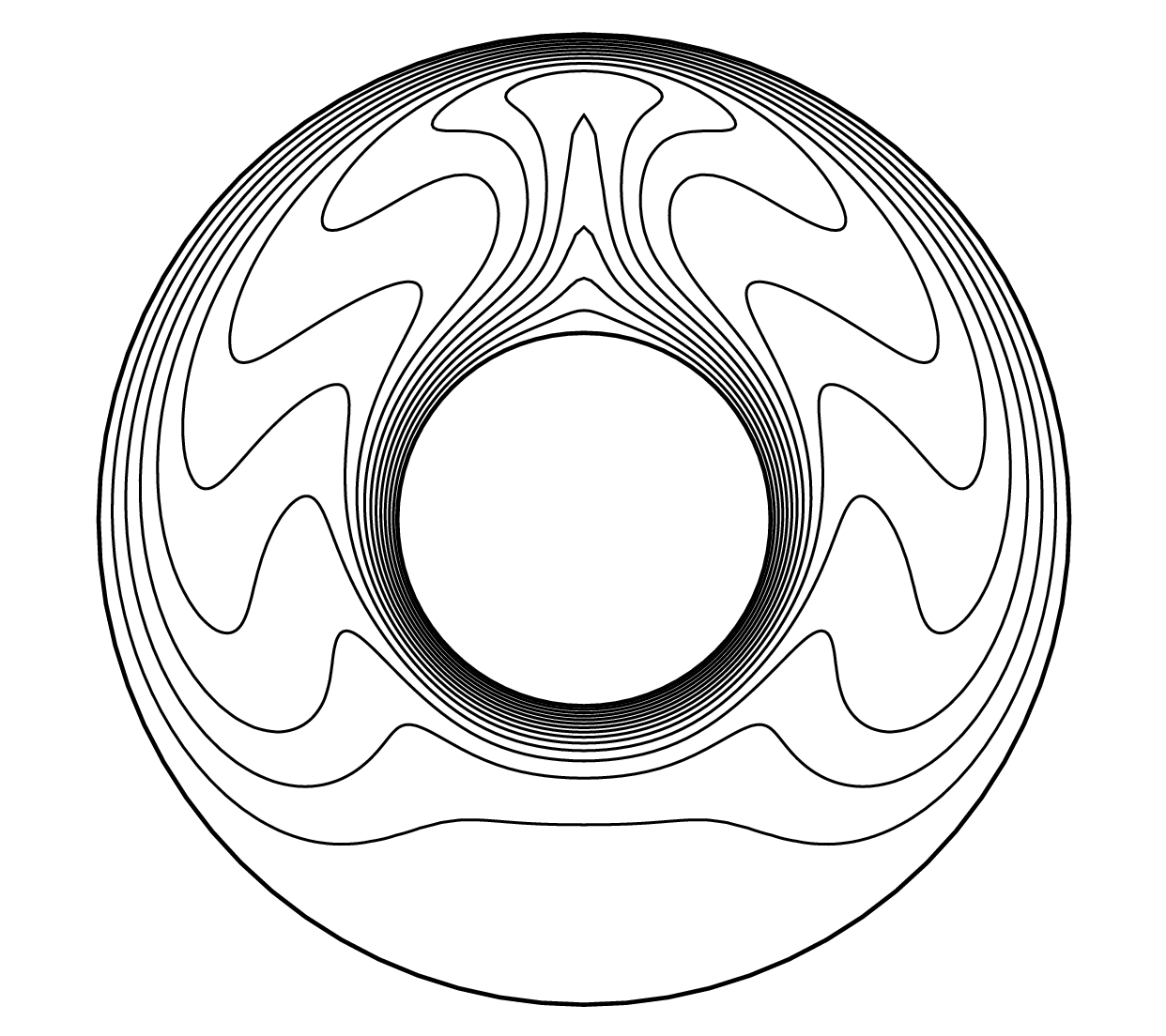}
        \end{subfigure}
        \\
        \begin{subfigure}[]{0.24\textwidth}
                \centering
                \includegraphics[width=1.1\linewidth]{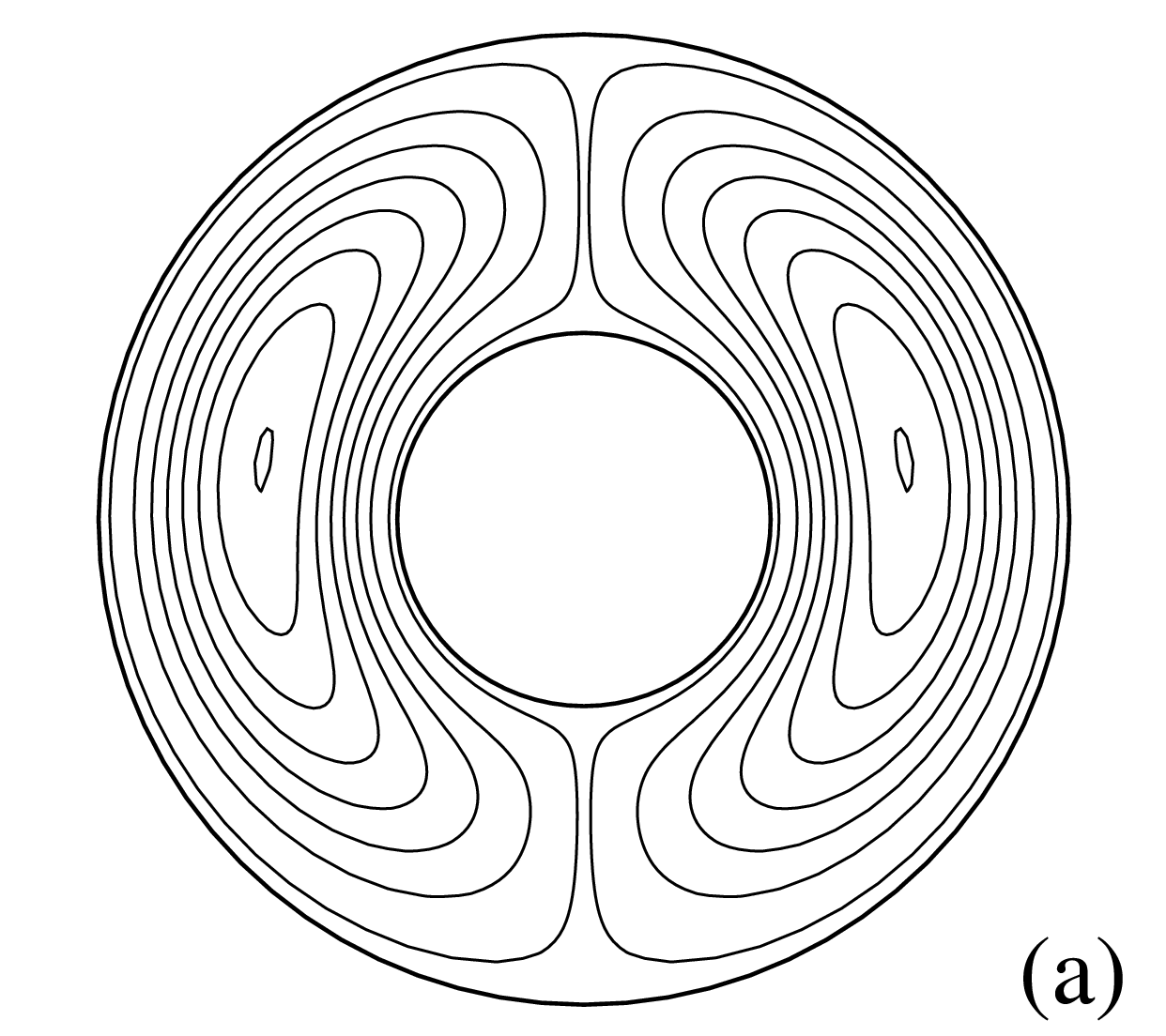}
        \end{subfigure}
        \begin{subfigure}[]{0.24\textwidth}
                \centering
                \includegraphics[width=1.1\linewidth]{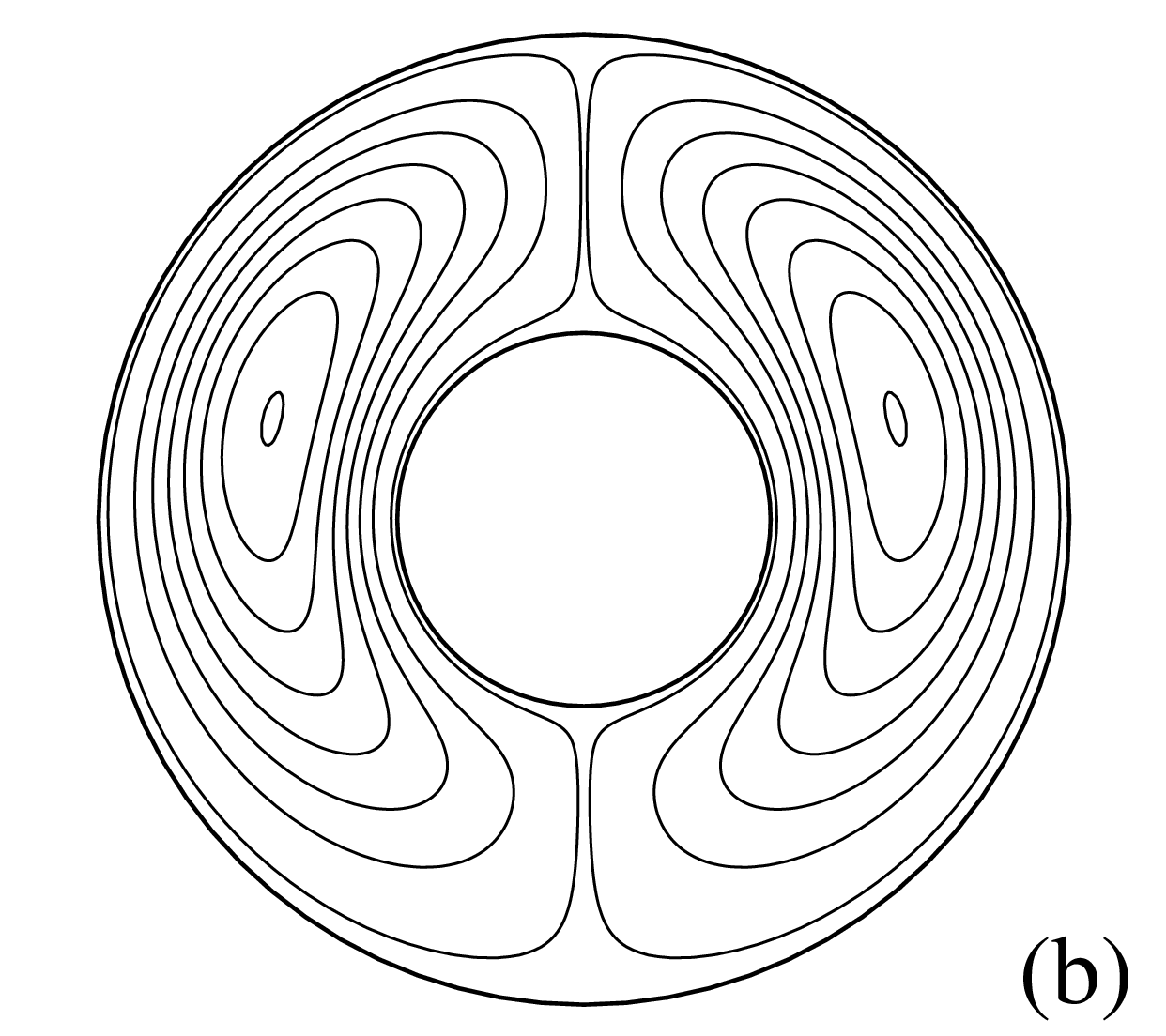}
        \end{subfigure}
        \begin{subfigure}[]{0.24\textwidth}
                \centering
                \includegraphics[width=1.1\linewidth]{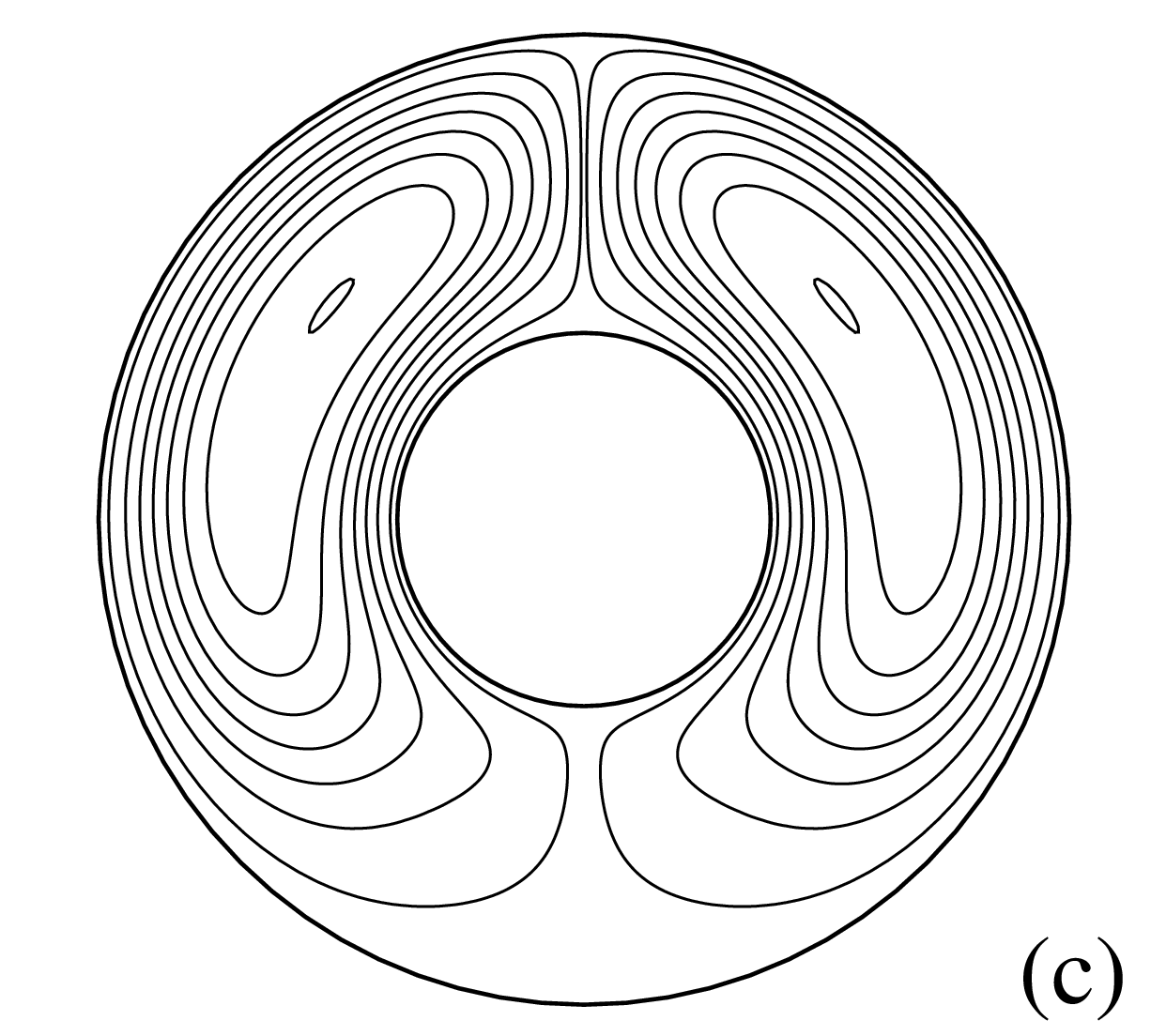}
        \end{subfigure}
        \begin{subfigure}[]{0.24\textwidth}
                \centering
                \includegraphics[width=1.1\linewidth]{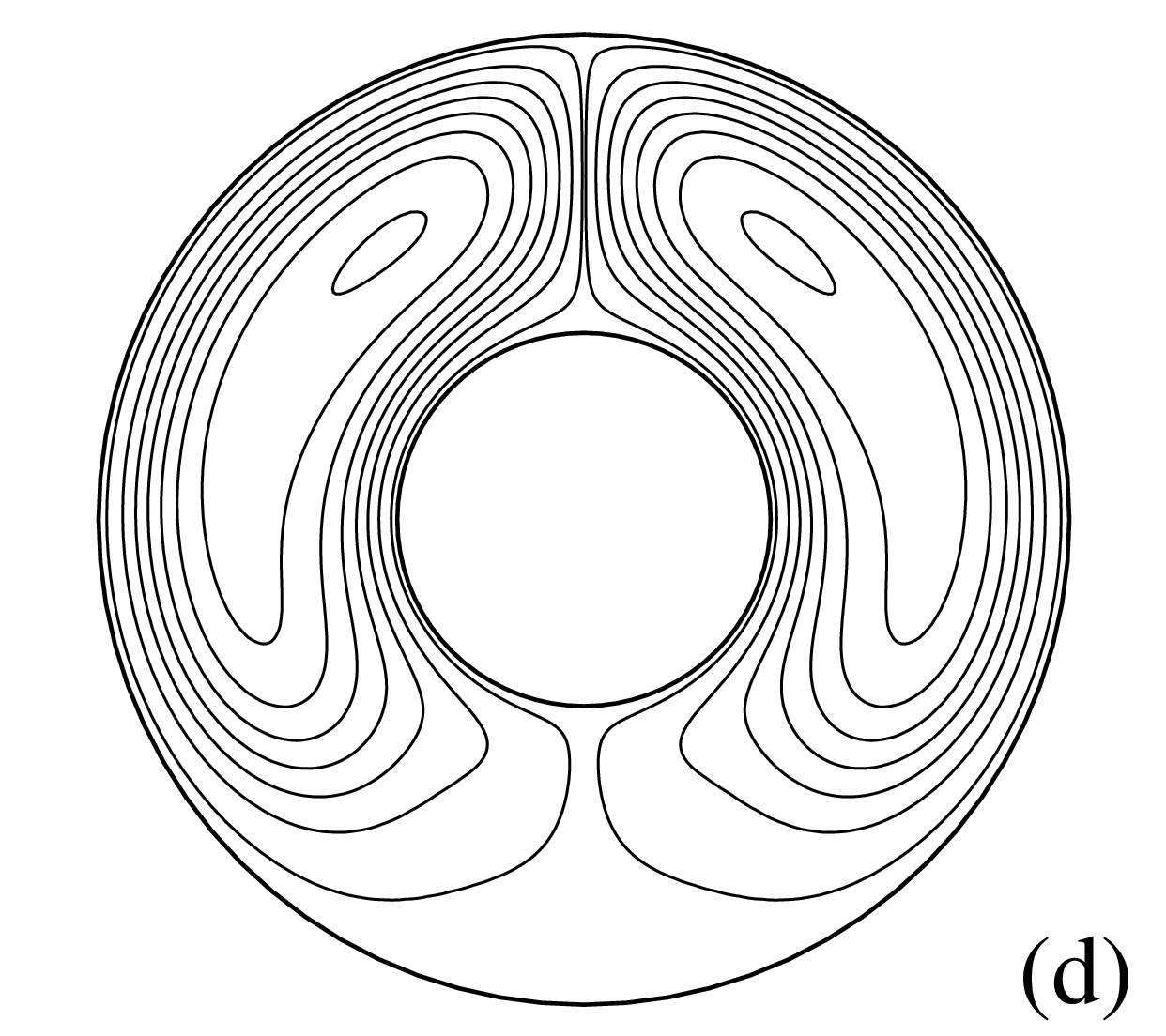}
        \end{subfigure}
\caption{Problem 3: Steady-state isotherms and streamlines for $Ra$ values (a) 2.38e3, (b) 9.50e4, (c) 6.19e4 and (d) 1.02e5.}\label{fig:P3_steady}
\end{figure}

The numerical results of the present simulation have been presented in \cref{fig:P3_expnum,fig:P3_steady}. We carry out a qualitative comparison in \cref{fig:P3_expnum} where the numerically obtained steady-state isotherm contours for $Ra=4.7e4$, $Pr$=0.706 are depicted alongside that of the experimental study \citep{KuehnGoldstein1976}. The figure shows that the fluid near the hot inner wall of the annulus flows up along the inner boundary and a strong plume emerges above the inner boundary at $\theta=\dfrac{\pi}{2}$, while the fluid near the cold outer wall flows down to the bottom of the annuli as an effect of buoyancy caused by the temperature difference. Once the flow reaches a steady state, the temperature field becomes symmetric. The maximum temperature gradient appears in the raised plume. Good agreements are obtained between the experimental and numerical solutions. \Cref{fig:P3_steady} contains the steady-state isotherms alongside the streamfunctions for different $Ra$ values. The figure shows that for lower $Ra$ values natural convection is dominated by heat conduction. Almost concentric isotherms can be seen for $Ra=2.38e3$ while two symmetric vortices can be seen caused by the weak natural convection. As the value of $Ra$ increases the fluid motion driven by buoyancy force also increases, leading to a stronger convection. As a result, the isotherms start moving upward and change shape to form a plume. The plume becomes more prominent with a further increase in the $Ra$ value. 

Additionally, for this problem we have also evaluated the average Nusselt number $(\overbar{Nu})$ which is the arithmetic mean of the surface-averaged Nusselt numbers of the annulus walls that are defined as,
\begin{equation}\label{Nu_innner}
Nu_i=-\frac{1}{\pi} R_i \int_{-\frac{\pi}{2}}^{\frac{\pi}{2}} \left. \frac{\partial \mathcal{T}}{\partial r} \right|_{r=R_i} d\theta
\end{equation}
and 
\begin{equation}\label{Nu_outer}
Nu_o=-\frac{1}{\pi} R_o \int_{-\frac{\pi}{2}}^{\frac{\pi}{2}} \left. \frac{\partial \mathcal{T}}{\partial r} \right|_{r=R_o} d\theta.
\end{equation} 
The average Nusselt number thus obtained is compared to the existing studied in \cref{table:P3_nusselt}. Our computed values agree well with those available in the literature at $Ra=2.38e3$, $Pr=0.716$. But it seen to differ by about 10\% as $Ra$ is increased.   

\begin{table}[!h]
\caption{Problem 3: Comparison of average nusselt number $(\overbar{Nu})$ computed using different schemes for different Rayleigh numbers. }\label{table:P3_nusselt}
\vspace{0.1cm}
\renewcommand*{\arraystretch}{0.8}
\begin{tabular}
{ M{0.135\textwidth} M{0.135\textwidth} M{0.15\textwidth} M{0.135\textwidth} M{0.135\textwidth} M{0.135\textwidth} }
\hline
$Ra$     &  $Pr$ & \citep{KuehnGoldstein1976}(expt.)        & \citep{Shahraki2002} & \citep{ShiZhaoGuo2006} & Present  \\
\hline
2.38e3   & 0.716 &  1.38                                    &                      & 1.320                  & 1.3882   \\
9.50e3   & 0.717 &  2.01                                    & 1.9901               & 1.999                  & 1.6909   \\
6.19e4   & 0.718 &  3.32                                    & 3.3092               & 3.361                  & 2.9176   \\
1.02e5   & 0.718 &  3.66                                    & 3.6475               & 3.531                  & 3.2994   \\            
\hline
\end{tabular}
\end{table}

\subsection{Problem 4: Forced convection over a stationary heated circular cylinder}\label{ex4}
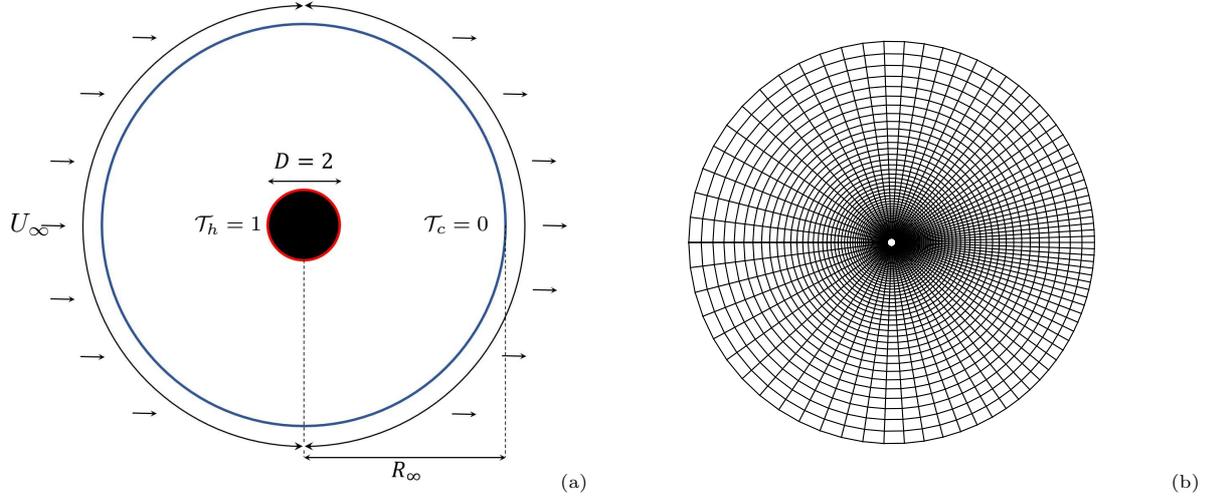
\begin{figure}[!h]
  \begin{subfigure}[b]{0.44\textwidth}
  \centering
  \begin{tikzpicture}
    \draw (0,0) node[inner sep=0] {\includegraphics[width=\linewidth]{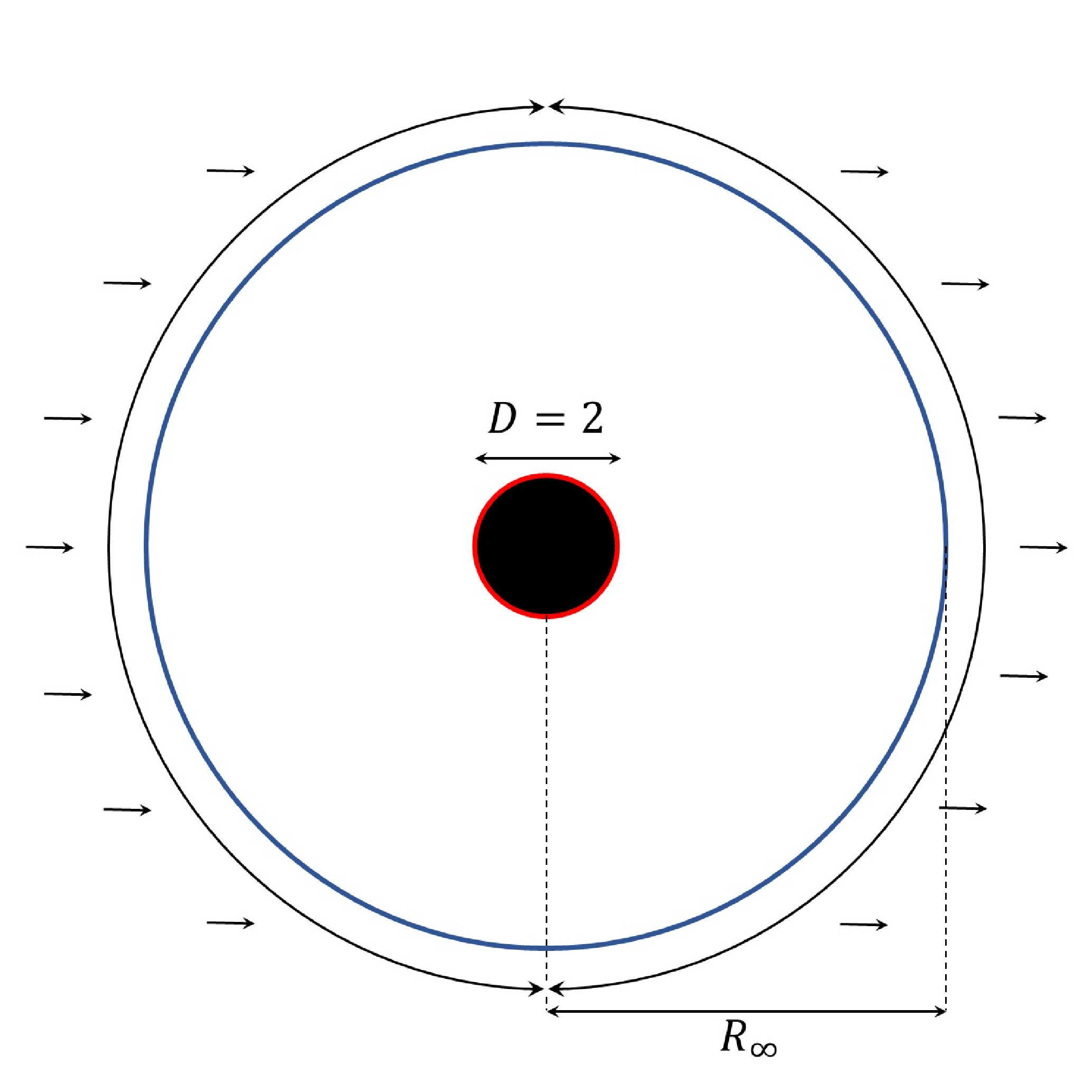}};
    \draw (-3.6,0) node {$U_\infty$};
    \draw (-1,0.0) node {\footnotesize $\mathcal{T}_h=1$};
    \draw (2.0,0.0) node {\footnotesize $\mathcal{T}_c=0$};
  \end{tikzpicture}
  \end{subfigure}
  \begin{minipage}[b]{0.02\textwidth}
  \subcaption{ }\label{fig:P4_setup}
  \end{minipage}
  \raisebox{5mm}[0pt][0pt]
  {
    \begin{subfigure}[b]{0.44\textwidth}
    \centering
     \includegraphics[width=0.9\linewidth]{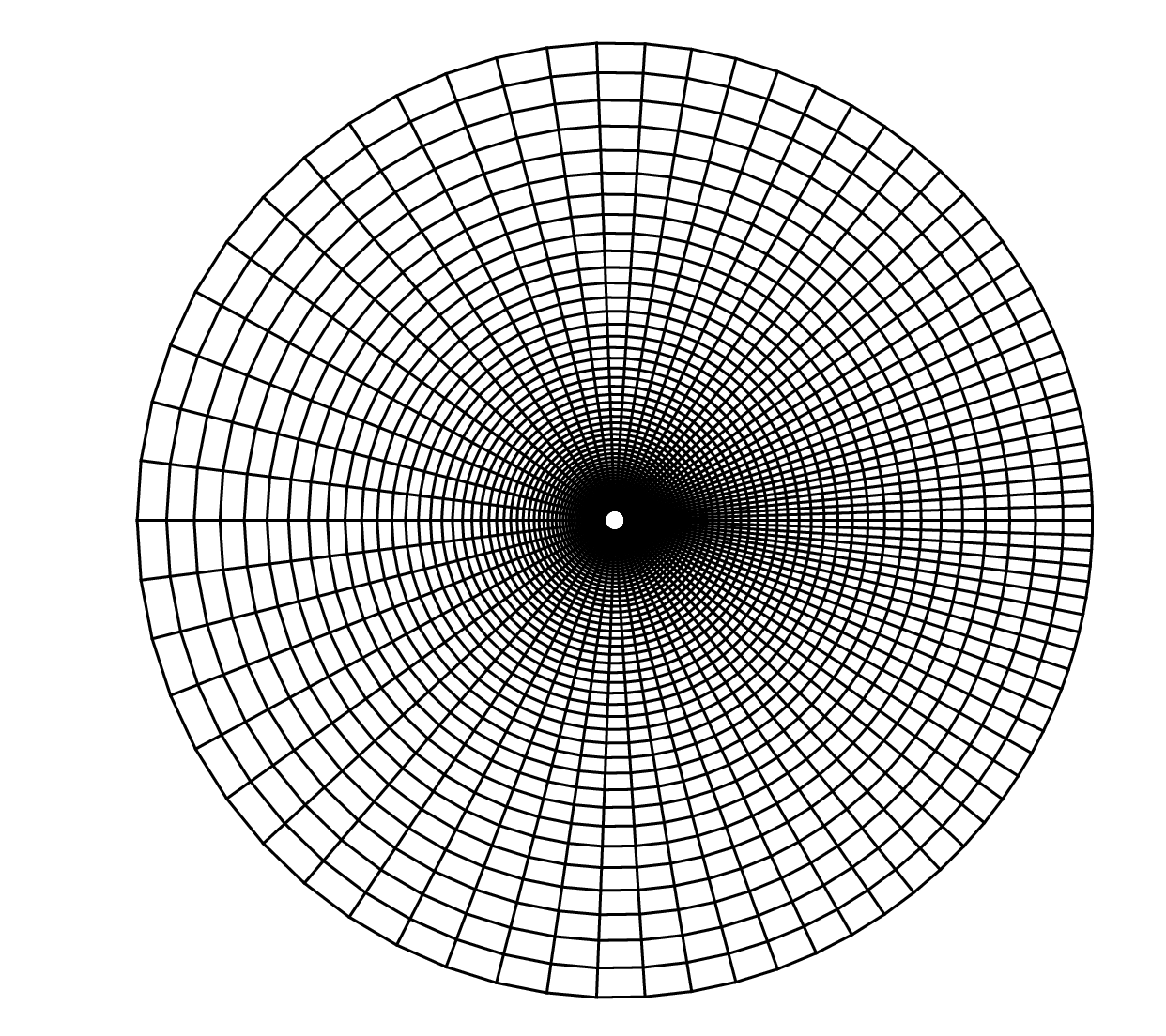}
    \end{subfigure}
  }
  \begin{minipage}[b]{0.02\textwidth}
  \subcaption{ }\label{fig:P4_grid}
  \end{minipage}
\caption{Problem 4: (a)Flow configuration and (b) typical $65\times97$ nonuniform grid.}\label{fig:P4_SetupGrid}
\end{figure}

Finally, we carry out numerical simulation for the classical problem of heat transfer in the fluid flow over the stationary heated circular cylinder. The stationary heated circular cylinder is presumed to be of unit radius ($R_0=1$) and immersed in a fluid of infinite domain maintained at unit nondimensional temperature \citep{DennisHudsonSmi1968,HeDoolen1997,LangeDurstBreuer1998,SparrowAbrahamTong2004,Niu_etal_2006,Soares_etal_2006,BhartiChhabraEswaran2007,SanyasiMishra2008,KalitaRay2009,WuShu2009,SenKalita2015,Chen_etal_2020,Kumarkalita2021}. The cylinder is kept in a cross-flow having uniform velocity $u=U_\infty=1$. The 2D flow configuration of the problem is shown in \cref{fig:P4_SetupGrid}. The cylinder is placed at the center of the circular domain. Following \citep{Frankeetal1990}, we have set the far field boundary at a distance $R_\infty=45R_0$. At the solid surface $r=R_0$, the boundary conditions for velocity components are those of no-slip type, i.e. $u=v=0 \implies \psi=0$. In the far stream $r=R_\infty$, in front of the cylinder the potential flow is prescribed a unit value $u=U_\infty$=1. For this problem, where the simulation results in periodic vortex shedding, at the downstream of the flow, we have imposed the convective boundary conditions $\phi_t+U_\infty \phi_r$=0 (where $\phi$ represents $\psi$, $u$ or $v$) to capture the shedding process efficiently. The continuous shedding of vortices, when they leave the computational domain in the direction of the flow, can be best facilitated by the convective boundary conditions \citep{Breuer1998,KalitaSen2012b,LanVen_2010,WangFanCen2009}. In addition, at the far field vorticity value decays and becomes $\omega=0$. However, it possesses nonzero value at the inner boundary, which may be derived by making use of the fact that $\psi=0$, $\psi_r=0$ on $r=R_0$ in \cref{Bous_pol2}. The vorticity gradients at all the boundaries are computed using one-sided approximations.

The essential nondimensional parameters for this problem include the Reynolds number ($Re$) and the Prandtl number ($Pr$). In this present investigation, the Prandtl number is fixed at $Pr=0.7$ for all the values of $Re$ under consideration. As shown in \cref{fig:P4_SetupGrid}, the computational domain is discretized using a nonuniform grid of size $129\times193$ for all the combinations of $Pr$ and $Re$. With the circular cylinder placed at the center of the domain, grid is clustered in the vicinity of the cylinder wall.

\begin{figure}[!h]
\begin{subfigure}[b]{0.44\textwidth}
    \centering
    \includegraphics[width=\linewidth]{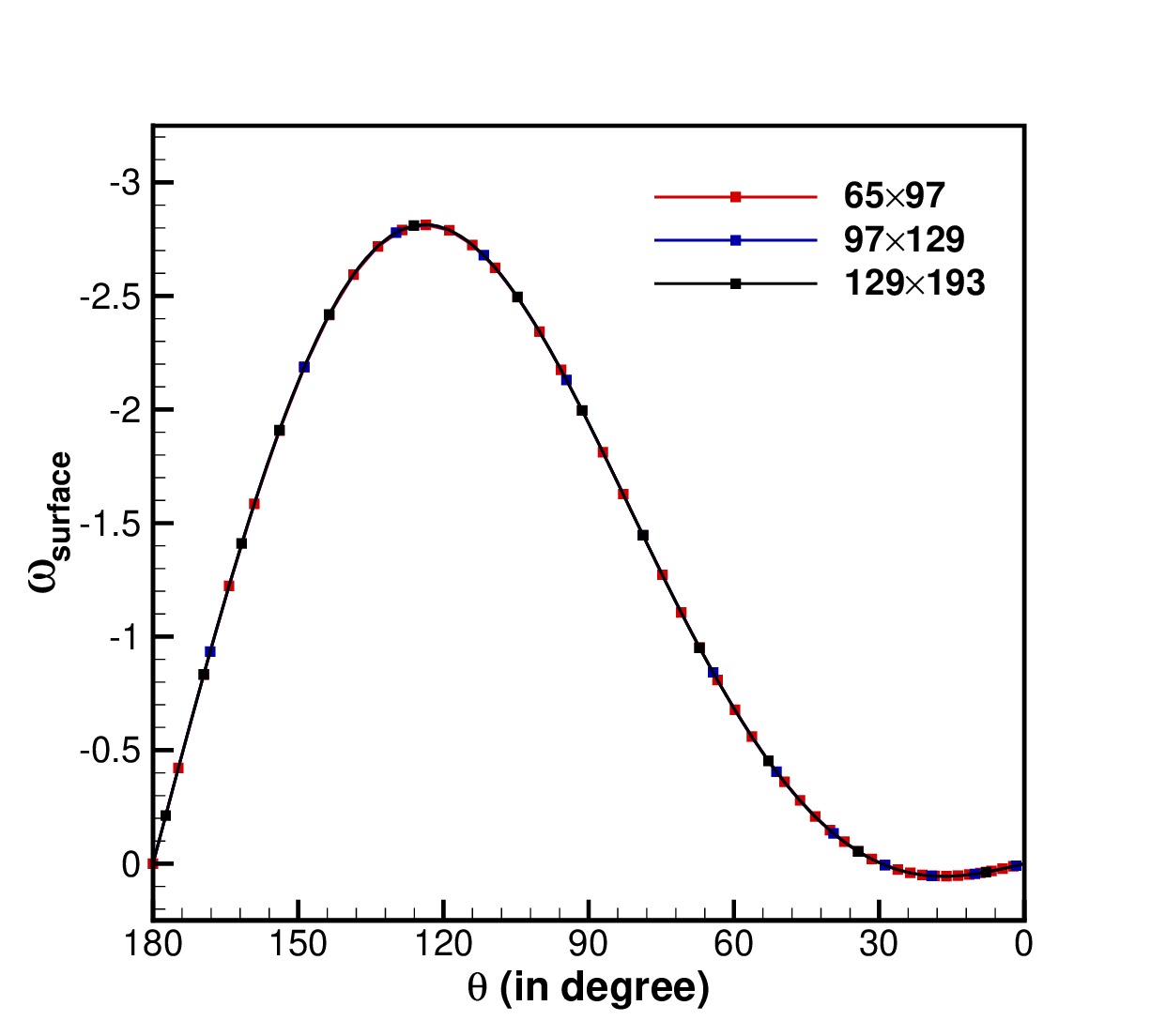}
\end{subfigure}%
\begin{minipage}[b]{0.05\textwidth}
\subcaption{ }\label{fig:P4_SV_Re10}
\end{minipage}
\begin{subfigure}[b]{0.44\textwidth}
    \centering
    \includegraphics[width=\linewidth]{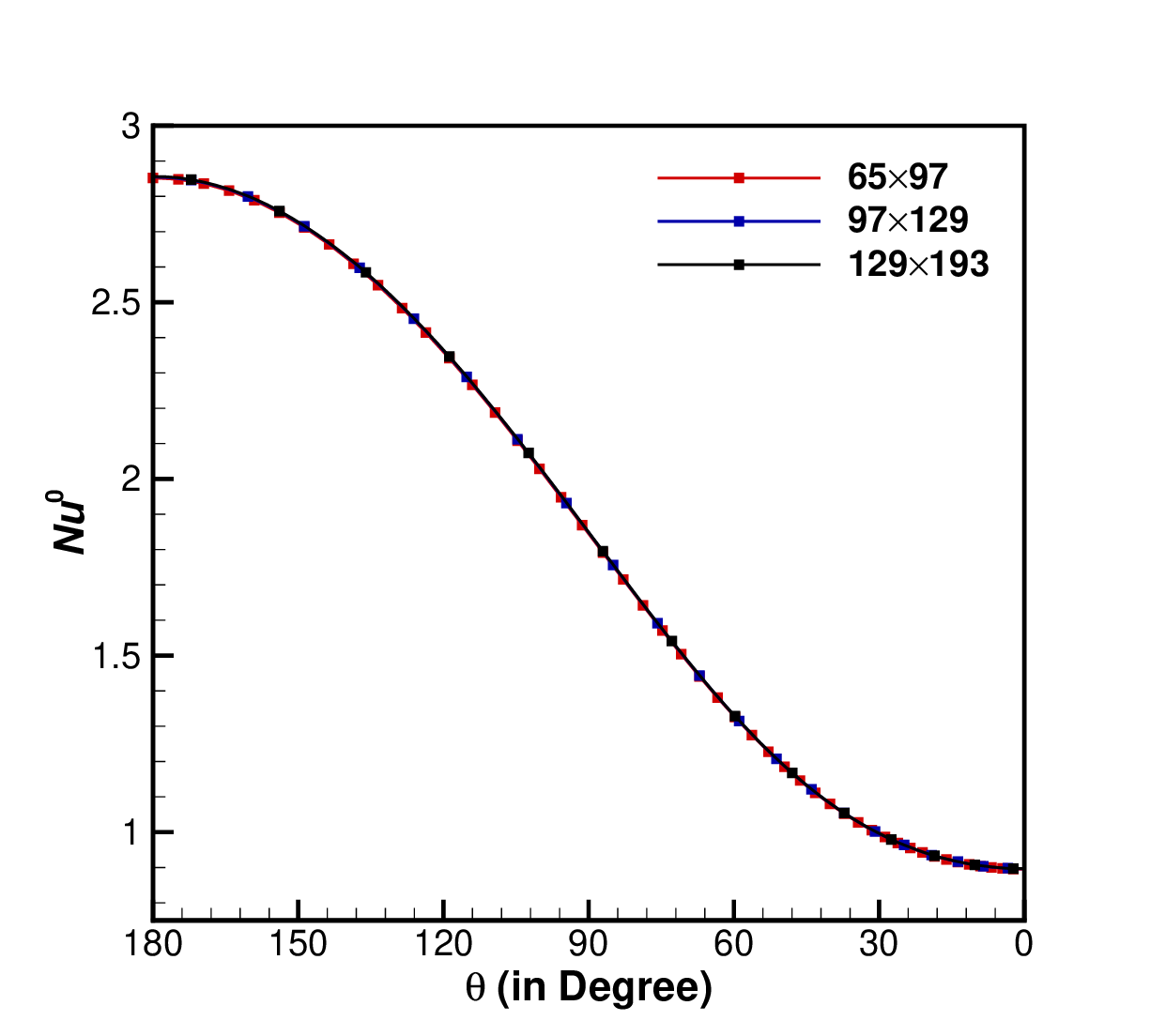}
\end{subfigure}%
\begin{minipage}[b]{0.05\textwidth}
\subcaption{ }\label{fig:P4_Nu_Re10}
\end{minipage}
\caption{Problem 4: Distribution of (a) vortcity and (b) local Nusselt number at steady state along the surface of the cylinder for $Re=10$.} \label{fig:P4_SV_Nu_Re10}
\end{figure}

We start by verifying the grid-independence of the present numerical solution. Here, we compute for $Re=10$ and present the distribution of surface vorticity and local Nusselt number $(Nu^0)$ and vorticity along the surface of the cylinder for three grids of different sizes \textit{viz.} $65\times97$, $97\times129$ and $129\times193$. Here, angle $\theta$ is measured in the counterclockwise direction starting from the rear stagnation point of the cylinder. One can see from the figures that the numerical results are unresponsive to the change in grid points and a grid of size $129\times193$ is sufficient to capture the flow accurately. 

\begin{figure}[!h]
\centering
        \begin{subfigure}[b]{0.44\textwidth}
                \centering
                \includegraphics[width=\linewidth]{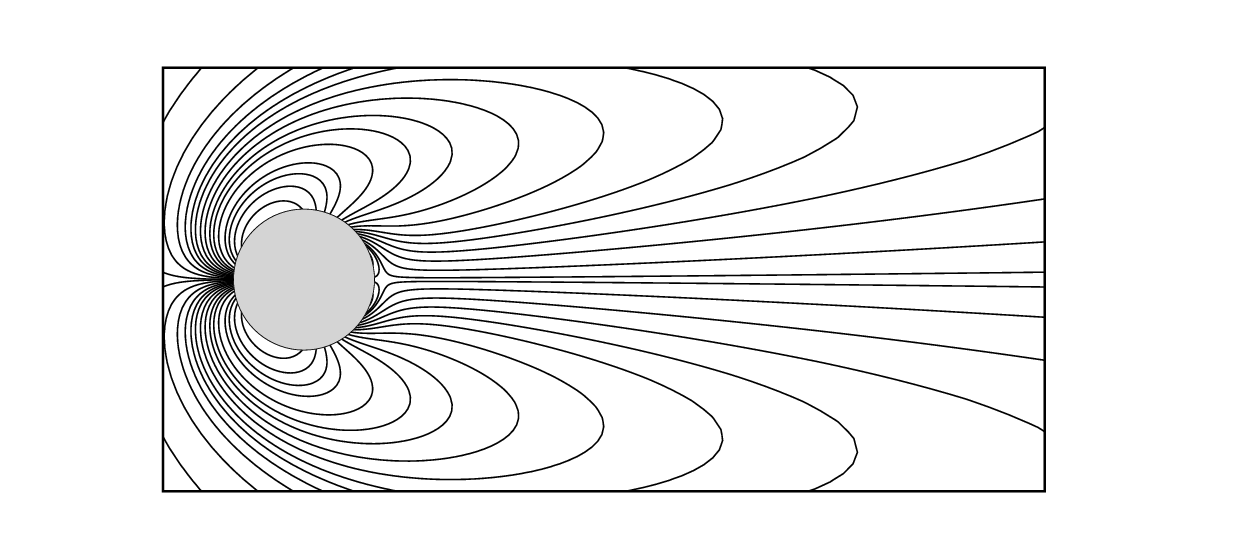}
        \end{subfigure}%
        \begin{subfigure}[b]{0.44\textwidth}
                \centering
                \includegraphics[width=\linewidth]{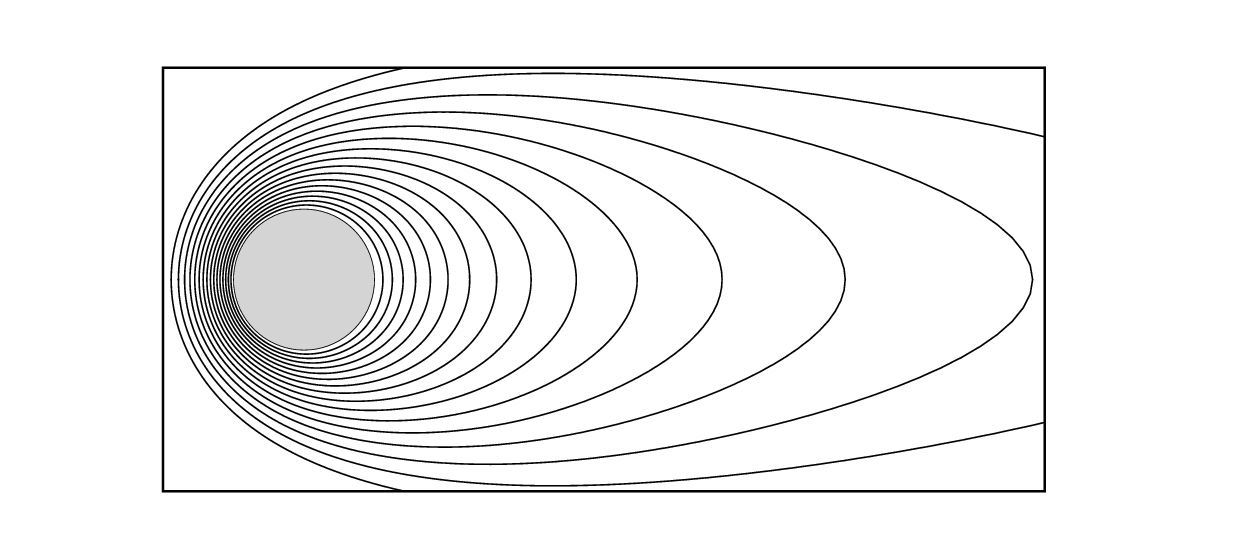}
        \end{subfigure}%
        \begin{minipage}[b]{0.04\textwidth}
            \subcaption{}\label{fig:P4_therm10}
        \end{minipage}
        \\
        \begin{subfigure}[b]{0.44\textwidth}
                \centering
                \includegraphics[width=\linewidth]{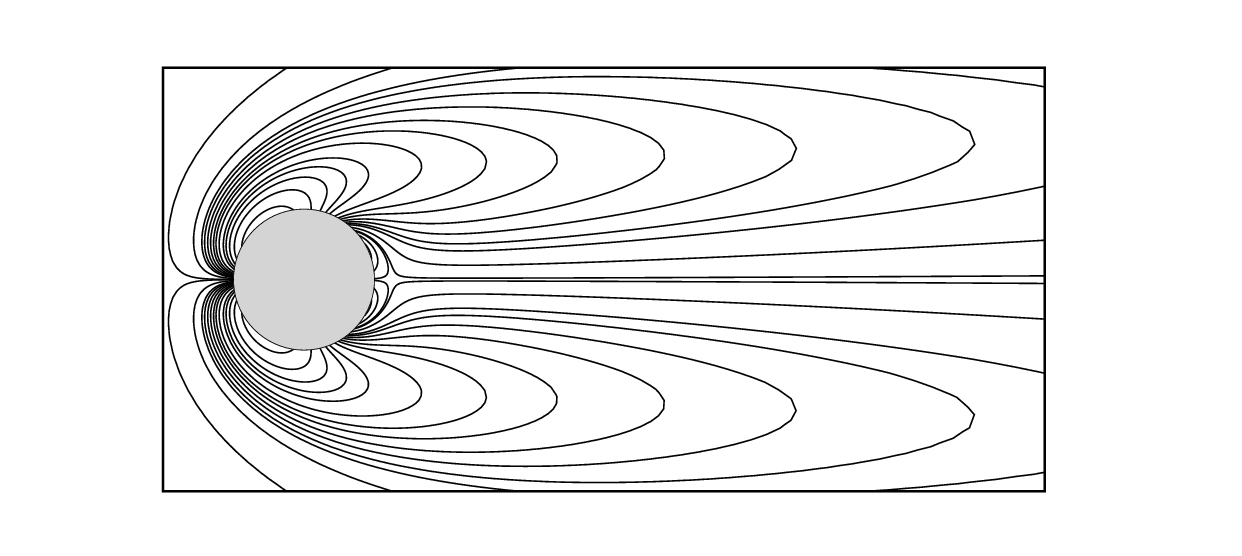}
        \end{subfigure}%
        \begin{subfigure}[b]{0.44\textwidth}
                \centering
                \includegraphics[width=\linewidth]{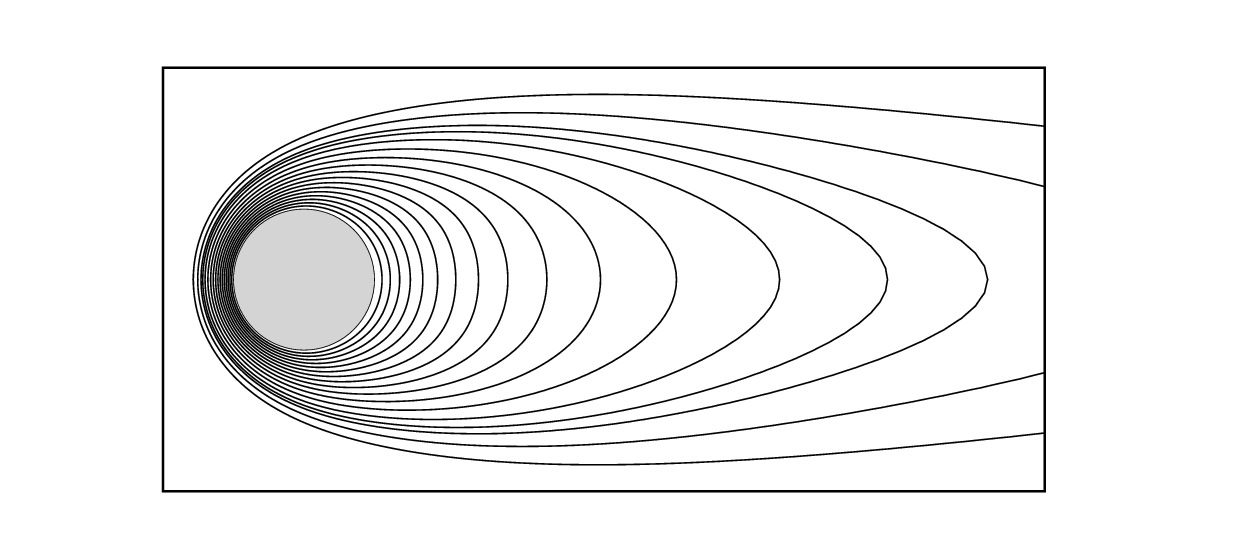}
        \end{subfigure}%
        \begin{minipage}[b]{0.04\textwidth}
            \subcaption{}\label{fig:P4_therm20}
        \end{minipage}
        \\
        \begin{subfigure}[b]{0.44\textwidth}
                \centering
                \includegraphics[width=\linewidth]{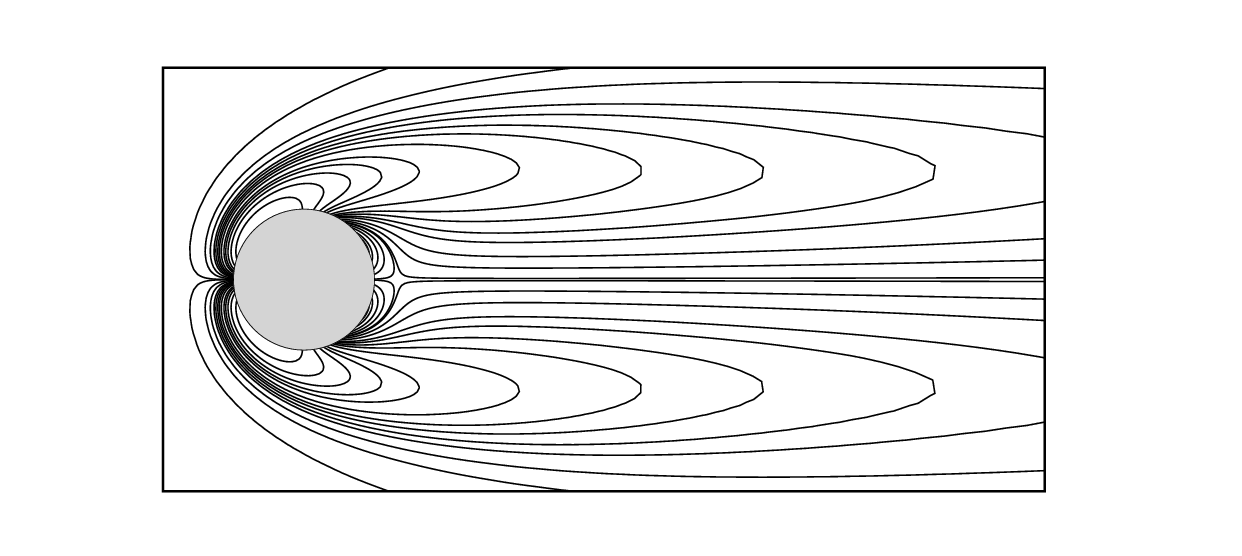}
        \end{subfigure}%
        \begin{subfigure}[b]{0.44\textwidth}
                \centering
                \includegraphics[width=\linewidth]{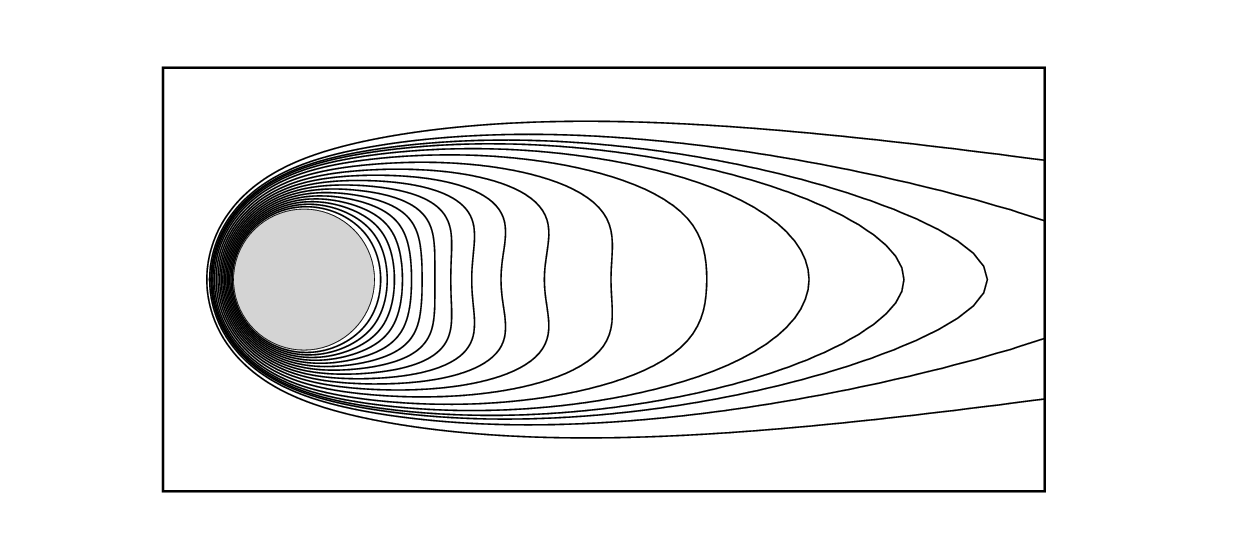}
        \end{subfigure}%
        \begin{minipage}[b]{0.04\textwidth}
            \subcaption{}\label{fig:P4_therm40}
        \end{minipage}        
\caption{Problem 4: Steady state vortifity contours (left) and isotherms (right) for (a) $Re=10$, (b) $Re=20$ and (c) $Re=40$.}\label{fig:P4_Steadycontour}
\end{figure}

\begin{figure}[!h]
\begin{subfigure}[b]{0.44\textwidth}
    \centering
    \includegraphics[width=\linewidth]{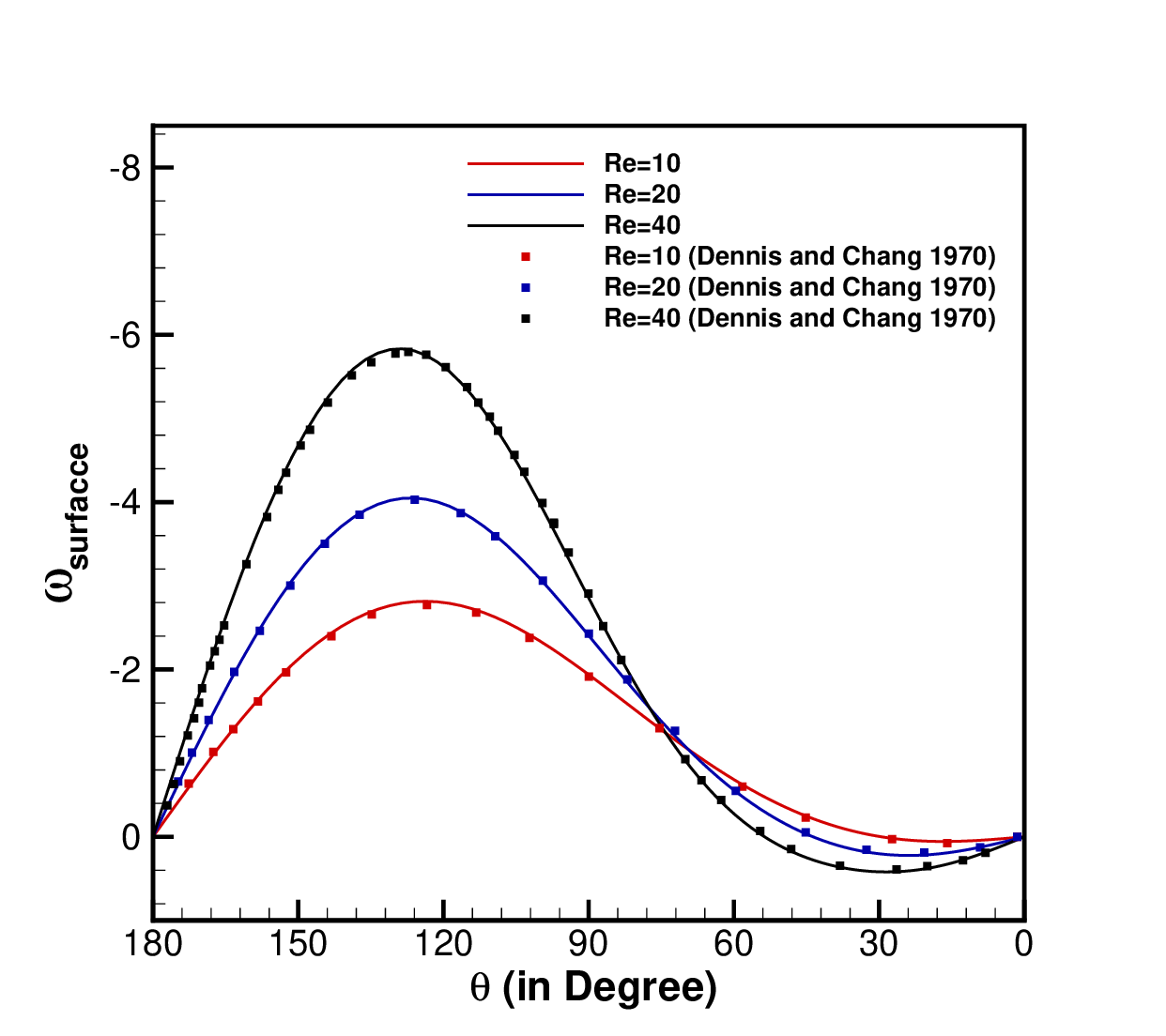}
\end{subfigure}%
\begin{minipage}[b]{0.05\textwidth}
\subcaption{ }\label{fig:P4_SV_all}
\end{minipage}
\begin{subfigure}[b]{0.44\textwidth}
    \centering
    \includegraphics[width=\linewidth]{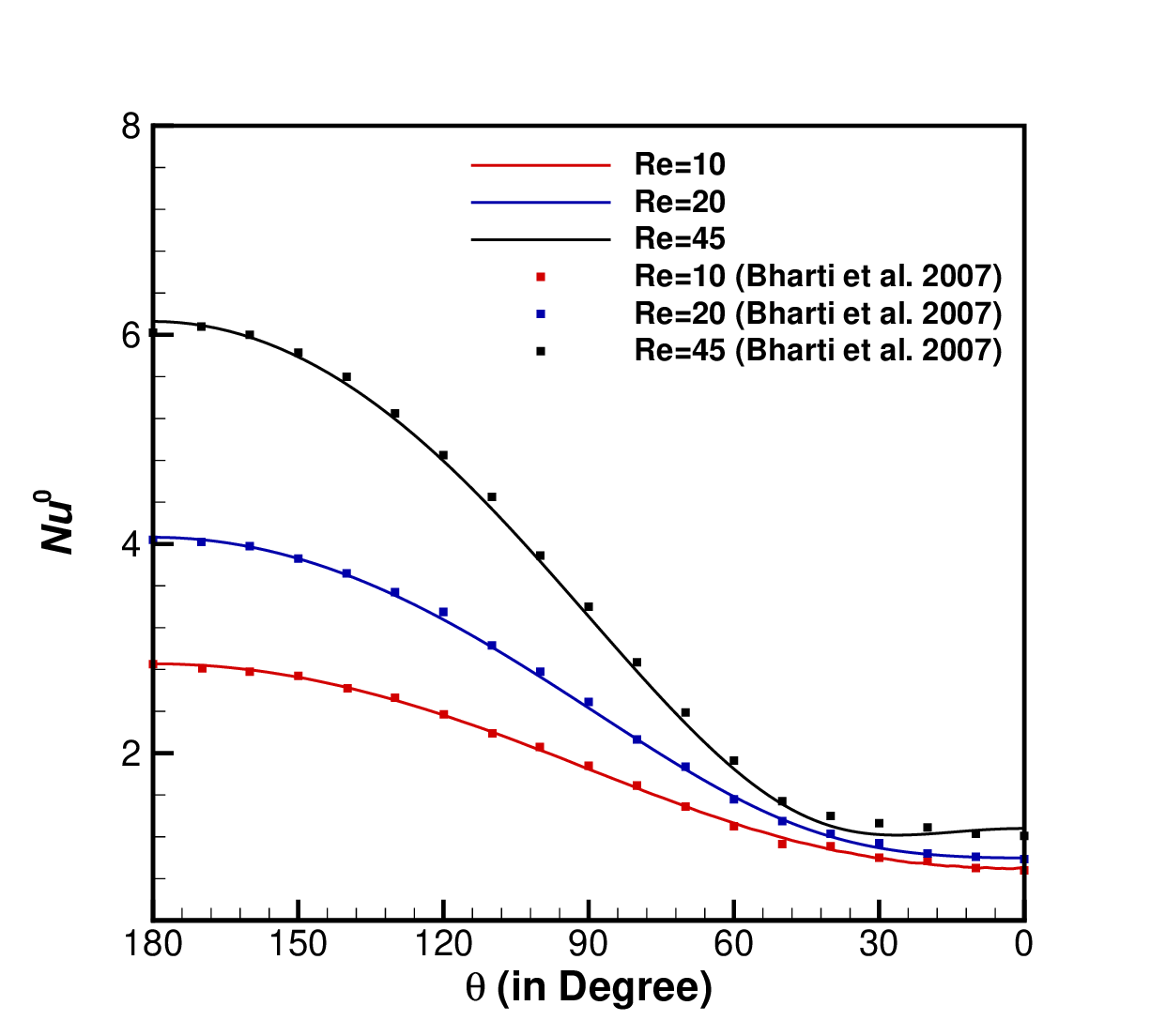}
\end{subfigure}%
\begin{minipage}[b]{0.05\textwidth}
\subcaption{ }\label{fig:P4_SNu_all}
\end{minipage}
\caption{Problem 4: Comparison of (a) surface vorticity distribution for $Re$=10, 20 and 40 and (b) local Nusselt number distribution for $Re$=10, 20 and 45 with existing numerical results \citep{DennisChang1970,BhartiChhabraEswaran2007}.} \label{fig:P4_SV_Nu_Re102040}
\end{figure}

\begin{table}[!h]
\caption{Problem 4: Comparison of steady state wake length, separation angle and drag coefficient for different $Re$.}\label{table:P4_steady}
\vspace{0.1cm}
\renewcommand*{\arraystretch}{0.8}
\begin{tabular}
{ M{0.075\textwidth} M{0.095\textwidth} M{0.075\textwidth} M{0.095\textwidth} M{0.02\textwidth} M{0.075\textwidth} M{0.095\textwidth} M{0.02\textwidth} M{0.075\textwidth} M{0.095\textwidth} }
\hline
                 &                               & $Re$=10 &$\%$-diff.     & &  $Re$=20   &$\%$-diff.     & &  $Re$=40  &$\%$-diff.          \\
\hline
$L/D$            & \citep{HeDoolen1997}          &  0.237  &   4.22        & & 0.921      &  0.54         & &  2.245    &  0.36              \\        
                 & \citep{Niu_etal_2006}         &         &               & & 0.945      &  3.07         & &  2.26     &  0.31              \\    
                 & \citep{SanyasiMishra2008}     &         &               & & 0.885      &  3.50         & &  2.105    &  7.03              \\ 
                 & \citep{WuShu2009}             &         &               & & 0.93       &  1.51         & &  2.31     &  2.47              \\
                 & \citep{KalitaRay2009}         &         &               & & 0.917      &  0.11         & &  2.207    &  2.08              \\                
                 & \citep{SenKalita2015}         &  0.252  &   1.98        & & 0.926      &  1.08         & &  2.323    &  3.01              \\            
                 & \citep{Chen_etal_2020}        &         &               & & 0.95       &  3.58         & &  2.39     &  5.73              \\
                 & \citep{Kumarkalita2021}       &  0.266  &   7.14        & & 0.937      &  2.24         & &  2.139    &  5.33              \\         
                 & Present                       &  0.247  &               & & 0.916      &               & &  2.253    &                    \\ 
\hline
$C_D$            & \citep{HeDoolen1997}          &  3.170  &   10.50       & & 2.152      &  4.23         & &  1.499    &  2.33              \\        
                 & \citep{Niu_etal_2006}         &         &               & & 2.144      &  3.68         & &  1.589    &  3.46              \\    
                 & \citep{SanyasiMishra2008}     &         &               & & 2.0597     &  0.06         & &  1.5308   &  0.21              \\ 
                 & \citep{WuShu2009}             &         &               & & 2.091      &  1.43         & &  1.565    &  1.98              \\
                 & \citep{KalitaRay2009}         &         &               & & 2.0193     &  2.07         & &  1.5145   &  1.29              \\                
                 & \citep{SenKalita2015}         &  2.699  &   5.11        & & 1.949      &  5.75         & &  1.439    &  6.60              \\            
                 & \citep{Chen_etal_2020}        &         &               & & 2.119      &  2.74         & &  1.582    &  3.03              \\
                 & \citep{Kumarkalita2021}       &  2.690  &   5.46        & & 2.160      &  4.58         & &  1.576    &  2.66              \\         
                 & Present                       &  2.837  &               & & 2.061      &               & &  1.534    &                    \\ 
\hline
$\overbar{Nu}$   & \citep{DennisHudsonSmi1968}   &  1.8673 &   0.37        & & 2.5216     &  2.32         & &  3.4317   &  4.48              \\
                 & \citep{LangeDurstBreuer1998}  &  1.8101 &   2.57        & & 2.4087     &  2.26         & &  3.2805   &  0.08              \\
                 & \citep{SparrowAbrahamTong2004}&  1.6026 &   15.86       & & 2.2051     &  11.70        & &  3.0821   &  6.35              \\
                 & \citep{Soares_etal_2006}      &  1.8600 &   0.18        & & 2.4300     &  1.36         & &  3.2000   &  2.43              \\     
                 & \citep{BhartiChhabraEswaran2007}&1.8623 &   0.30        & & 2.4653     &  0.09         & &  3.2825   &  0.14              \\
                 & \citep{Chen_etal_2020}        &  1.8671 &   0.56        & & 2.4718     &  0.35         & &  3.2912   &  0.41              \\
                 & Present                       &  1.8567 &               & & 2.4631     &               & &  3.2778   &                    \\
\hline        
\end{tabular}
\end{table}

A validation study is first carried out for low Reynolds number flow. At the low Reynolds number range $Re\leq 50$, the vortex structure in the wake remains steady and symmetric, thus the temperature field shows similar steady and symmetric characteristics as well. Here we select to work with $Re=10$, 20, 40, and 45. \Cref{fig:P4_Steadycontour} depicts the symmetric vorticity and isotherm contours at steady state for various Reynolds numbers. Similar vorticity and isotherm patterns can be found in various studies available in the literature. 

Present results of vorticity distribution on the cylinder surface for $Re=10$, 20 and 40 and local Nusselt number distribution for $Re=10$, 20 and 45 are portrayed in \cref{fig:P4_SV_all} and \ref{fig:P4_SNu_all} respectively; these values are compared with the previous numerical results from Dennis and Chang \citep{DennisChang1970} and Bharti et al. \citep{BhartiChhabraEswaran2007}. It can be seen that the two sets of data show excellent agreement with each other. The largest value of the local Nusselt number is obtained at the front stagnation point of the cylinder; the value of the Nusselt number then reduces gradually towards the rear stagnation point and attains its lowest value thereat (see \cref{fig:P4_SNu_all}). In \cref{table:P4_steady}, we have compiled our computed values of $L/D$, where $L$ is the length of the recirculation bubbled formed behind the cylinder, the drag coefficient $C_D$ and the average Nusselt number $\overbar{Nu}$ for $Re$=10, 20 and 40. As can be seen from the figures and table, the length of the recirculation region and the average Nusselt number depend directly on the value of the Reynolds number, while the drag coefficient varies inversely. A quantitative comparison is also carried out in \cref{table:P4_steady}. Computational results are compared with reference data \citep{DennisHudsonSmi1968,HeDoolen1997,LangeDurstBreuer1998,SparrowAbrahamTong2004,Niu_etal_2006,Soares_etal_2006,BhartiChhabraEswaran2007,SanyasiMishra2008,KalitaRay2009,WuShu2009,SenKalita2015,Chen_etal_2020,Kumarkalita2021}. Once again, the efficiency of the present scheme is evident from the proximity of our results with the well-established results available in the literature.

\begin{figure}[!h]
\centering
        \begin{subfigure}[b]{0.44\textwidth}
                \centering
                \includegraphics[width=1.10\linewidth]{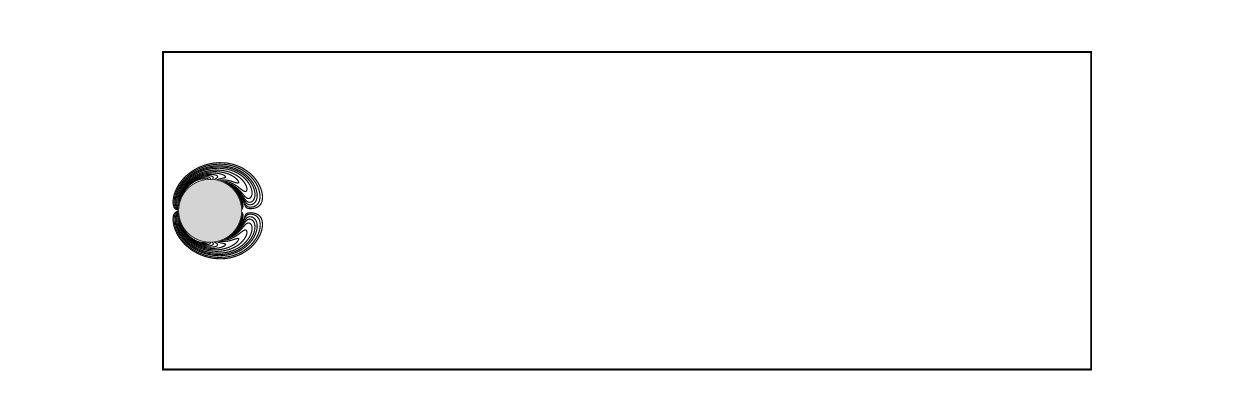}
        \end{subfigure}%
        \begin{subfigure}[b]{0.44\textwidth}
                \centering
                \includegraphics[width=1.10\linewidth]{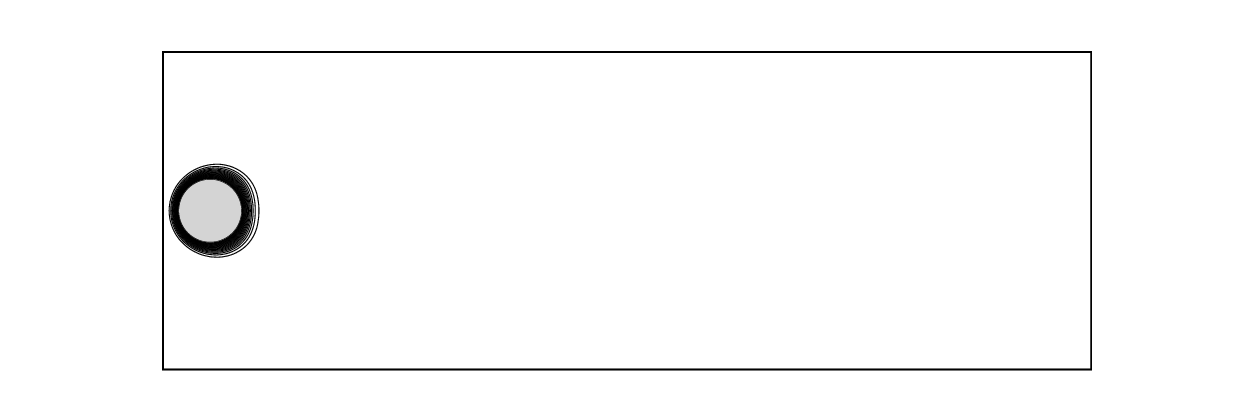}
        \end{subfigure}%
        \begin{minipage}[b]{0.05\textwidth}
        \subcaption{ }\label{fig:P4_100t05}
        \end{minipage}
        \\
        \begin{subfigure}[b]{0.44\textwidth}
                \centering
                \includegraphics[width=1.10\linewidth]{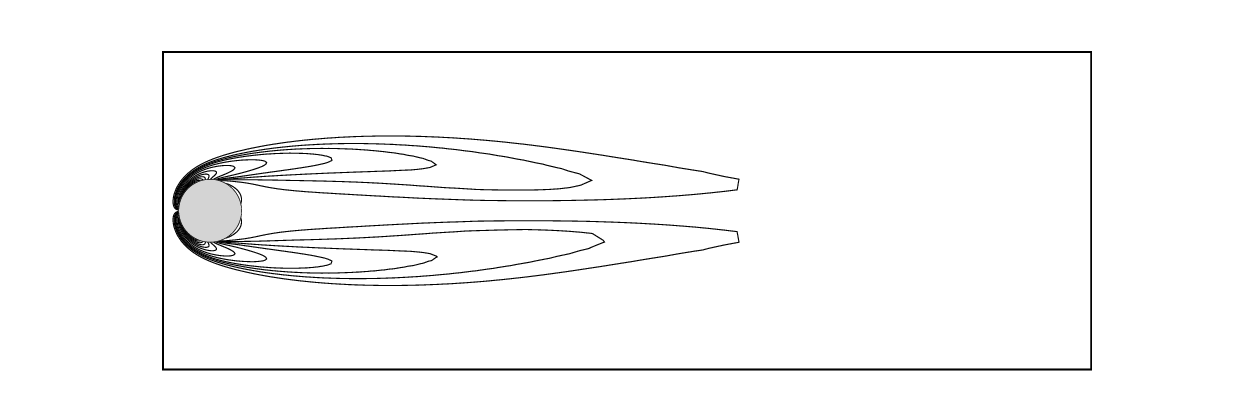}
        \end{subfigure}%
        \begin{subfigure}[b]{0.44\textwidth}
                \centering
                \includegraphics[width=1.10\linewidth]{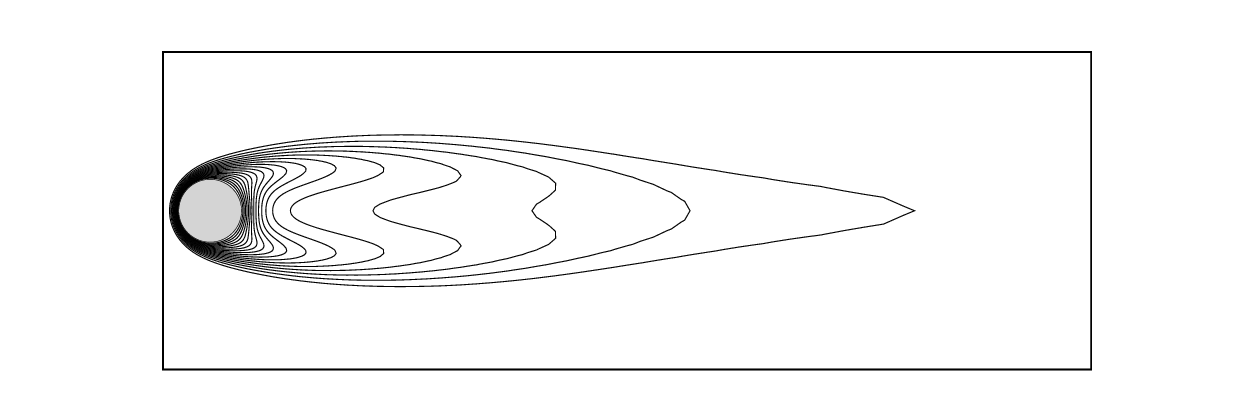}
        \end{subfigure}%
        \begin{minipage}[b]{0.05\textwidth}
        \subcaption{ }\label{fig:P4_100t25}
        \end{minipage}
        \\
        \begin{subfigure}[b]{0.44\textwidth}
                \centering
                \includegraphics[width=1.10\linewidth]{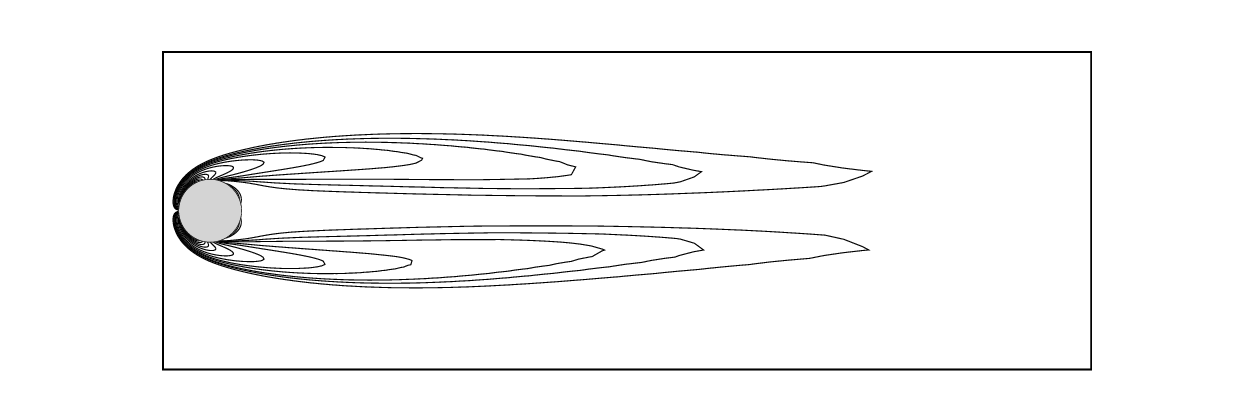}
        \end{subfigure}%
        \begin{subfigure}[b]{0.44\textwidth}
                \centering
                \includegraphics[width=1.10\linewidth]{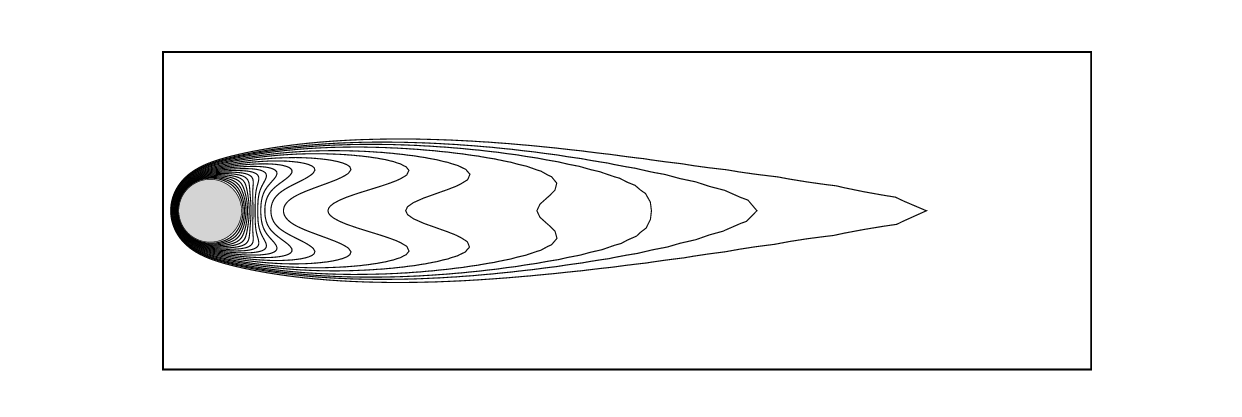}
        \end{subfigure}%
        \begin{minipage}[b]{0.05\textwidth}
        \subcaption{ }\label{fig:P4_100t50}
        \end{minipage}
        \\
        \begin{subfigure}[b]{0.44\textwidth}
                \centering
                \includegraphics[width=1.10\linewidth]{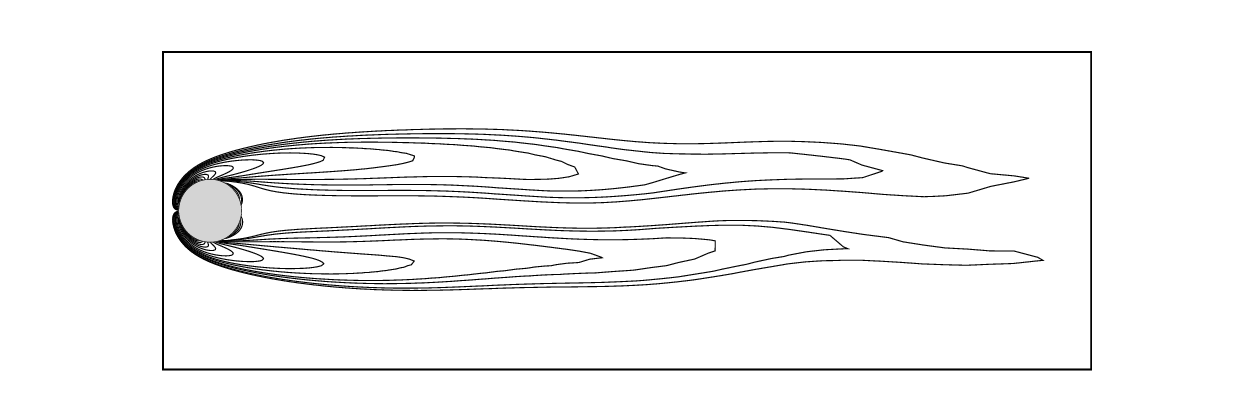}
        \end{subfigure}%
        \begin{subfigure}[b]{0.44\textwidth}
                \centering
                \includegraphics[width=1.10\linewidth]{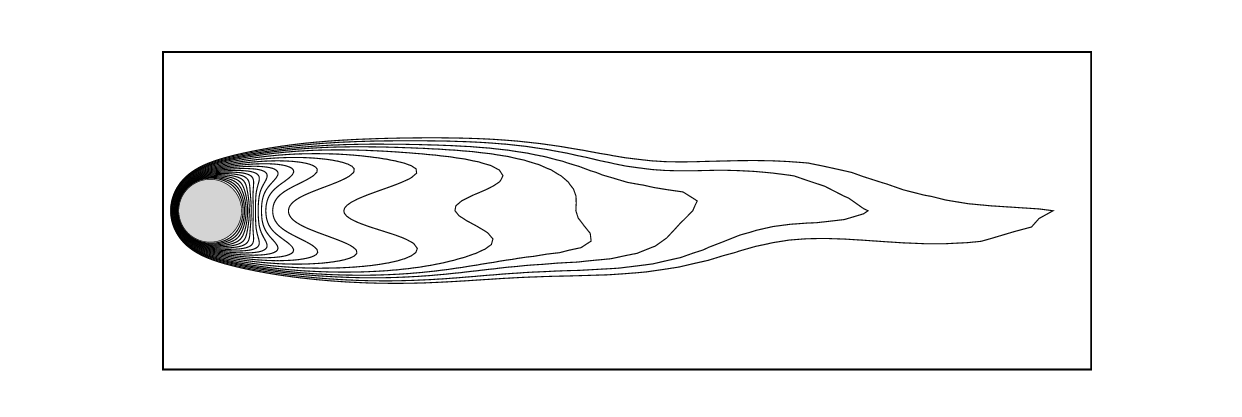}
        \end{subfigure}%
        \begin{minipage}[b]{0.05\textwidth}
        \subcaption{ }\label{fig:P4_100t75}
        \end{minipage}
        \\
        \begin{subfigure}[b]{0.44\textwidth}
                \centering
                \includegraphics[width=1.10\linewidth]{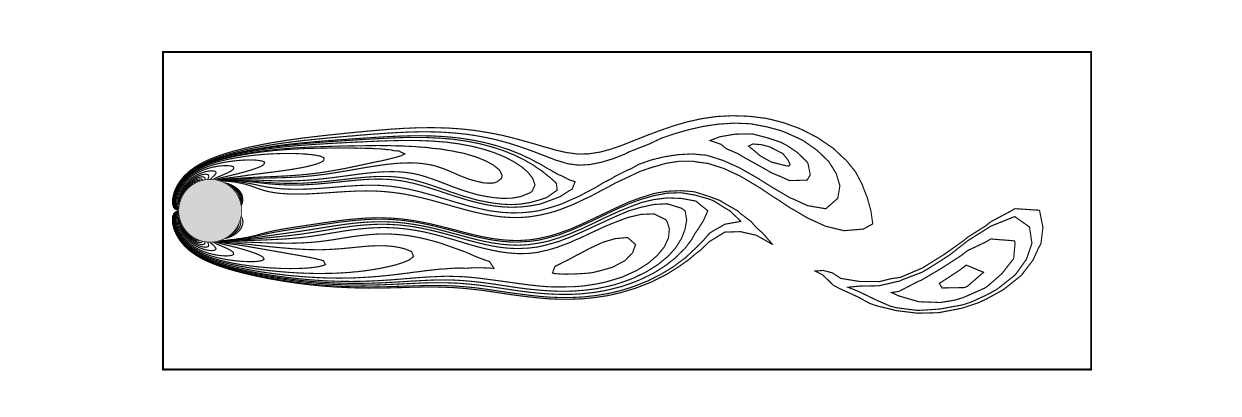}
        \end{subfigure}%
        \begin{subfigure}[b]{0.44\textwidth}
                \centering
                \includegraphics[width=1.10\linewidth]{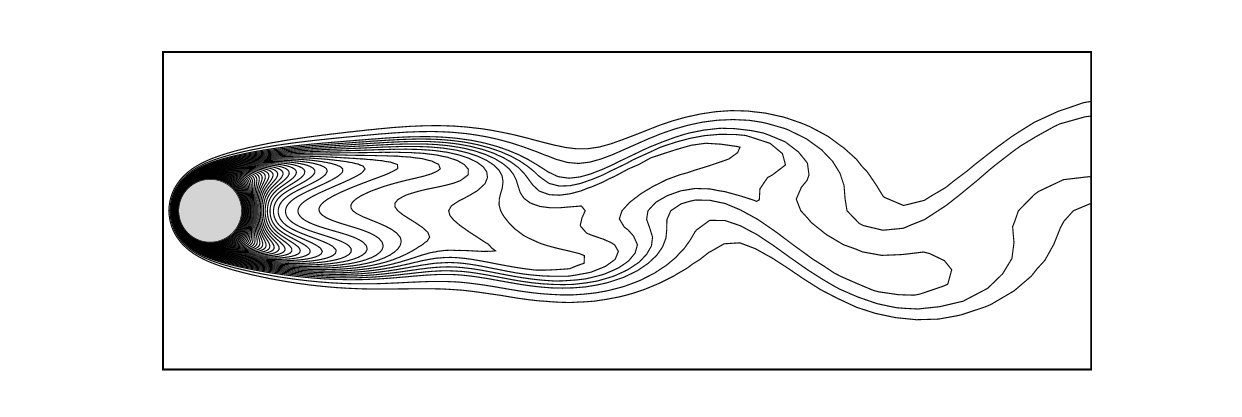}
        \end{subfigure}%
        \begin{minipage}[b]{0.05\textwidth}
        \subcaption{ }\label{fig:P4_100t90}
        \end{minipage}
        \\
        \begin{subfigure}[b]{0.44\textwidth}
                \centering
                \includegraphics[width=1.10\linewidth]{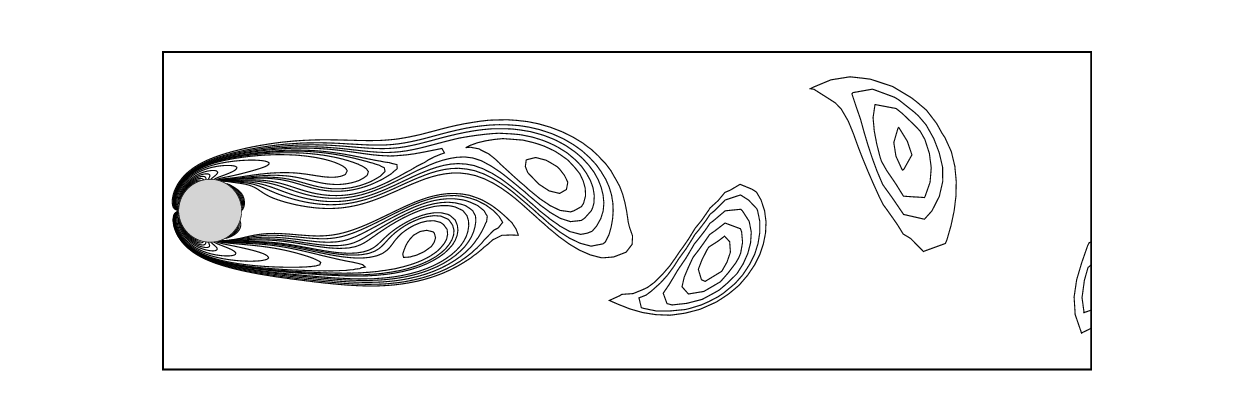}
        \end{subfigure}%
        \begin{subfigure}[b]{0.44\textwidth}
                \centering
                \includegraphics[width=1.10\linewidth]{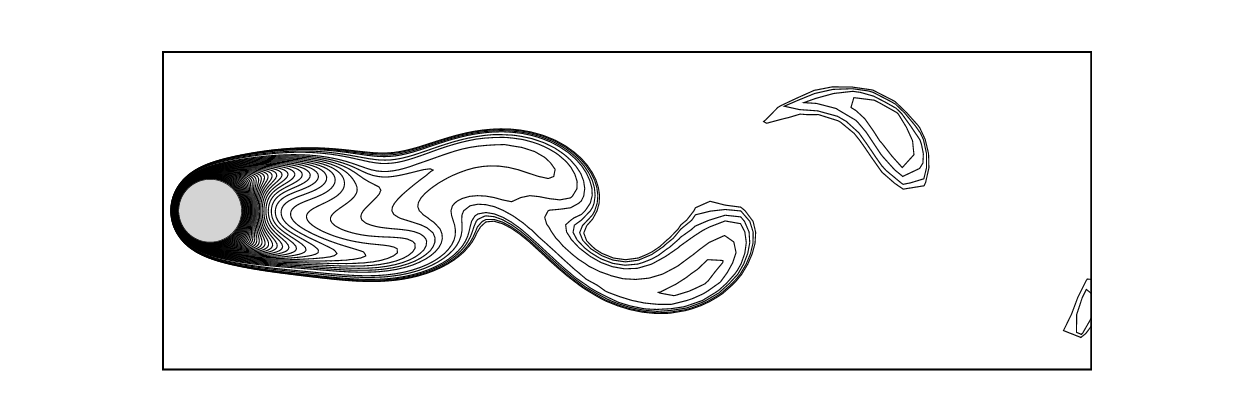}
        \end{subfigure}%
        \begin{minipage}[b]{0.05\textwidth}
        \subcaption{ }\label{fig:P4_100t100}
        \end{minipage}
        \\
        \begin{subfigure}[b]{0.44\textwidth}
                \centering
                \includegraphics[width=1.10\linewidth]{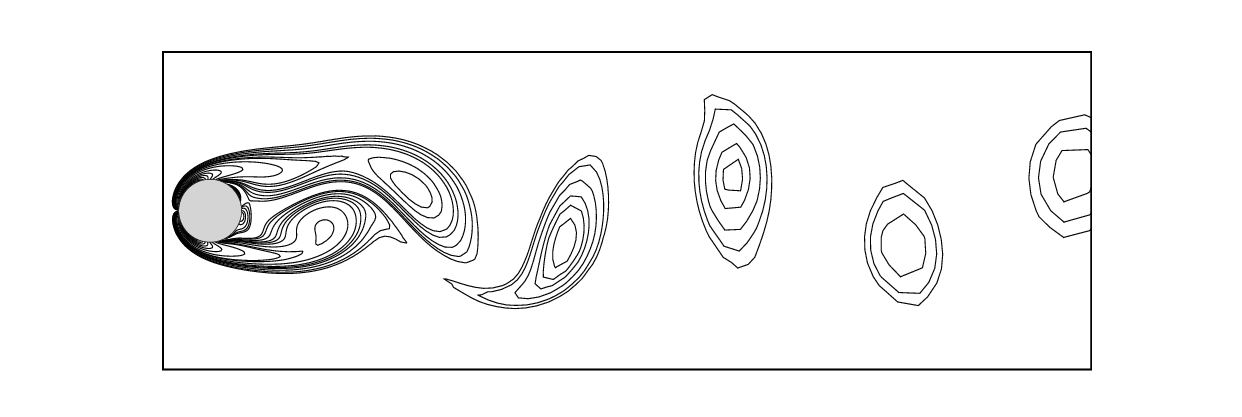}
        \end{subfigure}%
        \begin{subfigure}[b]{0.44\textwidth}
                \centering
                \includegraphics[width=1.10\linewidth]{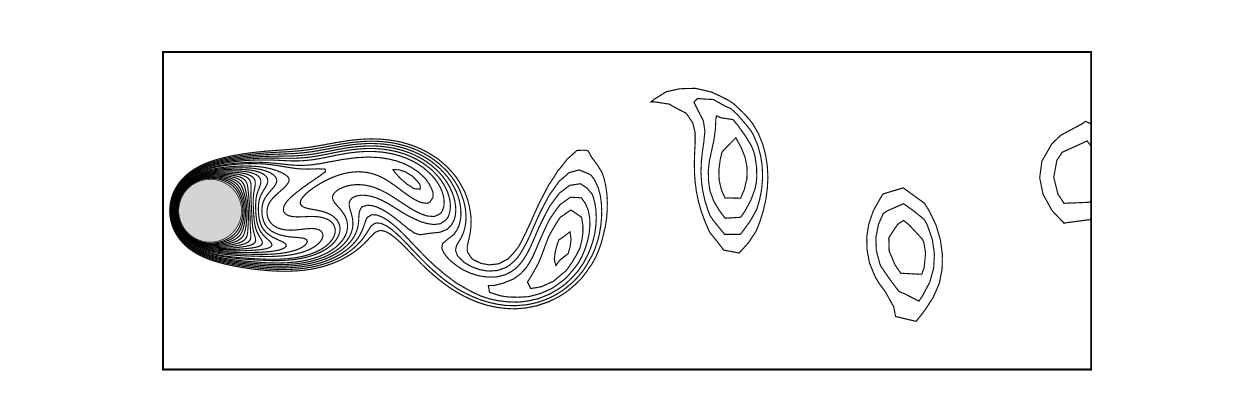}
        \end{subfigure}%
        \begin{minipage}[b]{0.05\textwidth}
        \subcaption{ }\label{fig:P4_100t250}
        \end{minipage}
\caption{Problem 4: Evolution of vorticity (left) and isotherms  (right) for $Re=100$ at (a) $t=0.5$, (b) $t=25$, (c) $t=50$, (d) $t=75$, (e) $t=90$, (f) $t=100$, (g) $t=250$.}\label{fig:P4_VortTemp100}
\end{figure}

\begin{figure}[!h]
\centering
\begin{subfigure}[b]{0.44\textwidth}
    \centering
    \includegraphics[width=\linewidth]{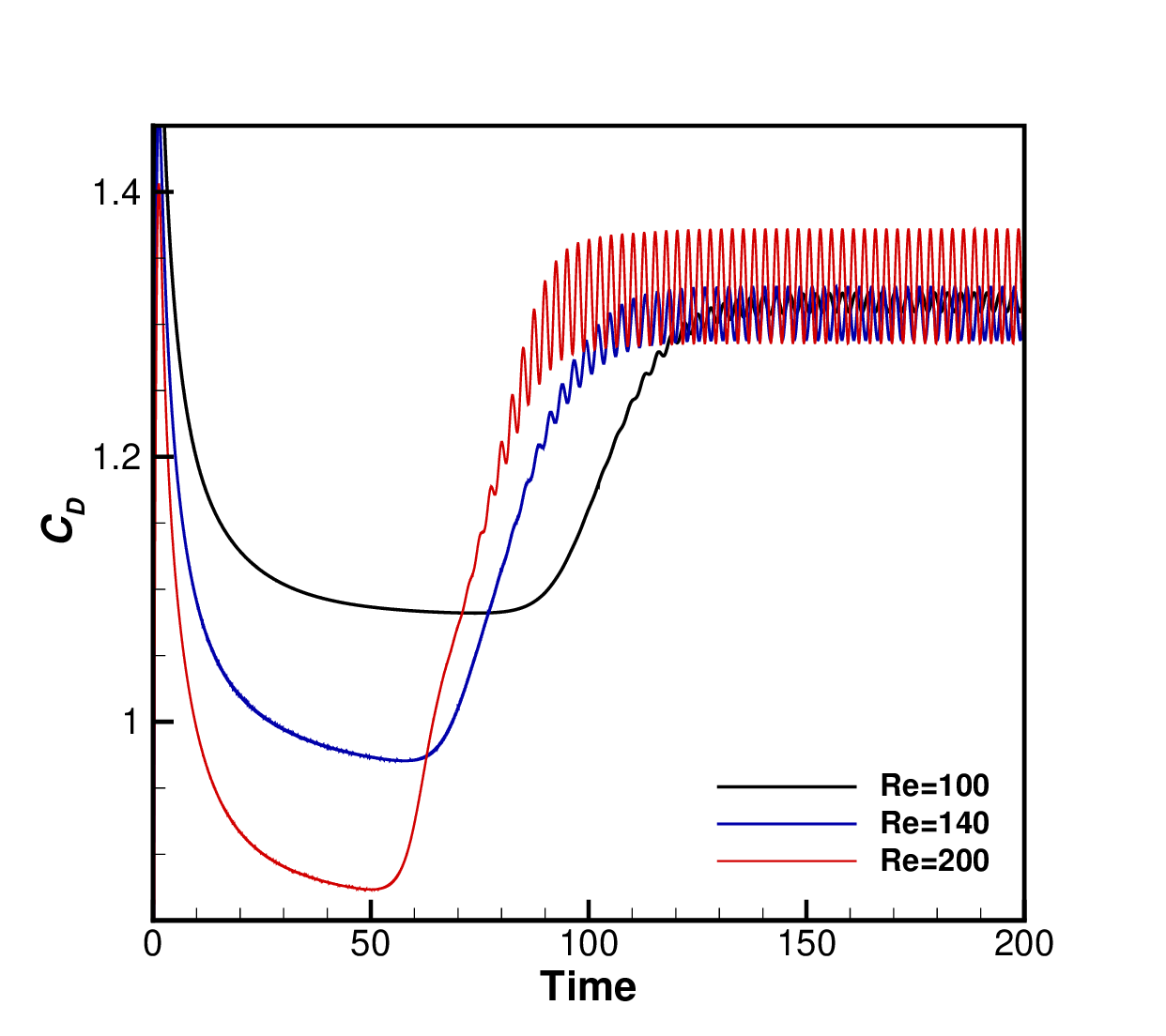}
\end{subfigure}%
\begin{minipage}[b]{0.05\textwidth}
\subcaption{ }\label{fig:P4_Cd_all}
\end{minipage}
\begin{subfigure}[b]{0.44\textwidth}
    \centering
    \includegraphics[width=\linewidth]{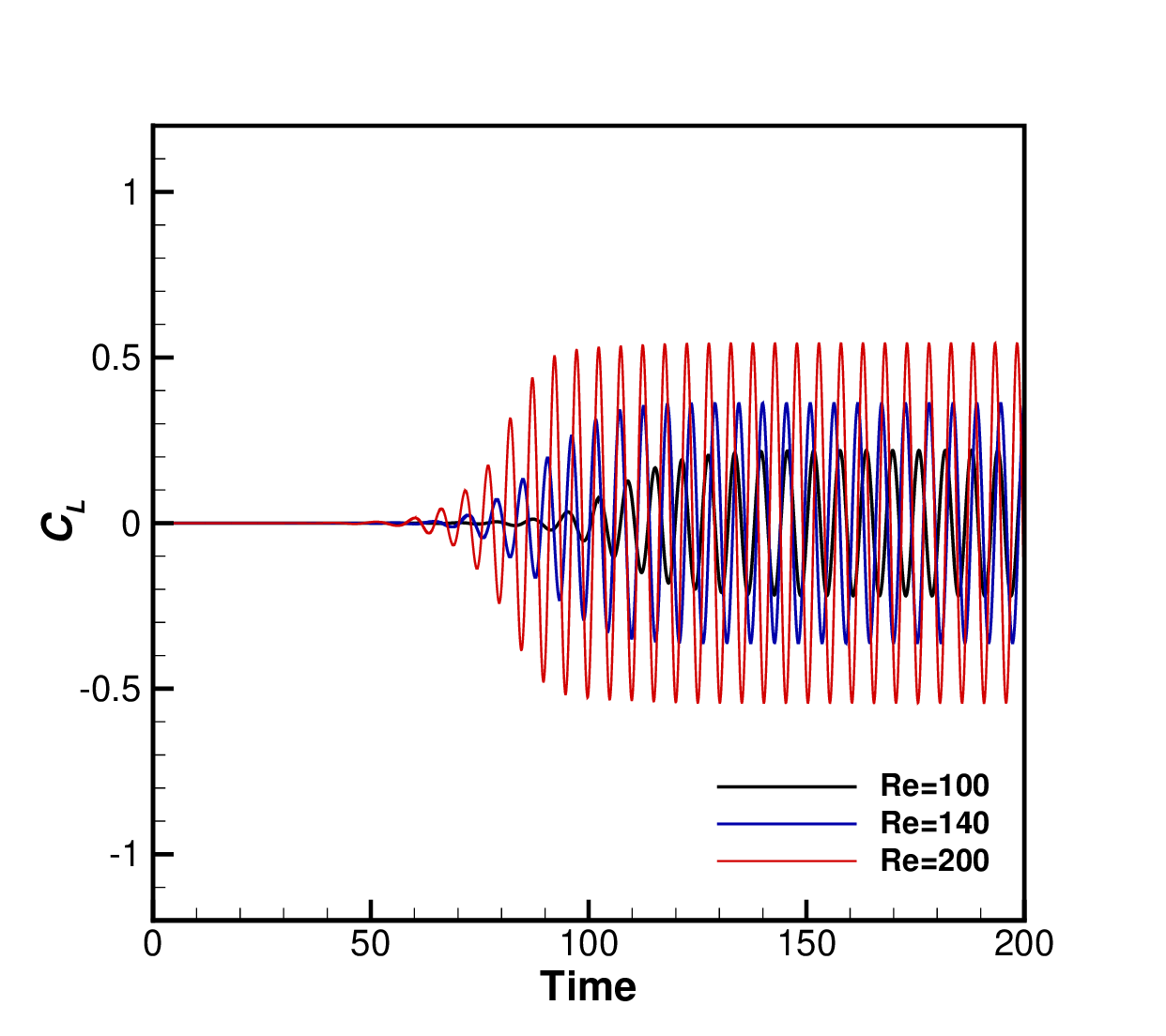}
\end{subfigure}%
\begin{minipage}[b]{0.05\textwidth}
\subcaption{ }\label{fig:P4_Cl_all}
\end{minipage}
\begin{subfigure}[b]{0.44\textwidth}
    \centering
    \begin{tikzpicture}
  \draw (0,0) node[inner sep=0] {\includegraphics[width=\linewidth]{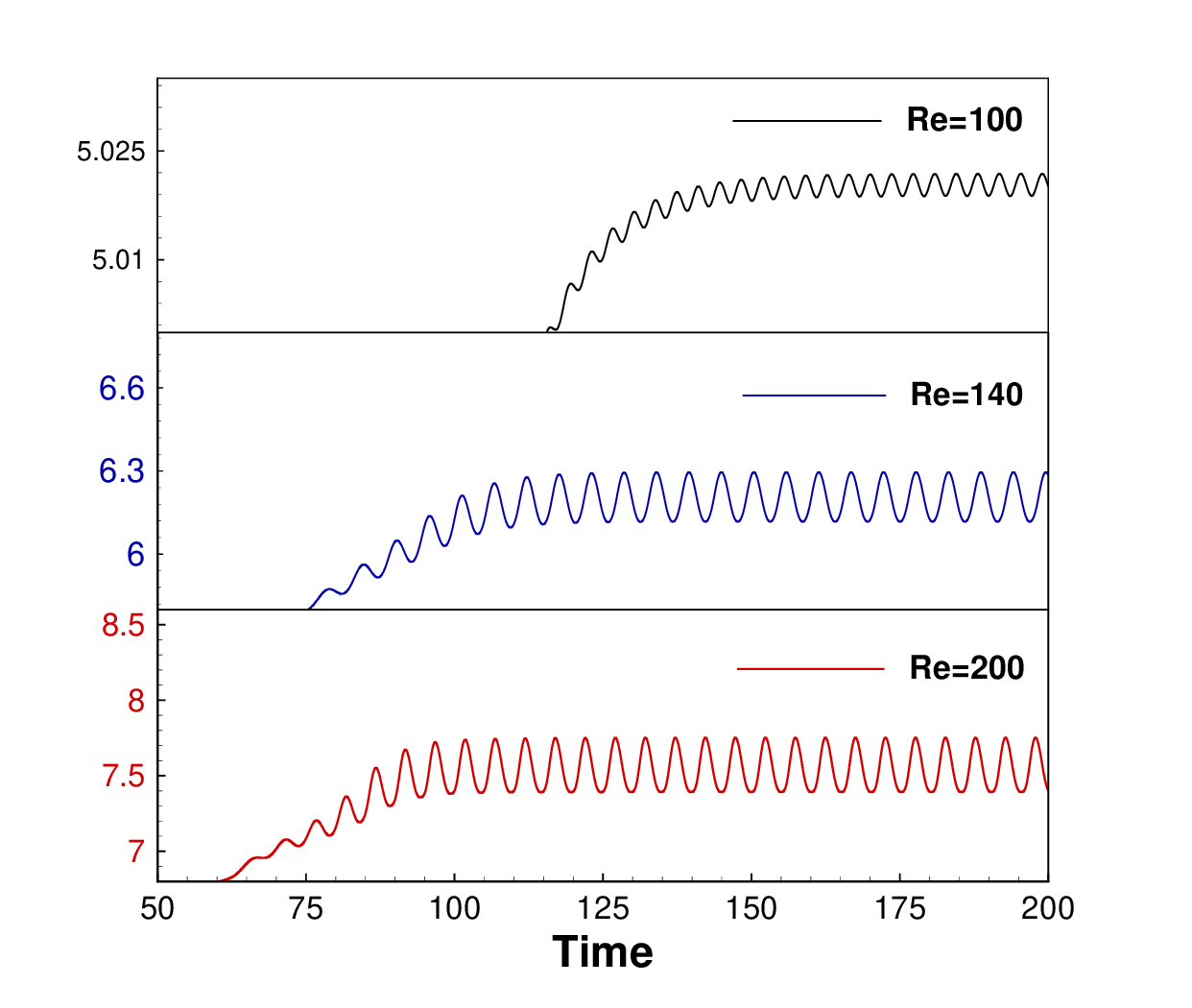}};
  \draw (-3.35,0) node {\rotatebox[origin=c]{90}{\footnotesize \textbf{$\overbar{Nu}$}}};
  \end{tikzpicture}
\end{subfigure}%
\begin{minipage}[b]{0.05\textwidth}
\subcaption{ }\label{fig:P4_Nu_all}
\end{minipage}
\begin{subfigure}[b]{0.44\textwidth}
    \centering
    \includegraphics[width=\linewidth]{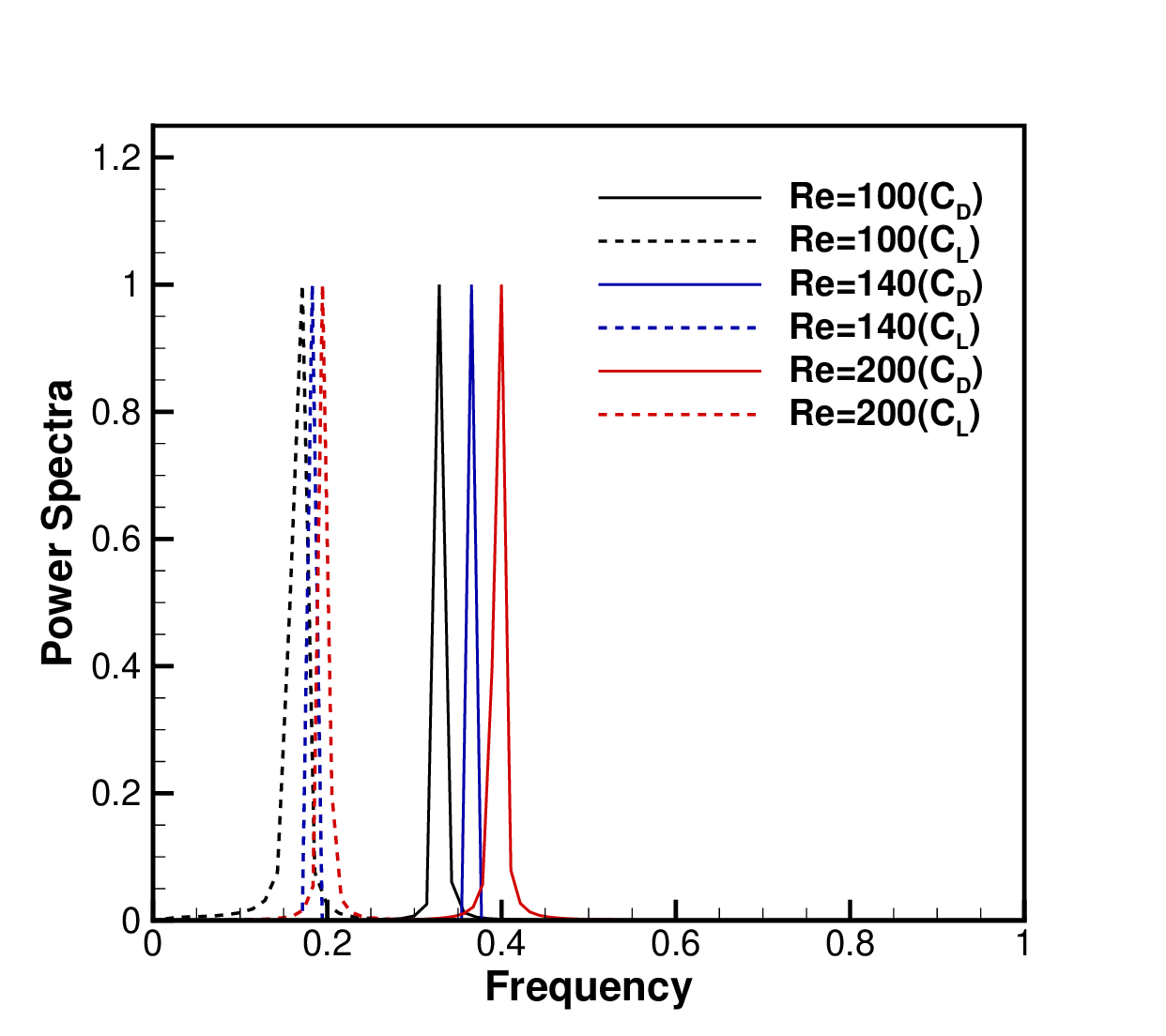}
\end{subfigure}%
\begin{minipage}[b]{0.05\textwidth}
\subcaption{ }\label{fig:P4_Power}
\end{minipage}
\caption{Problem 4: Temporal evolution of (a) drag coefficient, (b) lift coefficient, (c) average Nusselt number at the cylinder surface, (d) power spectra of drag and lift coefficient for $Re=100$, 140 and 200.} \label{fig:P4_CdClNu}
\end{figure}

Next, we are interested in studying the periodic flow for $Re$=100, 140, and 200. The transition of the vorticity and isotherm contours from steady to periodic state for $Re$=100 is depicted in \cref{fig:P4_VortTemp100}. This may be considered as the representation of the flow for the other $Re$ values considered, although each requires a different time to attain periodicity. As can be seen in \cref{fig:P4_100t05}, no vortex structure formed in the flow field, and the heated cylinder generates a thermal boundary layer near its surface at the beginning. As time marches, two opposite symmetrical vortices are formed simultaneously which grow in width and length with time, and remain symmetrically stable for a certain period as shown in \cref{fig:P4_100t25,fig:P4_100t50}. During this period, the thickness of the thermal boundary layer increases; the temperature distribution also remains symmetric about the line $\theta=0$. After a certain point of time, a fluctuation develops in the flow which destroys the symmetry in the vortices and temperature distribution and they start to oscillate downstream which can be noticed in \cref{fig:P4_100t75}. As the flow develops further, the vortices at the rear of the cylinder grow in size and ultimately are shed from the cylinder towards downstream while heat is carried away. It is heartening to see that the present scheme could capture the characteristic feature of the periodic flow, the so-called von K\'arm\'an vortex street very efficiently. With the vortex shedding taking place, isolated hot fluid clusters are observed in the flow (see \cref{fig:P4_100t90}). Subsequently, \cref{fig:P4_100t100,fig:P4_100t250} indicate the synchronized variation of flow and temperature fields, where vortex shedding phenomena play a determinant role in the heat transfer at the downstream.

\begin{figure}[!h]
\centering
  \begin{subfigure}[b]{0.44\textwidth}
  \centering
  \includegraphics[width=\linewidth]{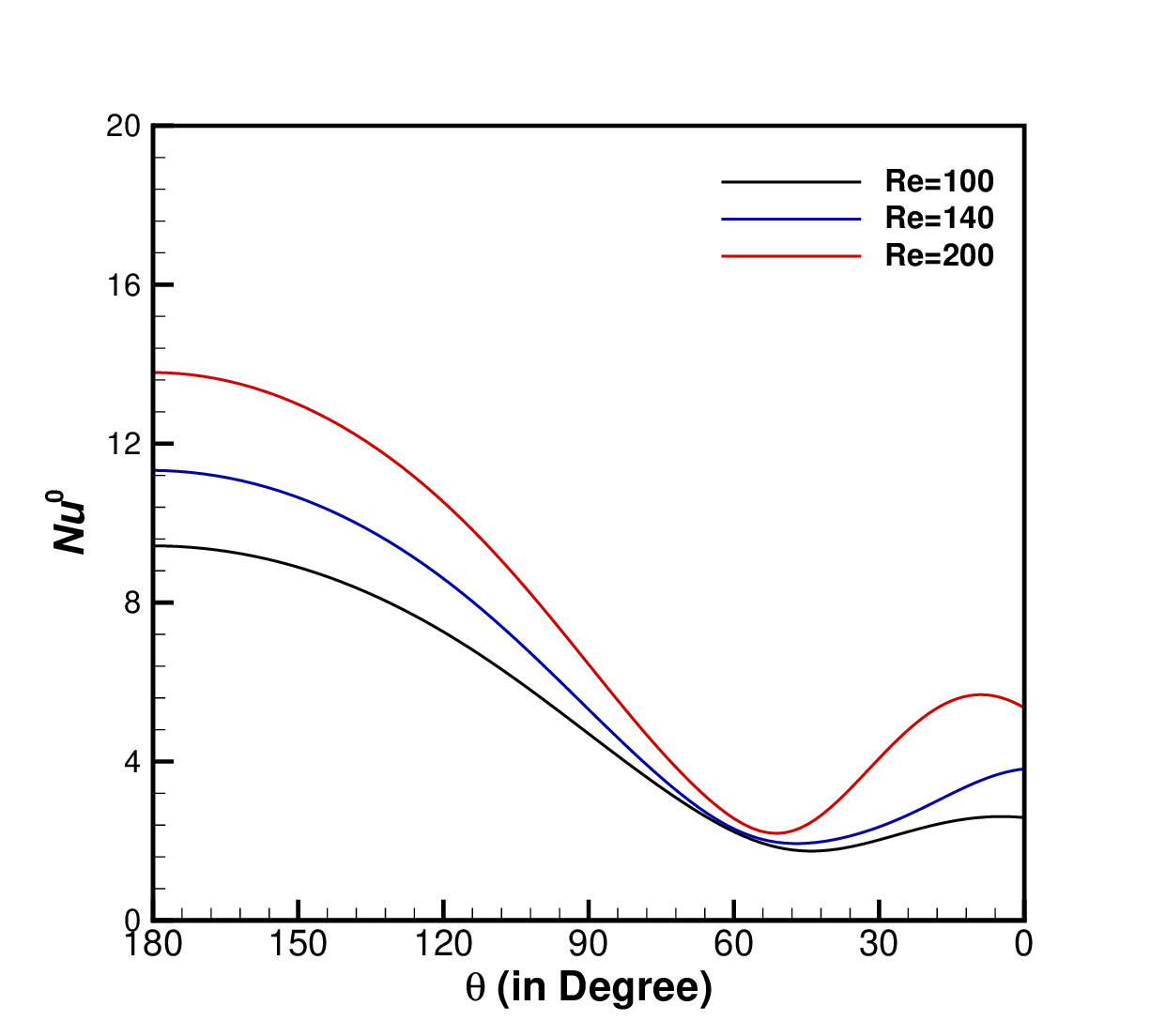}
  \end{subfigure}
  \begin{minipage}[b]{0.02\textwidth}
  \subcaption{ }\label{fig:P4_LocNuall}
  \end{minipage}
  \begin{subfigure}[b]{0.44\textwidth}
  \centering
  \includegraphics[width=\linewidth]{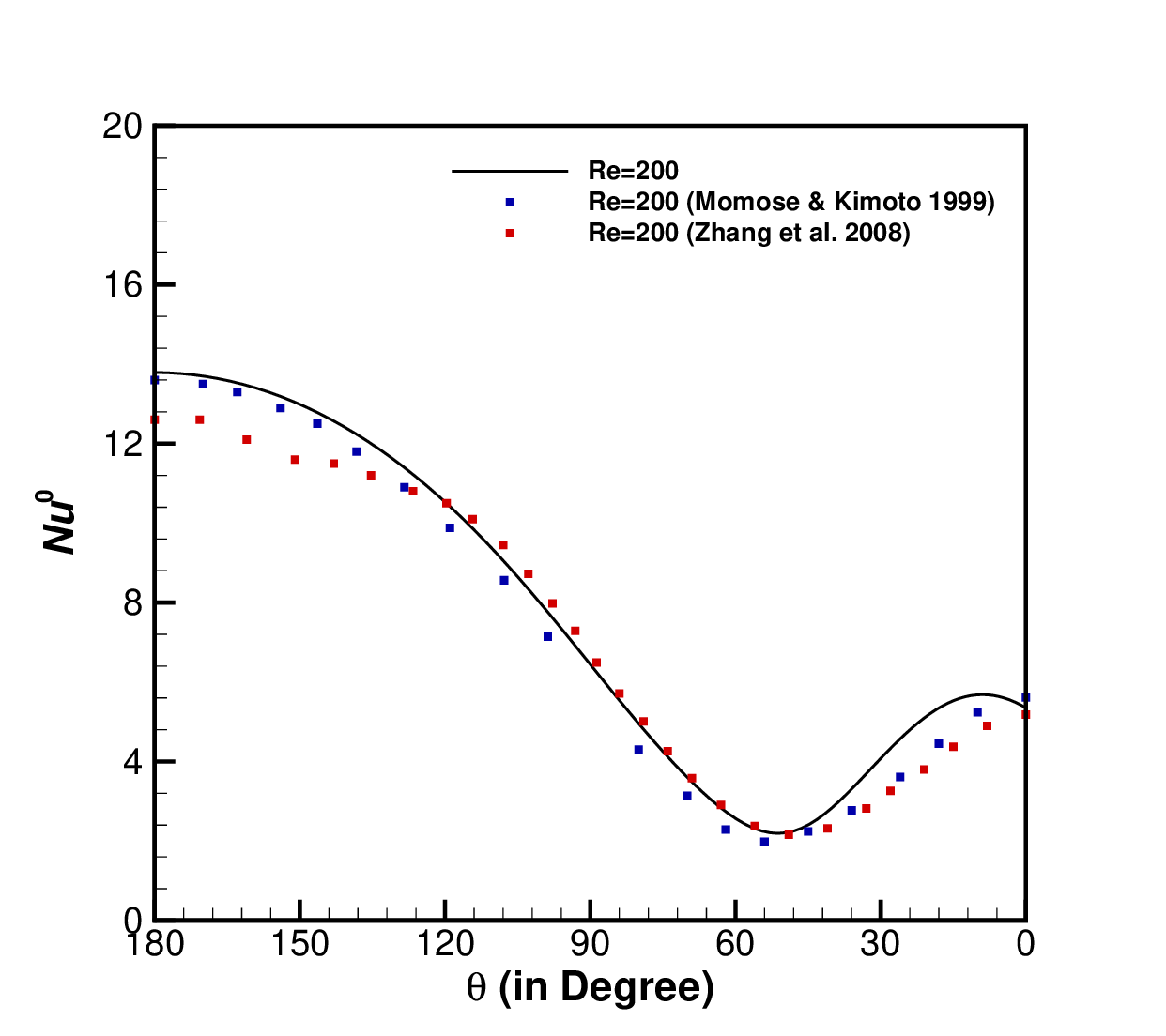}
  \end{subfigure}
  \begin{minipage}[b]{0.02\textwidth}
  \subcaption{ }\label{fig:P4_LocNu200}
  \end{minipage}
\caption{Problem 4: (a) Distribution of local Nusselt number on the cylinder surface for $Re=100$, 140, 200 and (b) comparison of local Nusselt number distribution over the cylinder surface for $Re=200$ with results from \citep{MomoseKimoto1999,Zhang_etal_2008}.}\label{fig:P4_LNusselt}
\end{figure}

The time history of the drag coefficient $(C_D)$, lift coefficient $(C_L)$ and average Nusselt number $(\overbar{Nu})$ displayed in \cref{fig:P4_Cd_all,fig:P4_Cl_all,fig:P4_Nu_all} pertinently implies that the newly developed scheme has accurately captured the periodic state for all the Reynolds numbers under consideration. The periodicity is further justified by the single dominating peak for $C_D$ and $C_L$ in the spectral density analysis (see \cref{fig:P4_Power}). The distribution of local Nusselt number $(Nu^0)$ over the surface of the cylinder for different $Re$ values are represented in \cref{fig:P4_LocNuall}. In \cref{fig:P4_LocNu200} we carry out a qualitative comparison between the surface distribution of $Nu^0$ for $Re$=200 computed in the present study with those of \citep{MomoseKimoto1999,Zhang_etal_2008}; which reveals close agreement of our results with those taken from the literature. We then compare our numerical values of the flow parameters $C_D$, $C_L$, $St$ and $\overbar{Nu}$ with existing studies in \cref{table:P4_periodic_flow}. The streakline from our simulation depicting the well-known von K\'arm\'an vortex street for $Re$=140 is presented in \cref{{fig:P4_streak140}} along with the streakline reported in the experimental work of Taneda \citep{Taneda1979}. From the comparison we note that vortex creation and dissipation as captured in the experimental work is effectively computed by the current formulation. Striking similarity between our numerical results and the existing numerical and experimental results for all the cases indeed verify and validate the scheme. 

\begin{table}[!h]
\caption{Problem 4: Comparison flow parameters $C_D$, $C_L$, $St$ and $\overbar{Nu}$ for different $Re$.}\label{table:P4_periodic_flow}
\vspace{0.1cm}
\renewcommand*{\arraystretch}{0.8}
\begin{tabular}
{ M{0.075\textwidth} M{0.095\textwidth} M{0.175\textwidth} M{0.05\textwidth} M{0.175\textwidth}  M{0.05\textwidth} M{0.175\textwidth}}
\hline
                 &                               & $Re$=100           & &  $Re$=140         & &  $Re$=200          \\
\hline
$C_D$            & \citep{Leetal2006}            & 1.37 $\pm$ 0.009   & &                   & & 1.34  $\pm$ 0.030  \\
                 & \citep{BerFal2008}            & 1.38 $\pm$ 0.010   & &                   & & 1.37  $\pm$ 0.046  \\
                 & \citep{SenThesis}             & 1.394 $\pm$ 0.007  & &                   & & 1.357 $\pm$ 0.038  \\
                 & \citep{KumarKalita2019}       & 1.325 $\pm$ 0.026  & &                   & & 1.333 $\pm$ 0.046  \\
                 & Present                       & 1.317 $\pm$ 0.008  & & 1.308 $\pm$ 0.020 & & 1.329 $\pm$ 0.043  \\ 
\hline
$C_L$            & \citep{Leetal2006}            &     $\pm$ 0.323    & &                   & &     $\pm$ 0.430    \\
                 & \citep{BerFal2008}            &     $\pm$ 0.340    & &                   & &     $\pm$ 0.700    \\
                 & \citep{SenThesis}             &     $\pm$ 0.191    & &                   & &     $\pm$ 0.453    \\
                 & \citep{KumarKalita2019}       &     $\pm$ 0.306    & &                   & &     $\pm$ 0.351    \\
                 & Present                       &     $\pm$ 0.22     & &    $\pm$ 0.361    & &     $\pm$ 0.543    \\
\hline
$St$             & \citep{Leetal2006}            &       0.160        & &                   & &        0.187       \\
                 & \citep{BerFal2008}            &       0.169        & &                   & &        0.200       \\
                 & \citep{SenThesis}             &       0.165        & &                   & &        0.197       \\
                 & \citep{KumarKalita2019}       &       0.162        & &                   & &        0.200       \\
                 & Present                       &       0.172        & &      0.183        & &        0.195       \\
\hline
$\overbar{Nu}$   & \citep{ChurchillBernstein1977}&        5.12        & &       5.87        & &        7.15        \\
                 & \citep{ChengHong1997}         &        5.26        & &                   & &        7.67        \\
                 & \citep{Zhang_etal_2008}       &                    & &                   & &        7.23        \\
                 & \cite{Cao_etal_2021}          &        5.07        & &       6.08        & &                    \\     
                 & Present                       &        5.02        & &       6.20        & &        7.55        \\
\hline
\end{tabular}
\end{table}

\begin{figure}[!h]
\centering
    \includegraphics[width=\linewidth]{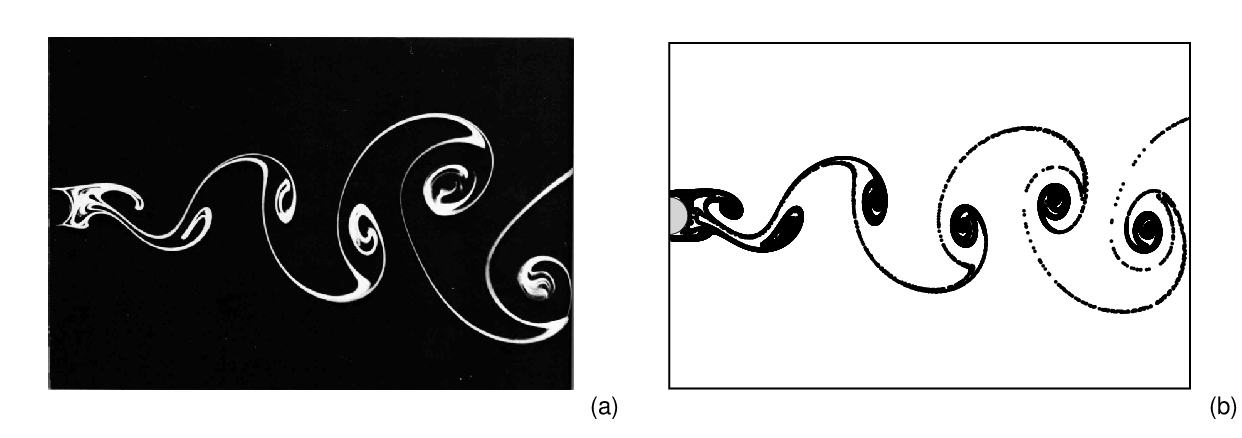}
\caption{Problem 4: Comparison between the instantaneous streaklines for $Re$=140 captured in the (a) experimental study of Taneda \citep{Taneda1979} and (b) present computation.}\label{fig:P4_streak140}
\end{figure}

\section{Conclusion}\label{conclusion}
In this work, we extend the philosophy of recent work on 2D transient convection and diffusion equation on Cartesian coordinates \citep{DekaSen2021} to polar coordinates to tackle non-rectangular geometries. The transformation-free higher-order compact finite difference scheme proposed here can be applied directly on polar grids, unlike most of the schemes in the literature on polar grids that use transformation between physical and computational domains. We have also combined the virtue of nonuniformity with the present scheme. This levitates the efficiency and robustness of the scheme as it acquires the advantage of accumulating and scattering grid points as necessary. The scheme is theoretically third-order accurate in space and second-order accurate in time. It is worthwhile to mention that for linear problems with analytical solution the scheme exhibits a spatial convergence of order four, which is higher than the theoretical value. The robustness of the scheme is examined by applying it to as many as three benchmark problems of fluid flow and heat flow, \textit{viz.} driven polar cavity, natural convection in a horizontal concentric annulus, and forced convection around a heated stationary cylinder. In order to resolve the Neumann type boundary conditions, we have also introduced one-sided approximations for first order derivatives. For the problem of forced convection around a heated stationary cylinder, both steady and periodic solutions are being accurately captured by the present scheme. Additionally, important aspects such as the von K\'arm\'an vortex phenomenon and the effect of vortex shedding on heat transfer are also studied comprehensively for this fluid body interaction problem. The accuracy of the computed solutions is estimated from the results obtained in all the test problems, which are in excellent agreement with the existing results both qualitatively and quantitatively. We have reported the perceived order of convergence for all the variables in flow problems that are not supplemented with analytical solutions. It is heartening to see that our solutions could attain the theoretical order of convergence in both space and time in all cases. 

\section*{Acknowledgement}
The second author is thankful to Science \& Engineering Research Board, India for assistance under Core Research Grant (Project File Number: CRG/2022/000668).

\bibliographystyle{unsrtnat}
\bibliography{dharmaraj}
\newpage

\end{document}